\documentclass[10pt,aps,showpacs,nofootinbib,notitlepage]{revtex4}
\usepackage{epsfig}
\usepackage{amsmath}
\usepackage{verbatim}
\usepackage{psfrag}
\usepackage{bm}
\usepackage{bbm}
\usepackage{graphicx}
\usepackage{latexsym}
\usepackage[english]{babel}
\usepackage{amsthm}
\usepackage{color}
\usepackage{amssymb}

\newcounter{fig}

\begin{document}

%\begin{titlepage}

\begin{flushright} {Date: \today\\} \end{flushright}
\rightline{ITP-UU-12/50, SPIN-12/47, NIKHEF-2012-027}

\vskip .5cm

\title{\Large\bf An exact solution of the Dirac equation with CP violation}

\author{Tomislav Prokopec$^{{\rm a}*}$, Michael G. Schmidt$^{{\rm b}*}$
              and Jan Weenink$^{\rm a,c}$}
\email[]{T.Prokopec@uu.nl; M.G.Schmidt@thphys.uni-heidelberg.de; J.G.Weenink@uu.nl}
\affiliation{$^{\rm a}$ Institute for Theoretical
Physics (ITP) \& Spinoza Institute, Utrecht University, Postbus
80195, 3508 TD Utrecht, The Netherlands}

\affiliation{$^{\rm b}$
Institut f\"ur Theoretische Physik, Heidelberg University,
Philosophenweg 16, D-69120 Heidelberg, Germany}

\affiliation{$^{\rm c}$
Nikhef, Science Park Amsterdam 105,
1098 XG Amsterdam, The Netherlands}

\begin{abstract}
We consider Yukawa theory in which the fermion mass is induced by a Higgs like scalar.
In our model the fermion mass exhibits a temporal dependence,
which naturally occurs in the early Universe setting.
Assuming that the complex fermion mass changes as a tanh-kink,
we construct an exact, helicity conserving, CP-violating
solution for the positive and negative frequency fermionic mode functions,
which is valid both in the case of weak and strong CP violation.
Using this solution we then study the fermionic currents
both in the initial vacuum and finite density/temperature setting.
Our result shows that, due to a potentially large state squeezing, fermionic currents
can exhibit a large oscillatory magnification.
Having in mind applications to electroweak baryogenesis,
we then compare our exact results with those obtained in a gradient
approximation.
Even though the gradient approximation does not capture the oscillatory effects
of squeezing, it describes quite well the averaged current,
obtained by performing a mode sum.
Our main conclusion is: while the agreement with the semiclassical force is
quite good in the thick wall regime,
the difference is sufficiently significant to motivate a more
detailed quantitative study of baryogenesis sources in the thin wall regime
in more realistic settings.

\end{abstract}


\maketitle

\section{Introduction}
\label{Introduction}

 Electroweak baryogenesis~\cite{Morrissey:2012db}
is a very appealing idea, and yet the mechanism for
dynamical baryon creation at the electroweak scale has suffered some serious blows.
Firstly, in the mid 90s it was found that
the electroweak phase transition in the standard model is
a crossover~\cite{Kajantie:1996qd,Rummukainen:1998as,Csikor:1998eu}.
While at first supersymmetric extensions looked promising,
the most popular supersymmetric model - the MSSM - is almost ruled out on two grounds
(a) it cannot give a strong enough phase transition for the observed
Higgs mass~\cite{Carena:2008vj} and
(b) it cannot produce enough baryons consistent with electric dipole
moment~\cite{Pospelov:2005pr}
bounds~\cite{Carena:2012np,Cirigliano:2009yd,Blum:2008ym,Konstandin:2005cd,Cline:2000nw,Huet:1995sh}
(albeit in some models resonance between fermionic flavors
can be helpful to increase baryon
production~\cite{Kozaczuk:2012xv,Cirigliano:2011di,Cirigliano:2009yd,Konstandin:2004gy}).
The models that are still viable are the supersymmetric models with additional
Higgs singlet(s)~\cite{Huber:2000mg,Kang:2004pp}
both because they allow for
a stronger phase transition~\cite{Carena:2011jy,Konstandin:2006nd,Huber:1998ck}
and generate more baryons~\cite{Cline:2012hg,Huber:2006ma,Huber:2006wf,Balazs:2007pf,Menon:2004wv}.
In addition, general two Higgs doublet models~\cite{Cline:2011mm,Fromme:2006cm}
and composite Higgs models~\cite{Espinosa:2011eu,Konstandin:2011ds,Cline:2008hr}
are still viable. Works on cold electroweak
baryogenesis~\cite{Tranberg:2012jp,Tranberg:2012qu,Tranberg:2009de}
are also worth mentioning.
In summary, while electroweak
baryogenesis has been a very attractive proposal,
precisely because it is testable by contemporary accelerators,
recent experiments have cornered it to models where most researchers have
not focused their attention during the pre-LHC era.
Hence, at this stage
theoretical work, that will refine our ability to make a quantitative assessment
of electroweak baryogenesis in different models, is still a worthy pursuit.

 One of the most important unsolved problems in dynamical modeling of electroweak
baryogenesis is a reliable calculation of the CP-violating sources that bias
sphaleron transitions~\cite{Manton:1983nd,Klinkhamer:1984di},
which at high temperatures violate baryon number.
In the fermionic sector the most prominent CP-violating source is the fermionic
axial vector current~\cite{Prokopec:2003pj,Prokopec:2004ic},
since that current directly couples to sphalerons,
and can thus bias baryon production.
There are essentially two approximations used in literature to calculate axial vector
currents:
\begin{itemize}
 \item[(a)] the quantum-mechanical reflection~\cite{Cohen:1990it,Joyce:1994bi,Joyce:1994zn}
used in the {\it thin wall} case, and
 \item[(b)] the semiclassical
force~\cite{Joyce:1994zt,Joyce:1994fu,Cline:2000nw,Joyce:1999fw,Prokopec:2003pj,Prokopec:2004ic}
 used in the {\it thick wall} case.
\end{itemize}
In general thin wall baryogenesis is more efficient in producing baryons.
Its main drawback is that the calculational methods used are unreliable:
one calculates the CP-violating reflected current ignoring the plasma,
and then inserts it into a transport equation
in an intuitive (but otherwise rather arbitrary) manner~\cite{Joyce:1994zn}.
How bad the situation can get is witnessed by
the controversy that developed around the work of
Farrar and Shaposhikov~\cite{Farrar:1993hn} (who used a quantum mechanical reflection
to calculate the source). The subsequent works~\cite{Gavela:1994ds,Gavela:1994dt,Huet:1994jb}
came up with an orders of magnitude smaller answer for baryon production.
And yet these latter works used unreliable methods that {\it e.g.} violate unitarity,
such that the issue remained unsettled
\footnote{Research on the topic subsided
not because the problem was resolved, but because standard model baryogenesis was ruled
out based on equilibrium considerations alone~\cite{Kajantie:1996qd}.}.
So, the problem of the source calculation in the thin wall regime remains still
to a large extent open.

 In the thick wall case the situation became much more satisfactory after
the works of Joyce, Kainulainen, Prokopec, Schmidt and
Weinstock~\cite{Joyce:1999fw,Kainulainen:2001cn,Kainulainen:2002th}. It was shown that
one can calculate the semiclassical force
(which rather straightforwardly sources the axial vector current)
from first principles and in a controlled approximation
from the Kadanoff-Baym (KB) equations for Wightman functions. These KB equations
are the quantum field theoretic generalization of kinetic equations.
The positive and negative frequency Wightman functions
represent the quantum field theoretic generalization of
the Boltzmann distribution function,
that provide statistical information on both on-shell and off-shell phase space flow.
In a certain limit, when integrated over energies, the Wightman functions yield Boltzmann's
distribution function. When written in a gradient approximation,
the KB equations can be split into the constraint equations (CE)
and the kinetic equations (KE). The authors of
Refs.~\cite{Kainulainen:2001cn,Kainulainen:2002th} have rigorously shown that,
in the presence of a moving planar interface,
in which fermions acquire a mass that depends on one spatial coordinate,
single fermions live on a shifted energy shell, which to
first order in gradients (linear in $\hbar$) and in the wall frame equals
\begin{equation}
 \omega_{\pm s} = \omega_0
\mp \hbar s\frac{|m|^2\partial_z \theta}{2\omega_0\sqrt{k_\perp^2+|m|^2}}
 \,,\qquad \omega_0=\sqrt{\vec k^2 +|m|^2}
\,,
\label{shifted energy shell:planar wall}
\end{equation}
where $m(z)=m_R(z)+\imath m_I(z) =|m|(z){\rm e}^{\imath\theta(z)}$ is the fermion mass,
which varies in the $z$-direction in which the wall moves, $\vec k$ is particle's momentum,
$\vec k_\perp$ is the momentum orthogonal to the wall, and $s=\pm 1$
is the corresponding spin.
This energy shift acts as a pseudo-gauge field (also known from condensed matter studies),
which lowers or increases particle's energy.
Relation~(\ref{shifted energy shell:planar wall}) clearly shows that
particles with a positive spin orthogonal to the wall and a positive frequency
(as well as particles with a negative spin and a negative frequency) will feel
a semiclassical force that is proportional to the gradient of $\theta={\rm arg}[m]$.
Particles with a negative spin and a positive frequency will feel an opposite force.
This force appears in the kinetic equation for the Boltzmann-like distribution functions
$f_{\pm s}$, and reads
\begin{equation}
 F_{\pm s} = -\frac{\partial_z|m|^2}{2\omega_{\pm s}}
    \pm\hbar s\frac{\partial_z(|m|^2\partial_z \theta)}{2\omega_0\sqrt{k_\perp^2+|m|^2}}
\,.
\label{semiclassical force:planar wall}
\end{equation}
It was also shown that this force then sources an axial vector current,
which in turn can bias sphalerons.

 The work of~\cite{Huet:1994jb,Huet:1995sh,Riotto:1997vy,Carena:1997gx,Carena:2000id}
has shown that, in the case that fermions mix through a mass matrix,
there is an additional CP-violating source resulting from flavor
mixing. This was put on a more formal ground by~\cite{Konstandin:2004gy}, where
a flavor independent formalism was developed, and where it was shown that
flavor non-diagonal source is subject to flavor oscillations induced by a commutator term
of the form $\imath [M,f]$, not unlike the famous flavor (vacuum) oscillations of neutrinos.
This idea was further developed by~\cite{Cirigliano:2009yt}.
Since we do not deal with flavor mixing
in this work, we shall not further dwell on this mechanism, which should not
diminish its importance. In passing we just mention that in most of the relevant parameter
space of {\it e.g.} the chargino mediated baryogenesis in the MSSM,
the semiclassical force induces the dominant CP-violating source
current~\cite{Konstandin:2005cd}.

 We shall now present a qualitative argument which suggests that in many situations
thin wall sources can dominate over the thick wall sources
(calculated in a gradient approximation). If true
this means that any serious attempt to make a quantitative assessment of baryon production
cannot neglect the thin wall contribution. To see why this is so, recall that
a gradient approximation applies for those plasma excitations whose
orthogonal momentum, $k_\perp = 2\pi/\lambda_\perp$, satisfies:
\begin{equation}
    k_\perp \gg \frac{2\pi}{L} \qquad {\tt (THICK\;\;WALL)}
\,,
\label{thick wall criterion}
\end{equation}
where $L$ is the typical thickness of the bubble wall.
On the other hand, the thin wall approximation belongs to the realm of momenta which satisfy
\begin{equation}
    k_\perp \leq \frac{2\pi}{L} \qquad {\tt (THIN\;\;WALL)}
\,.
\label{thick wall criterion:2}
\end{equation}
Typical momenta of particles in a plasma (per direction) is $k_\perp\sim  T$.
Now, unless $LT\gg 2 \pi$, we have a larger or comparable number of particles in the thin and
thick wall regimes! But, since the thin wall source is typically stronger, unless
thermal scattering significantly suppress the thin wall source, it will dominate
over the thick wall source. It is often incorrectly stated in literature
that the number of particles to which thin wall calculation applies is largely phase space
suppressed, {\it i.e.} that their number is small when compared to the number of particles
to which the semiclassical treatment applies. So, to conclude, it is of essential importance
to get the thin wall source right if we are to claim that we can reliably calculate
baryon production at the electroweak transition in a model.

 We believe that this represents a good motivation for what follows: a complete analytic
treatment of fermion tree-level dynamics for a time dependent mass.
The time dependence has been chosen such to correspond to a $\tanh$-kink wall,
because it is known that this represents a good approximation
to a realistic bubble wall~\cite{Moore:1995ua,Moore:1995si},
and equally importantly, in this case one can construct
exact solutions for mode functions. Before we begin our quantitative analysis,
we recall that a related study for the CP even case and planar wall has been conducted
by Ayala, Jalilian-Marian, McLerran and~Vischer~\cite{Ayala:1993mk},
while a semianalytic, perturbative treatment of the CP-violating
case has been conducted in~Ref.~\cite{Funakubo:1994yt}.
The main advantage of the latter study is that it allows for a general profile
of the CP-violating mass parameter, the drawbacks
are that the method is semianalytic (the final expression for the source is in terms of
an integral), and furthermore it is perturbative, such that it can be applied to
small CP violation only. To conclude, an exact treatment of fermion dynamics
in the presence of a strong CP violation
is highly desirable, and this is precisely what we do in this paper.

%%%%%%%%%%%%%%%%%%%%%%%%%%%%%%%%%%%%%%%%%%%%%%%%%%%%
%%%%%%%%%%%%%% The Model %%%%%%%%%%%%%%%%%%%%%%%%%%%
%%%%%%%%%%%%%%%%%%%%%%%%%%%%%%%%%%%%%%%%%%%%%%%%%%%%

\section{The model}
\label{The model}

 Here we consider the free fermionic lagrangian of the form,
\begin{equation}
{\cal L}_0 =  \bar \psi \imath \gamma^\mu \partial_\mu \psi - m^* \bar \psi_R \psi_L - m\bar\psi_L\psi_R
\,,
\label{Lagrangian:free}
\end{equation}
where $\psi_L = P_L \psi$ and $\psi_R = P_R\psi$ are the left and right handed
single fermionic fields, $P_L=(1-\gamma^5)/2$ and $P_R=(1+\gamma^5)/2$ are
the left and right handed projectors, and $\gamma^\mu$ and $\gamma^5$ are the
Dirac gamma matrices. We shall assume that the fermion mass $m$
is complex and space-time dependent. This can be generated {\it e.g.}
when a Yukawa interaction term, ${\cal L}_y = -y\phi \bar\psi_L\psi_R + {\rm h.c.}$,
is approximated by $-y\langle\hat \phi\rangle \bar\psi_L\psi_R + {\rm h.c.}$,
where $\langle\hat \phi\rangle$ stands for a Higgs-like scalar field condensate
which can generate a space-time dependent fermion mass,
\begin{equation}
  m(x) = y \langle\hat \phi(x)\rangle
\,,
%\nonumber
\label{space time mass}
\end{equation}
where $y$ is a (complex) Yukawa coupling.
The Dirac equation implied by~(\ref{Lagrangian:free}) is
\begin{equation}
\imath \gamma^\mu \partial_\mu \psi - m^* \psi_L - m\psi_R = 0
\,.
\label{Dirac equation}
\end{equation}
In this paper we consider the simplest case: a single fermion in a time dependent,
but spatially homogeneous, background. Such situations can occur, for example
in expanding cosmological backgrounds~\cite{Garbrecht:2006jm}, or during
second order phase transitions and crossover transitions in the early Universe.
In this case helicity is conserved~\cite{Garbrecht:2002pd,Garbrecht:2003mn,Garbrecht:2005rr}.
We shall perform the usual canonical quantization procedure, according to which
the spinor operator $\hat\psi(x)$ satisfies the following anti-commutator
($\hbar=1$),
\begin{equation}
 \{\hat\psi_\alpha(\vec x,t),\hat\psi_\beta^\dagger(\vec x^\prime,t)\}
   = \delta_{\alpha\beta}\delta^3(\vec x-\vec x^\prime)
\,.
\label{quantization:psi}
\end{equation}
In the free case under consideration, the Dirac equation~(\ref{Dirac equation}) is linear,
and consequently $\hat\psi(x)$ can be expanded in terms of
the creation and annihilation operators, which in the helicity basis reads,
\begin{equation}
 \hat\psi(\vec x,t) = \int \frac{d^3k}{(2\pi)^3}\sum_{h=\pm}
   \Big[{\rm e}^{\imath \vec k\cdot \vec x} \hat a_{\vec k h} \chi_h(\vec k,t)
      +{\rm e}^{-\imath \vec k\cdot \vec x}\hat b^\dagger_{\vec k h} \nu_h(\vec k,t)
    \Big]
\,,
\label{quantization:psi:2}
\end{equation}
where $\chi_h(\vec k,t)$ and $\nu_h(\vec k,t)$ are particle and antiparticle
four-spinors. $\hat a_{\vec k h}$ and $\hat b_{\vec k h}$ are the annihilation
operators that destroy the fermionic vacuum state $|\Omega\rangle$,
   $\hat a_{\vec k h}|\Omega\rangle = 0 = \hat b_{\vec k h}|\Omega\rangle$, while
$\hat a^\dagger_{\vec k h}$ and $\hat b^\dagger_{\vec k h}$ are
the creation operators that create a particle and an antiparticle with momentum
$\vec k$ and helicity $h$.
These operators obey the following anticommutator algebra,
\begin{eqnarray}
 \{\hat a_{\vec k h},\hat a^\dagger_{\vec k^\prime h^\prime}\}
    &=& \delta_{hh^\prime} (2\pi)^3\delta^3(\vec k-\vec k^\prime)
\,,\qquad
 \{\hat a_{\vec k h},\hat a_{\vec k^\prime h^\prime}\}  = 0
\,,\qquad
 \{\hat a^\dagger_{\vec k h},\hat a^\dagger_{\vec k^\prime h^\prime}\}  = 0
\nonumber\\
 \{\hat b_{\vec k h},\hat b^\dagger_{\vec k^\prime h^\prime}\}
    &=& \delta_{hh^\prime}(2\pi)^3 \delta^3(\vec k-\vec k^\prime)
\,,\qquad
 \{\hat b_{\vec k h},\hat b_{\vec k^\prime h^\prime}\}  = 0
\,,\qquad
 \{\hat b^\dagger_{\vec k h},\hat b^\dagger_{\vec k^\prime h^\prime}\}  = 0
\,,
\label{quantization:psi:3}
\end{eqnarray}
where all mixed anticommutators are zero.
The momentum space quantization conditions~(\ref{quantization:psi:3})
and the position space quantization rule~(\ref{quantization:psi})
have to be mutually consistent.
This imposes the following consistency condition on the
positive and negative frequency spinors,
\begin{equation}
 \sum_{h=\pm}[\chi_{h\alpha}(\vec k,t)\chi^*_{h\beta}(\vec k,t)
         + \nu_{h\alpha}(-\vec k,t)\nu^*_{h\beta}(-\vec k,t)] = \delta_{\alpha\beta}
\,.
\label{consistency condition:mode functions}
\end{equation}
This is usually supplied by the mode orthogonality conditions,
\begin{equation}
 \bar\chi_h(\vec k,t) \cdot\nu_h(\vec k,t)  = 0
    = \bar\nu_h(\vec k,t) \cdot\chi_h(\vec k,t)
\,.
\label{mode orthogonality}
\end{equation}
and by the mode normalization conditions,
\begin{equation}
 \chi_h^\dagger(\vec k,t)\cdot \chi_h(\vec k,t) = 1
   = \nu_h^\dagger(\vec k,t)\cdot \nu_h(\vec k,t)
\,,
\label{mode normalization conditions}
\end{equation}
which -- as we will see below -- are chosen to be consistent with the more general
requirement~(\ref{consistency condition:mode functions}). Although
the orthogonality condition~\eqref{mode orthogonality} is usually met,
it is however not a necessity. Important is that the mode functions
span all of the Hilbert space, which is true in this case.
Because we consider a system which is time-translationally invariant,
helicity is conserved, and it is thus convenient to work with
helicity conserving spinors
\begin{equation}
\chi_h(\vec k,t) = \left(\begin{array}{c} L_h(\vec k,t) \cr R_h(\vec k,t) \cr
                         \end{array}\right)\otimes
                    \xi_h(\vec k)
\,,\qquad
\nu_h(\vec k,t) = \left(\begin{array}{c} \bar L_h(\vec k,t)\cr \bar R_h(\vec k,t) \cr
                         \end{array}\right)\otimes
                    \xi_h(\vec k)
\,,
\label{helicity spinors}
\end{equation}
where $\xi_h(\vec k)$ is the helicity two eigen-spinor, satisfying
$\hat h \xi_h = h \xi_h$, where $\hat h = \hat k\cdot {\vec\sigma}$ is the helicity
operator and $h=\pm 1$ are its eigenvalues.

 We shall work here with the Dirac matrices in the chiral representation, in  which
\begin{equation}
\gamma^0 = \left(\begin{array}{cc} 0 & I \cr I & 0 \cr
                         \end{array}\right)
         = \rho^1\otimes I
\,,\qquad
\gamma^i = \left(\begin{array}{cc} 0 & \sigma^i \cr - \sigma^i & 0 \cr
                         \end{array}\right)
         = \imath\rho^2\otimes \sigma^i
\,,\qquad
\gamma^5 \equiv \imath \gamma^0\gamma^1\gamma^2\gamma^3
         = \left(\begin{array}{cc} -I & 0 \cr 0 & I \cr
                         \end{array}\right)
          = -\rho^3\otimes I
\,,
\label{chiral representation}
\end{equation}
where the last equalities follow from the usual direct product (Bloch) representation
of the Dirac matrices. Here
$\rho^i$ and $\sigma^i$ are the Pauli matrices obeying,
$\rho^i\rho^j=\delta^{ij}+\imath \epsilon^{ijl}\rho^l$ and
$\sigma^i\sigma^j=\delta^{ij}+\imath \epsilon^{ijl}\sigma^l$.
The left and right projectors are then,
\begin{equation}
 P_L= \frac{1-\gamma^5}{2}
         = \left(\begin{array}{cc} I & 0 \cr 0 & 0 \cr
                         \end{array}\right)
         = \frac{1+\rho^3}{2}\otimes I
\,,\qquad
 P_R= \frac{1+\gamma^5}{2}
         = \left(\begin{array}{cc} 0 & 0 \cr 0 & I \cr
                         \end{array}\right)
         = \frac{1-\rho^3}{2}\otimes I
\,,
\label{LR projectors}
\end{equation}
which can be used to write, $\psi_L=P_L\psi$ and $\psi_R=P_R\psi$ as it is done
in~(\ref{Lagrangian:free}--\ref{Dirac equation}).
Now, making use of Eqs.~(\ref{quantization:psi:2}--\ref{LR projectors}) in
the Dirac equation~(\ref{Dirac equation}) one gets the following four equations
for the component functions
\begin{eqnarray}
 \imath \dot L_h + h k L_h &=& m R_h
 \nonumber
 \\
 \imath \dot R_h - h k R_h &=& m^* L_h
\label{eom:LhRh}
\end{eqnarray}
and
\begin{eqnarray}
 \imath \dot {\bar L}_h - h k \bar L_h &=& m \bar R_h
 \nonumber
 \\
 \imath \dot {\bar R}_h + h k \bar R_h &=& m^* \bar L_h
\,,
\label{eom:barLhRh}
\end{eqnarray}
where the mass can be complex and time dependent, $m=m(t)$,
and the modes are normalized to unity,
\begin{equation}
 |L_h|^2+|R_h|^2 = 1 =  |\bar L_h|^2+|\bar R_h|^2
\label{normalization Lh and Rh}
\end{equation}
 The equations of motion for $L_h$ and $R_h$ can be decoupled,
resulting in the second order equations,
\begin{eqnarray}
 \ddot L_h + \omega^2 L_h - \frac{\dot m}{m} (\dot L_h - \imath h k L_h) &=& 0
 \nonumber
 \\
 \ddot R_h + \omega^2 R_h - \frac{\dot m^*}{m^*} (\dot R_h + \imath h k R_h) &=& 0
\,,
\label{eom:LhRh:2}
\end{eqnarray}
where $\omega^2 = k^2 + |m(t)|^2$.
For the case at hand a better way of
proceeding is to go to the positive and negative frequency basis, defined by:
\begin{equation}
u_{\pm h} = \frac{1}{\sqrt{2}}(L_h\pm R_h)
\,,\qquad
v_{\pm h} = \frac{1}{\sqrt{2}}(\bar L_h\pm \bar R_h)
\,,
\label{u+-}
\end{equation}
since then the equation of motion can be reduced to the Gauss' hypergeometric
equation. Indeed, from~(\ref{eom:LhRh}--\ref{eom:barLhRh})
and~(\ref{u+-}) it follows,
\begin{eqnarray}
 \imath \dot u_{\pm h} \mp m_R(t)u_{\pm h} &=& -(h k\pm\imath m_I) u_{\mp h}
 \nonumber
 \\
 \imath \dot v_{\pm h} \mp m_R(t)v_{\pm h} &=& (h k\mp\imath m_I) v_{\mp h}
\,,
\label{eom:u+-hv+-h}
\end{eqnarray}
which, when decoupled, yields a second order equation,
\begin{equation}
 \ddot u_{\pm h} \mp\imath \frac{\dot m_I}{hk \pm\imath m_I}\dot u_{\pm h}
 +\Bigg(k^2+|m|^2 \pm\imath \dot m_R + \frac{m_R\dot m_I}{hk \pm\imath m_I}
   \Bigg)u_{\pm h} =  0
\,.
\label{eom:u+-h:decoupled}
\end{equation}
So far our analysis has been general, in the sense that we have assumed
no special time dependence in $m(t)$.
In order to make progress however, we have to make a special choice for
$m(t)$, which is what we do next.

%%%%%%%%%%%%%%%%%%%%%%%%%%%%%%%%%%%%%%%%%%%%%%%%%%%%
%%%%%%%%%%%%%% Mode functions %%%%%%%%%%%%%%%%%%%%%%
%%%%%%%%%%%%%%%%%%%%%%%%%%%%%%%%%%%%%%%%%%%%%%%%%%%%

\section{Mode functions for the kink profile}
\label{Mode functions for the kink profile}

 In Ref.~\cite{Ayala:1993mk} an exact solution of the Dirac equation
was found for a wall of arbitrary
thickness with a kink wall profile $\propto\tanh(-z/L)$, where $L\equiv 1/\lambda$
characterizes the wall thickness. Here we generalize this solution
to include CP violation. While in this paper we consider only a time dependent
mass profile, the generalization to the planar ($z$-dependent) case is straightforward,
and will be considered separately.
Constructing an exact solution is important for baryogenesis since
one can then consider in detail how the CP-odd quantities that
source baryogenesis (directly or indirectly) depend on
the mass profile, and in particular investigate what is the optimal profile and its duration.
Unfortunately, analytic solutions cannot include plasma scattering and width effects,
whose treatment will be therefore typically left to numerical simulations.

 Here we assume the following `wall' profile
\begin{equation}
m(t) = m_1 + m_2 \tanh\Big(-\frac{t}{\tau}\Big)
\,,
\label{mass tanh}
\end{equation}
where $\tau\equiv 1/\gamma$ represents the time scale over which
the wall varies (for convenience we shall use the terms `wall' and `profile'
interchangeably).
Both $m_1$ and $m_2$ are complex mass parameters.
In the case when a single Higgs field is responsible for
the phase transition, one expects that both real and imaginary part of
$m(t)$ exhibit a similar behaviour, which is reflected in the
{\it Ansatz}~(\ref{mass tanh}). Moreover, we do not know how to
construct exact solutions when different time scales govern the rate of
change of the real and imaginary masses. Nevertheless, we believe
that the {\it Ansatz}~(\ref{mass tanh}) represents quite well realistic
walls for a wide variety of single stage phase transitions, {\it cf.}
Refs.~\cite{Moore:1995ua,Moore:1995si}.

Note that the thin wall limit is $\tau\rightarrow 0$
($\gamma\rightarrow \infty$). In that limit the mass function
becomes the step function {\it Ansatz}~\eqref{mass:1}, whereby $m_{\pm}=m_1 \mp m_2$.
In appendix \ref{Thin wall} we construct the normalized
fundamental solutions of Eqs. \eqref{eom:LhRh} for a constant
mass. The thin wall case is treated explicitly in appendix
\ref{Mode functions for a step fermion mass}. The thin wall results
serve as a check for the kink wall case in the appropriate limits.
Moreover it allows for a quantitative comparison of the thick wall
to the thin wall results.

 Since the ratio of the real and imaginary parts of the mass
$m_I(t)/m_R(t)$ is time dependent, the {\it Ansatz}~(\ref{mass tanh})
contains CP violation (which can be either small or large, depending on
how much the ratio $m_I(t)/m_R(t)$ changes.
Since the physical CP-violating phase is in the relative phase
between $m_1$ and $m_2$, one can perform a global rotation
of the left- and right-handed spinors that does not
affect CP violation. It turns out that the equations of motion simplify
if one performs a global rotation that removes the imaginary part of
$m_2$. The constant rotation that does that is
\begin{equation}
m(t)\rightarrow m(t)e^{\imath \chi}\,,\qquad \chi=\arctan\left(-\frac{m_{2I}}{m_{2R}}\right)
\,.
\label{mass: real imaginary}
\end{equation}
In that case
\begin{equation}
 m_1 = m_{1R}+m_{1I}\,,\qquad m_2 = m_{2R}
\,.
\nonumber
\end{equation}
 This rotation is important, because the mode equations~(\ref{eom:u+-h:decoupled})
significantly simplify to become
\begin{equation}
 \ddot u_{\pm h} +(\omega^2(t) \pm\imath \dot m_R)u_{\pm h} =  0
\,,
\label{eom:u+-h:decoupled:2}
\end{equation}
where $\omega^2(t)=k^2+|m(t)|^2$.
Furthermore, from~(\ref{eom:u+-hv+-h}) one can infer that $v_{\pm h}$ obey the same
equations as $u_{\pm-h}$.
In what follows, we show that these equations
can be reduced to the Gauss' hypergeometric equation.

\medskip

To show this, it is instructive to introduce
a new variable,
\begin{equation}
 z = \frac12-\frac12 \tanh\left(-\frac{t}{\tau}\right)
\,,
\label{z:definition}
\end{equation}
in terms of which
\begin{equation}
 m(t) = m_1 +m_2 (1-2z)
 \,,\qquad \dot m_R(t) = -2 m_{2R} \dot{z}=-\frac{\gamma m_{2R}}{\cosh^2(-t/\tau)}
                       = - 4\gamma m_{2R}z(1-z)
\nonumber
\,,
\end{equation}
with $\gamma={1}/{\tau}$.
Eq.~(\ref{eom:u+-h:decoupled:2}) becomes,
\begin{eqnarray}
&&\bigg\{4\gamma^2[z(1\!-\!z)]^2\frac{d^2}{dz^2}+4\gamma^2 (1\!-\!2z)z(1\!-\!z)\frac{d}{dz}
\nonumber\\
&&\hskip 2cm  +\,\Big[k^2+m_I^2+(m_{1R}\!+\!m_{2R})^2-4z m_{1R}m_{2R}-4z(1\!-\!z)m_{2R}(m_{2R}\pm\imath\gamma)\Big]
\bigg\}u_{\pm h}
    =\, 0\,.
 \hskip 0.5cm
\label{eom:upm:z}
\end{eqnarray}
Now, performing a rescaling,
\begin{equation}
 u_{\pm h} = z^\alpha (1-z)^\beta \chi_{\pm h}(z)
\label{rescaling:u vs chi}
\end{equation}
and choosing
\begin{equation}
 \alpha = -\frac{\imath}{2}\frac{\omega_-}{\gamma}
\,,\qquad
\beta = -\frac{\imath}{2}\frac{\omega_+}{\gamma}
\,,
\label{alpha and beta}
\end{equation}
where
\begin{equation}
 \omega_{\mp}\equiv \omega(t\rightarrow \mp\infty)
  = \sqrt{k^2+m_I^2+(m_{1R}\pm m_{2R})^2}
\,,
\label{omega-+}
\end{equation}
yields the following Gauss' hypergeometric equation for $\chi_{\pm h}$,
\begin{equation}
\Bigg[ z(1-z)\frac{d^2}{dz^2} +[c-(a_\pm+b_\pm+1)z]\frac{d}{dz}-a_\pm b_\pm\Bigg]
  \chi_{\pm h}(z) = 0
\,,
\label{eom:hypergeometric}
\end{equation}
where
\begin{equation}
a_\pm = \alpha+\beta+1 \mp \imath\frac{m_{2R}}{\gamma}
\,,\qquad
b_\pm = \alpha+\beta \pm \imath\frac{m_{2R}}{\gamma}
\,,\qquad
c = 2\alpha+ 1
\,.
\label{a+- and b+-}
\end{equation}
Note that the rescaling~(\ref{rescaling:u vs chi}) was chosen
such to remove the terms $\propto 1/z$ and $\propto 1/(1-z)$
from Eq.~(\ref{eom:hypergeometric}).
Since $a_\pm,b_\pm,c$ are non-integer,
the two independent solutions for $\chi_{\pm h}$ are the usual ones.
A detailed normalization procedure is provided in Appendix~\ref{Mode function normalization}
and the result are the following normalized {\it early time} mode functions
\begin{eqnarray}
 u_{+h} &\equiv&  u^{(1)}_{+h} = \sqrt{\frac{\omega_-+(m_{1R}+m_{2R})}{2\omega_-}}
           \times z^\alpha (1-z)^\beta \times
   \phantom{,\!\!}_2F_1(a_+,b_+;c;z)
\nonumber\\
 u_{-h} &\equiv&  u^{(1)}_{-h} =
 -\frac{h k-\imath m_I}{\sqrt{k^2+m_I^2}}
 \times
\sqrt{\frac{\omega_--(m_{1R}+m_{2R})}{2\omega_-}}
 \times
                 z^{\alpha} (1-z)^\beta \times
   \phantom{,\!\!}_2F_1(a_-,b_-;c;z)
\,.
\label{u+-h:normalized}
\end{eqnarray}
These functions are valid of course for all times. They are called
{\it early time} mode functions because at
early times ($t\rightarrow -\infty$) they
reduce to the positive frequency mode functions~(\ref{early times mode functions}),
and they are normalized
as, $|u^{(1)}_{+h}|^2+|u^{(1)}_{-h}|^2 = 1$,
which follows from Eqs.~(\ref{normalization Lh and Rh}) and~(\ref{u+-}),
see also Eq.~(\ref{normalization:3}).

For completeness, we also quote the second pair~(\ref{u:two independent solutions})
of early time solutions,
\begin{eqnarray}
 u^{(2)}_{+h} &=& \sqrt{\frac{\omega_--(m_{1R}+m_{2R})}{2\omega_-}}
           \times z^{\alpha+1-c} (1-z)^{\beta+c-a_+-b_+} \times
   \phantom{,\!\!}_2F_1(1-a_+,1-b_+;2-c;z)
\nonumber\\
 u^{(2)}_{-h} &=&
 \frac{h k-\imath m_I}{\sqrt{k^2+m_I^2}}
 \times
\sqrt{\frac{\omega_-+(m_{1R}+m_{2R})}{2\omega_-}}
 \times
                 z^{\alpha+1-c} (1-z)^{\beta+c-a_--b_-} \times
   \phantom{,\!\!}_2F_1(1-a_-,1-b_-;2-c;z)
\,.
\label{u+-h:normalized:2}
\end{eqnarray}
Just as before, at early times ($t\rightarrow -\infty$) these solutions
reduce to the negative frequency mode functions~(\ref{early times mode functions}),
and they are also normalized
as, $|u^{(2)}_{+h}|^2+|u^{(2)}_{-h}|^2 = 1$.

\bigskip

An analogous procedure as above yields the following
normalized fundamental solutions suitable for {\it late times},
\begin{eqnarray}
 \tilde u^{(1)}_{+h} &=& \sqrt{\frac{\omega_++(m_{1R}-m_{2R})}{2\omega_+}}
           \times z^{\alpha+1-c} (1-z)^{\beta+c-a_+-b_+} \times
   \phantom{,\!\!}_2F_1(1-a_+,1-b_+;2-c\tilde ;1-z)
\nonumber\\
 \tilde u^{(1)}_{-h} &=&
 -\frac{h k-\imath m_I}{\sqrt{k^2+m_I^2}}
 \times
\sqrt{\frac{\omega_+-(m_{1R}-m_{2R})}{2\omega_+}}
 \times
                 z^{\alpha+1-c} (1-z)^{\beta+c-a_--b_-} \times
   \phantom{,\!\!}_2F_1(1-a_-,1-b_-;2-\tilde c;1-z)
\,
\label{tilde u+-h:normalized}
\end{eqnarray}
and
\begin{eqnarray}
 \tilde u^{(2)}_{+h} &=& \sqrt{\frac{\omega_+-(m_{1R}-m_{2R})}{2\omega_+}}
           \times z^{\alpha} (1-z)^{\beta} \times
   \phantom{,\!\!}_2F_1(a_+,b_+;\tilde c;1-z)
\nonumber\\
 \tilde u^{(2)}_{-h} &=&
 \frac{h k-\imath m_I}{\sqrt{k^2+m_I^2}}
 \times
\sqrt{\frac{\omega_++(m_{1R}-m_{2R})}{2\omega_+}}
 \times
                 z^{\alpha} (1-z)^{\beta} \times
   \phantom{,\!\!}_2F_1(a_-,b_-;\tilde c;1-z)
\,,
\label{tilde u+-h:normalized:2}
\end{eqnarray}
while the late time solutions~(\ref{tilde u+-h:normalized:2}) reduce at
asymptotically late times to positive and negative frequency solutions
$\propto {\rm e}^{\mp \imath\omega_+t}$, respectively,
see Eq.~(\ref{u:late time:asymptotic}).

 Now, a general early time solution can be written as a linear combination
 of the fundamental solutions~(\ref{u+-h:normalized}--\ref{u+-h:normalized:2});
 for simplicity we shall take here~(\ref{u+-h:normalized}) for the early time solutions.
Similarly, general late time solutions are a linear combination of the fundamental
late time solutions~(\ref{tilde u+-h:normalized}--\ref{tilde u+-h:normalized:2}),
\begin{equation}
 \tilde u_{\pm h}
 = \alpha_{\pm h}\tilde u^{(1)}_{\pm h} + \beta_{\pm h} \tilde u^{(2)}_{\pm h}
\,,
\label{late time solutions:general}
\end{equation}
where $\alpha_{\pm h}$ and $\beta_{\pm h}$ are complex functions of $\vec k$
(for spatially homogeneous systems they are functions of the magnitude $\|\vec k\|$ only)
that satisfy the standard normalization condition,
\begin{equation}
   |\alpha_{\pm h}|^2+|\beta_{\pm h}|^2=1
\,.
\label{normalization on alpha and beta}
\end{equation}

Now, upon choosing~(\ref{u+-h:normalized}) as the early time solutions
and making use of the matching between the general early and late time solutions
\begin{equation}
 \tilde u_{\pm h}(k,t) = u_{\pm h}(k,t)
\label{late = early}
\end{equation}
and of the relation for the Gauss' hypergeometric
functions~(\ref{identity GR:9.131.2}) one gets,
\begin{eqnarray}
\alpha_{\pm h} &=&\sqrt{\frac{\omega_+[\omega_-\pm(m_{1R}+m_{2R})]}{\omega_-[\omega_+\pm(m_{1R}-m_{2R})]}}
                 \frac{\Gamma(c)\Gamma(a_\pm+b_\pm-c)}{\Gamma(a_\pm)\Gamma(b_\pm)}
\nonumber\\
\beta_{\pm h} &=&\pm\sqrt{\frac{\omega_+[\omega_-\pm(m_{1R}+m_{2R})]}{\omega_-[\omega_+\mp(m_{1R}-m_{2R})]}}
                 \frac{\Gamma(c)\Gamma(c-a_\pm-b_\pm)}{\Gamma(c-a_\pm)\Gamma(c-b_\pm)}
\,.
\label{alpha-beta}
\end{eqnarray}
It can be shown that
\begin{eqnarray}
\alpha_{+h}&=&\alpha_{-h}
\nonumber\\
\beta_{+h}&=&\beta_{-h}
\,.
\label{alphabetaequalities}
\end{eqnarray}
Useful identities here are
\begin{align}
\omega_{-}\mp (m_{1R}+m_{2R}) & = \frac{\pm \omega_{+}^2 \mp (\omega_{-} \mp 2 m_{2R})^2}{4 m_{2R}}
\nonumber \\
\omega_{+}\mp (m_{1R}-m_{2R}) & = \frac{\mp \omega_{-}^2 \pm (\omega_{+} \pm 2 m_{2R})^2}{4 m_{2R}}
\,.
\end{align}
Because $\alpha_{\pm h}$ and $\beta_{\pm h}$ are functions of
$a_\pm$, $b_\pm$ and $c$, (just as in the thin wall
case~(\ref{alphah+betah+}--\ref{number produced}))
there are no CP odd contributions
in the mode mixing (Bogoliubov) coefficients~(\ref{alpha-beta}).
$\alpha_{\pm h}$ and $\beta_{\pm h}$ are indeed the usual Bogoliubov coefficients
that transform an asymptotically early time vacuum state to a late time vacuum
state. Hence $n_{\pm h}=|\beta_{\pm h}|^2$ is the particle number observed
by a late time observer, in the late time state that evolves from
the early time positive frequency vacuum state.

To make contact with the thin wall case~(\ref{alphah+betah+}), we take the limit
$\gamma\rightarrow \infty$ in~(\ref{alpha-beta}) to get,
\begin{equation}
\beta_{\pm h} \stackrel{\gamma\rightarrow \infty}{\longrightarrow}
         \mp\sqrt{\frac{[\omega_-\pm(m_{1R}+m_{2R})]}{\omega_+\omega_-[\omega_+\mp(m_{1R}-m_{2R})]}}
         \left[\frac{\omega_--\omega_+}{2}\mp m_{2R}\right]
\,.
\label{beta:thin}
\end{equation}
It can be checked that Eq. \eqref{beta:thin} satisfies $\beta_{+h}=\beta_{-h}$.
Moreover, since $\omega_{-}-\omega_{+}-2 m_{2R}<0$, the $\beta_{+h}$ and $\beta_{-h}$
are always positive.
One can show that $\alpha_{\pm h}$ and $\beta_{\pm h}$ given in~(\ref{alpha-beta})
 obey $|\alpha_{\pm h}|^2+|\beta_{\pm h}|^2=1$, as they should. This equality follows from,
\begin{eqnarray}
 |\alpha_{\pm h}|^2 &=& \frac{\sinh\left(\frac{\pi[\omega_++\omega_-+2m_{2R}]}{2\gamma}\right)
                            \sinh\left(\frac{\pi[\omega_++\omega_--2m_{2R}]}{2\gamma}\right)}
                            {\sinh\left(\frac{\pi\omega_+}{\gamma}\right)
                            \sinh\left(\frac{\pi\omega_-}{\gamma}\right)}
 \nonumber\\
 n_{\pm h}=  |\beta_{\pm h}|^2 &=& \frac{\sinh\left(\frac{\pi[\omega_--\omega_++2m_{2R}]}{2\gamma}\right)
                            \sinh\left(\frac{\pi[\omega_+-\omega_-+2m_{2R}]}{2\gamma}\right)}
                            {\sinh\left(\frac{\pi\omega_+}{\gamma}\right)
                            \sinh\left(\frac{\pi\omega_-}{\gamma}\right)}
                            \,,
\label{alpha and beta squared}
\end{eqnarray}
from which it also follows that $|\alpha_{\pm h}|^2 =1 - |\beta_{\pm h}|^2$.
Now, taking a thin wall limit $\gamma\rightarrow \infty$ in~(\ref{alpha and beta squared})
yields
\begin{equation}
  n_{\pm h}\stackrel{\gamma\rightarrow \infty}{\longrightarrow}
               \frac{|m_--m_+|^2-(\omega_--\omega_+)^2}{4\omega_-\omega_+}
\,,
\label{n:thin wall limit}
\end{equation}
where we made use of Given that $4m_{2R}^2=|m_--m_+|^2$. This expression agrees with
the thin wall particle number Eq.~(\ref{number produced})
derived in appendix~\ref{Mode functions for a step fermion mass}.

It is interesting to note that, although particle number agrees,
the Bogoliubov coefficient $\beta_{\pm h}$ in the thin wall
limit \eqref{beta:thin} appears very different from the one
derived explicitly for the thin wall \eqref{alphah+betah+}.
For instance, the coefficients in \eqref{alphah+betah+}
are complex and depend explicitly on helicity, whereas
the limiting coefficient \eqref{beta:thin} is real and
helicity independent. A similar situation occurs
for $\alpha_{\pm h}$, see \eqref{alpha-beta-thinwall}.
The apparent discrepancy is caused by
an overall phase factor by which the coefficients in the thin
wall limit differ from those directly computed for the thin wall.
This phase factor does not affect particle number and can be removed
by a global rotation of the (anti)particle spinors.
In appendix \ref{Connecting the kink wall to the thin wall case}
we show explicitly how the kink wall case and thin wall case are
connected.

\begin{figure}[t!]
 \begin{minipage}[t]{.45\textwidth}
        \begin{center}
\includegraphics[width=\textwidth]{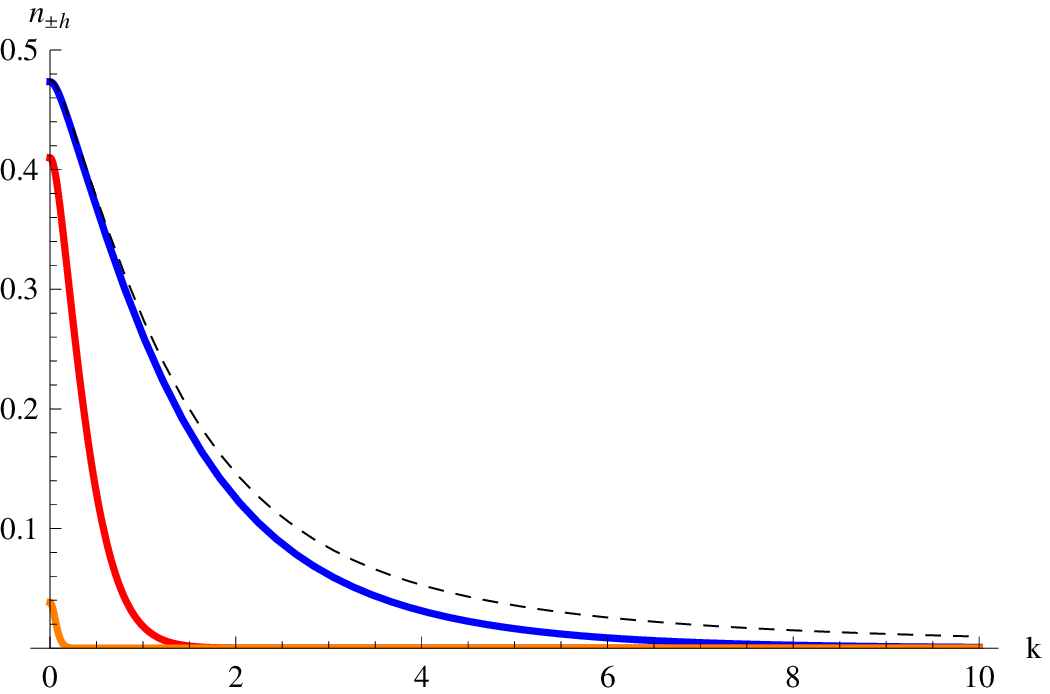}
   {\em \caption{Plot of $n_{\pm h}=|\beta_{\pm h}|^2$ as function of $k$
   in units of $m_{2R}$. Parameters are
   $m_{1R}=m_{2R}$, $m_{I}=0.1 m_{2R}$. The dashed line is the thin wall solution
   of $n_{\pm h}$. The other (solid) lines show -- from top to bottom -- the particle numbers for
   $\gamma=10m_{2R}$ (blue, dark), $\gamma=m_{2R}$ (red) and $\gamma=0.1m_{2R}$ (orange, light). In general
   particle number is suppressed for decreasing $\gamma$ (a thicker wall), but
   still a large particle number is reached when $k,m_I \ll m_{2R}$ and $m_{1R}-m_{2R} \simeq 0$.
   \label{fig:nhnormalpop} }}
        \end{center}
    \end{minipage}
\hfill
    \begin{minipage}[t]{.45\textwidth}
        \begin{center}
\includegraphics[width=\textwidth]{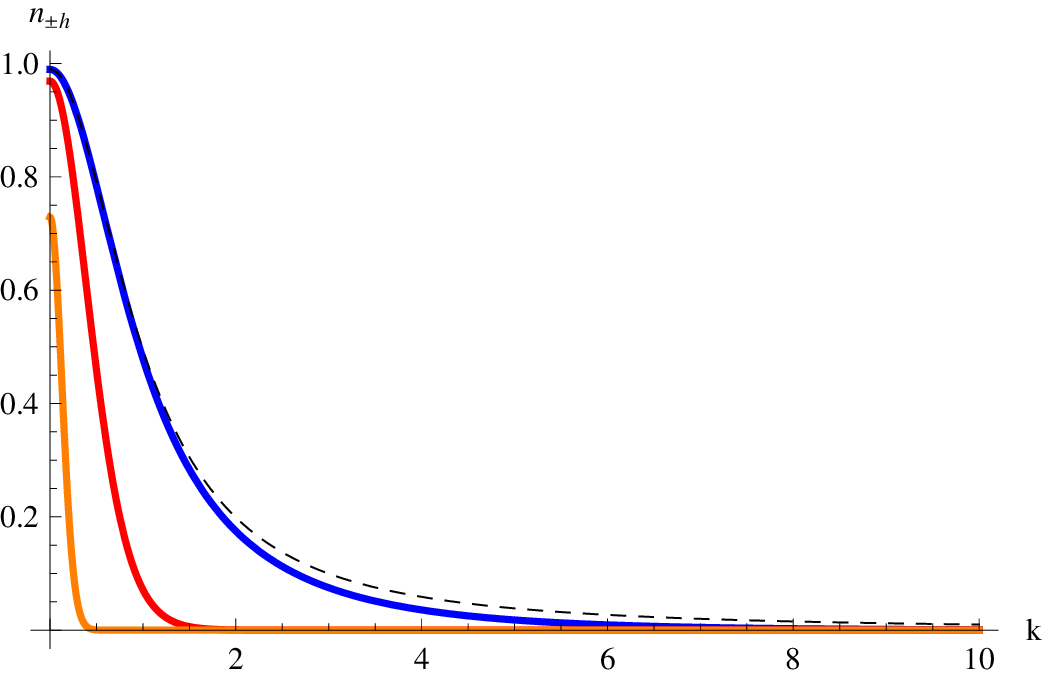}
   {\em \caption{Plot of $n_{\pm h}=|\beta_{\pm h}|^2$ as function of $k$
   in units of $m_{2R}$.
   Parameters are
   $m_{I}=0.1m_{2R},~m_{1R}=0.1 m_{2R}$. The dashed line is the thin wall solution
   of $n_{\pm h}$. The other (solid) lines  show -- from top to bottom -- the particle numbers for
    $\gamma=10m_{2R}$ (blue), $\gamma=m_{2R}$ (red) and $\gamma=0.1m_{2R}$ (orange). Because
   $m_{1R}-m_{2R}<0$, an inverse population is reached. The CP-violating phase is
   maximal because $m_{+}\simeq -m_{-}$.
   \label{fig:nhinversepop} }}
        \end{center}
    \end{minipage}

\end{figure}

The particle production can also be analyzed in the opposite limit,
$\gamma\rightarrow 0$. In this thick wall regime particle production is exponentially
suppressed as,
\begin{equation}
  n_{\pm h}\stackrel{\gamma\rightarrow 0}{\longrightarrow}
      \exp\left[-\frac{\pi(\omega_++\omega_--2m_{2R})}{\gamma}\right]
\,,
\label{n:thick wall limit}
\end{equation}
which is also what one expects. However, note that when $\pi(\omega_++\omega_--2m_{2R})\lesssim\gamma$,
the suppression is not large. This is demonstrated in figures
\ref{fig:nhnormalpop} and \ref{fig:nhinversepop}, where the particle
number is shown as a function of $k$ for several different wall thicknesses.
In figure \ref{fig:nhnormalpop} the mass parameters are $m_{1R}=m_{2R}$
and $m_I \ll m_{1R}, m_{2R}$. In this case CP violation is weak.
For these mass parameters the thin wall
particle number \eqref{n:thin wall limit}, represented by the dashed line,
reaches the maximal particle number $n_{\pm h}=\frac12$ as $k\rightarrow 0$.
For thicker walls (decreasing $\gamma$) the particle number is exponentially
suppressed with respect to the thin wall. For very small $k$ the
suppression is much smaller, since
$\pi(\omega_++\omega_--2m_{2R})/\gamma \sim \sqrt{k^2+m_I^2}/\gamma$.

In figure \ref{fig:nhinversepop} the mass parameters are chosen
such that $m_I,m_{1R}\ll m_{2R}$. In this case CP-violation is maximal
for the thin wall in the limit $k\rightarrow 0$,
see also \eqref{number produced:k=0:2}. The maximal particle number
in this limit is 1, which indicates an inverse population.
This inverse population, induced by large CP violation, is a novel result
and, as far as we know, not noticed in literature before.
For thicker walls the particle number is still suppressed, but much less
than for the mass parameters in figure \ref{fig:nhnormalpop}. In fact,
for $m_{1R}=m_{I}=0$ the particle number is
unsuppressed in the limit $k\rightarrow 0$.

A large late time Bogoliubov particle
number for a free fermionic system indicates large squeezing.
It is interesting to see what effect such a large squeezing may have
on the fermionic currents. In particular, we are interested in the CP-odd
fermionic axial vector current that couples to sphalerons. The next section is
devoted to computing these currents in the setting of a tanh-kink wall.

\bigskip

%%%%%%%%%%%%%%%%%%%%%%%%%%%%%%%%%%%%%%%%%%%%%%%%%%%%
%%%%%%%%%%%%%% Currents  %%%%%%%%%%%%%%%%%%%%%%%%%%%
%%%%%%%%%%%%%%%%%%%%%%%%%%%%%%%%%%%%%%%%%%%%%%%%%%%%

\section{The currents and CP violation}
\label{The Currents and CP violation}

 In this section we consider the evolution of the two point Wightman functions,
defined as the expectation values~\cite{Prokopec:2003pj,Garbrecht:2002pd}
\begin{equation}
\imath S^{+-}_{\alpha\beta}(u,v) \equiv \imath S^{<}_{\alpha\beta}(u,v)
        = - \langle \hat{\bar\psi}_{\beta}(v)\hat\psi_{\alpha}(u)\rangle
\,;\qquad
\imath S^{-+}_{\alpha\beta}(u,v) \equiv \imath S^{>}_{\alpha\beta}(u,v)
        = \langle \hat\psi_{\alpha}(v)\hat{\bar\psi}_{\beta}(v)\rangle
\,,
\label{Wightman fermions: definition}
\end{equation}
and which satisfy the homogeneous Dirac equations~(\ref{Dirac equation})
\begin{equation}
(\imath \gamma^\mu \partial_\mu - m_R -\imath m_I\gamma^5)
         \imath S^{\pm\mp}_{\alpha\beta}(u,v) = 0
\,.
\label{Dirac equation:wightman}
\end{equation}
For the problem at hand, when written in a Wigner mixed representation
\begin{equation}
\imath S^{\pm\mp}_{\alpha\beta}(u,v) = \int\frac{d^4k}{(2\pi)^4}{\rm e}^{\imath k\cdot (u-v)}
         \imath S^{\pm\mp}_{\alpha\beta}(k;x)
\,,\qquad \big(x=(u+v)/2\big)
\,,
\label{wigner transform}
\end{equation}
the fermionic Wightman function can be written in a helicity
block-diagonal form
\begin{equation}
\imath S^{+-}(x;k)\equiv \imath S^{<}= \sum_{h=+,-} \imath S^<_h
\,,\qquad
 - \imath \gamma^0S^{+-}_h = (\rho^a g_{ah})\otimes\frac14(1+h\hat k\cdot\vec \sigma)
\,,
\label{S+-}
\end{equation}
where $\sigma^a,\rho^a$ ($a=0,1,2,3$) are the Pauli matrices and
$g_{ah}$ are the (off-shell) distribution functions measuring
the vector, scalar, pseudo-scalar and pseudo-vector phase space densities of fermions,
respectively. Their on-shell version
\begin{equation}
  f_{ah} = \int \frac{dk_0}{2\pi}g_{ah} \,;\qquad (a=0,1,2,3)
\label{fah:def}
\end{equation}
satisfy the following equations of motion~\cite{Prokopec:2003pj,Garbrecht:2002pd},
\begin{eqnarray}
 \dot f_{0h} &=& 0
\nonumber \\
 \dot f_{1h} + 2hk f_{2h} - 2m_I f_{3h} &=& 0
\nonumber \\
 \dot f_{2h} - 2hk f_{1h} + 2m_R f_{3h} &=& 0
\nonumber \\
 \dot f_{3h} + 2m_I f_{1h} - 2m_R f_{2h} &=& 0
\,,
\label{eom:fah}
\end{eqnarray}
where here $k\equiv \|\vec k\|$.
To make the connection with section~\ref{Mode functions for the kink profile}
and Appendix~\ref{Mode functions for a step fermion mass},
we note that one can express $f_{ah}$ in terms of $u_{\pm h}$
or $L_h$ and $R_h$ as
follows~\footnote{Note that, due to difference in conventions, there are sign differences when compared
                   with~Ref.~\cite{Garbrecht:2002pd}.}:
\begin{eqnarray}
 f_{0h} &=& |u_{+h}|^2+ |u_{-h}|^2 = |R_h|^2 + |L_h|^2
\,; \qquad
 f_{3h} = 2\Re[u_{+h}u_{-h}^*] = |L_h|^2 - |R_h|^2
\nonumber\\
 f_{1h} &=& |u_{-h}|^2-|u_{+h}|^2 = -2\Re[L_h R_h^*]
\,;\qquad\quad\!\!
f_{2h} = 2 \Im[u_{+h}u_{-h}^*] = - 2\Im[L_h R_h^*]
\,.
\label{fah}
\end{eqnarray}
such that $f_{1h}+\imath f_{2h} = -2 L_hR_h^*$.
From Eq.~(\ref{vacuum mode solutions}) and~(\ref{fah})
we immediately obtain that for $t\rightarrow -\infty $ ($z\rightarrow 0$),
\begin{equation}
 f_{0h}^- = 1\,;\qquad
 f_{1h}^- = -\frac{\Re[m_-]}{\omega_-}\,;\qquad
 f_{2h}^- = -\frac{\Im[m_-]}{\omega_-}\,;\qquad
 f_{3h}^- = -\frac{hk}{\omega_-}
 \,,
\label{fah-}
\end{equation}
where % $m_R^-=\Re[m_-]$ and  $m_I^-=\Im[m_-]$ and
we took account of
$u_{+h}u_{-h}^* = -(kh+\imath m_I)/(2\omega_-)$,
$z = \exp(2t/\tau)/[1+\exp(2t/\tau)]\rightarrow \exp(2t/\tau)$ (as $t\rightarrow -\infty$)
and of ${}_2F_1(a,b;c;0)=1$.
Inserting Eqs.~(\ref{fah-}) into the particle number definition~\cite{Garbrecht:2002pd},
\begin{equation}
 n_h(k,t) = \frac{m_Rf_{1h}+m_If_{2h}+hkf_{3h}}{2\omega}+\frac12
\,,
\label{nh}
\end{equation}
yields that $n_h(k,t)=0$ for $t\rightarrow -\infty$, as it should be since we have prepared
the initial state to be in the pure free vacuum.

\bigskip

One can also consider the statistical particle number~\cite{Prokopec:2012xv},
\begin{equation}
\bar{n}_{h\pm}=\frac12 f_{0h}\pm \frac12 \sqrt{f_{1h}^2+f_{2h}^2+f_{3h}^2}
\,.
\label{statistical particle number}
\end{equation}
A statistical particle number is defined as the particle number associated with
the basis in which the density operator is diagonal~\cite{Prokopec:2012xv}.
Statistical particle numbers can be used as a quantitative measure of state impurity,
{\it i.e.} of how much a state deviates from a pure state.
From previous work we have learned
that the statistical particle number is constant in the absence of
interactions. This can also be seen from the kinetic equations
\eqref{eom:fah}, which give
\begin{equation}
\frac{d}{dt}\left(f_{1h}^2+f_{2h}^2+f_{3h}^2\right)=0
\,.
\label{conserved sum}
\end{equation}
Of course, when interactions are included, the righthand side of Eqs.~\eqref{eom:fah}
is in general nonzero.
Here we consider a free Dirac equation~(\ref{Dirac equation:wightman}),
and therefore the statistical
particle number should remain constant. Indeed, Eqs.~(\ref{fah-}) imply
\begin{equation*}
  f_{1h}^2+f_{2h}^2+f_{3h}^2
    =\big[|u_{+h}|^2+|u_{-h}|^2\,\big]^2 = 1
\,,
\end{equation*}
such that the statistical particle
numbers of a pure state are trivial,
\begin{equation}
  \bar n_{\pm h} = \frac{1}{2}\bigg[f_{0h}\pm \sqrt{f_{1h}^2+f_{2h}^2+f_{3h}^2}\,\bigg]
   =\begin{cases}  1, \cr
                   0  \cr
    \end{cases}
  \,.
\nonumber
\end{equation}
Thus the statistical particle number is either $0$ or $1$,
the latter corresponding to a fully occupied Dirac sea.

\bigskip

The exact solutions for the phase space densities $f_{ah}$ as in
Eq. \eqref{fah} are complicated functions containing
products of hypergeometric functions, which can be analyzed numerically.
Analytically, we can study the behavior of $f_{ah}$'s in certain asymptotic limits.
By making use of Eqs.~(\ref{tilde u+-h:normalized}--\ref{alpha-beta}),
we get at asymptotically late times,
\begin{eqnarray}
\tilde u_{+h} &\stackrel{t\rightarrow \infty}{\longrightarrow}&
   \alpha_{+h} \sqrt{\frac{\omega_++(m_{1R}-m_{2R})}{2\omega_+}} {\rm e}^{-\imath\omega_+t}
  +\beta_{+h} \sqrt{\frac{\omega_+-(m_{1R}-m_{2R})}{2\omega_+}} {\rm e}^{\imath\omega_+t}
\nonumber\\
\tilde u_{-h} &\stackrel{t\rightarrow \infty}{\longrightarrow}&
   -\alpha_{-h}\frac{hk-\imath m_I}{\sqrt{k^2+m_I^2}}
        \sqrt{\frac{\omega_+-(m_{1R}-m_{2R})}{2\omega_+}} {\rm e}^{-\imath\omega_+t}
  +\beta_{-h}\frac{hk-\imath m_I}{\sqrt{k^2+m_I^2}}
          \sqrt{\frac{\omega_++(m_{1R}-m_{2R})}{2\omega_+}} {\rm e}^{\imath\omega_+t}
\label{mode functions: kink wall:late}
\end{eqnarray}
From these and Eqs.~(\ref{normalization on alpha and beta}), (\ref{alphabetaequalities})
and~(\ref{fah}) we easily obtain,
\begin{eqnarray}
  f_{0h}^+  &=& 1
\nonumber\\
  f_{1h}^+ &=&  -\frac{m_R}{\omega_+}(1-2|\beta_{\pm}|^2)
  -\frac{2\sqrt{k^2+m_I^2}}{\omega_+ }
  \big[\Re[\alpha_{\pm h}\beta_{\pm h}^{*}] \cos(2\omega_+ t)
               +\Im[\alpha_{\pm h}\beta_{\pm h}^{*}]\sin(2\omega_+ t)\big]
\nonumber\\
  f_{2h}^+ &=&  -\frac{m_I}{\omega_+}(1-2|\beta_{\pm}|^2)+
  \frac{2\cos(2\omega_+ t)}{\omega_+ \sqrt{k^2+m_I^2}}
  \big[\Re[\alpha_{\pm h}\beta_{\pm h}^{*}] m_I\Re[m_{+}]
               +\Im[\alpha_{\pm h}\beta_{\pm h}^{*}]h k\omega_+ \big]
  \nonumber \\
  &~&+\, \frac{2hk\sin(2\omega_+ t)}{\sqrt{k^2+m_I^2}}
  \big[-\Re[\alpha_{\pm h}\beta_{\pm h}^{*}]
            +\Im[\alpha_{\pm h}\beta_{\pm h}^{*}]\big]
 \nonumber\\
  f_{3h}^+ &=&  -\frac{h k}{\omega_+}(1-2|\beta_{\pm}|^2)+
  \frac{2\cos(2\omega_+ t)}{\omega_+ \sqrt{k^2+m_I^2}}
  \big[\Re[\alpha_{\pm h}\beta_{\pm h}^{*}] h k\Re[m_{+}]
               -\Im[\alpha_{\pm h}\beta_{\pm h}^{*}]\omega_+ m_I\big]
  \nonumber \\
  &~&+\, \frac{2\sin(2\omega_+ t)}{\omega_+ \sqrt{k^2+m_I^2}}
  \big[\Re[\alpha_{\pm h}\beta_{\pm h}^{*}]\omega_+ m_I
            +\Im[\alpha_{\pm h}\beta_{\pm h}^{*}]h k\Re[m_{+}]\big]
  \,.
\label{f3h-late}
\end{eqnarray}
Since we are primarily interested in the CP-violating axial vector currents,
which can bias sphalerons, here we shall focus our attention primarily to $f_{3h}$.
We can compute the CP-odd and CP-even axial vector phase space densities
as follows,
\begin{eqnarray}
  \sum_{h=\pm}f_{3h}^+ &=&
  \frac{4 |\alpha_{\pm h}||\beta^{*}_{\pm h}| m_I }{\sqrt{k^2+m_I^2}}
  \sin(2\omega_{+} t -\Delta\varphi)
  \nonumber\\
   \sum_{h=\pm}h f_{3h}^+ &=&
  \frac{2k}{\omega_+}\left[-(1-2|\beta_{\pm}|^2)
  +\frac{2 |\alpha_{\pm h}||\beta^{*}_{\pm h}| m_{+R} }{\sqrt{k^2+m_I^2}}\cos(2\omega_{+} t -\Delta\varphi)\right]
  \,,
\label{CPoddevencurrent:tanhwall-late}
\end{eqnarray}
where
\begin{align}
\Delta\varphi &= \varphi_{\alpha}-\varphi_{\beta}
\,,
\label{phaseCPf3h-late}
\end{align}
$\varphi_{\alpha}$ and $\varphi_{\beta}$ are the phases
for $\alpha_{\pm h}$ and $\beta_{\pm h}$ in Eq. \eqref{alpha-beta}.
In the thin wall limit $\gamma\rightarrow \infty$ the phases
of $\alpha_{\pm h}$ and $\beta_{\pm h}$ are zero (see Eq. \eqref{alpha-beta-thinwall})
and the CP-odd and -even phase space densities coincide with those in
Eqs. \eqref{f3-thermal} (in the free vacuum).
In the opposite 'thick wall' limit, $\gamma\rightarrow 0$,
\begin{align}
\alpha_{\pm h} &\stackrel{\gamma\rightarrow 0}{\longrightarrow} e^{i\varphi_{\alpha}}
\nonumber\\
\beta_{\pm h} &\stackrel{\gamma\rightarrow 0}{\longrightarrow}
\exp\left[-\frac{\pi(\omega_++\omega_--2m_{2R})}{2\gamma}\right]e^{i\varphi_{\beta}}
\,,
\label{alpha-beta-thickwall}
\end{align}
with
\begin{align}
\varphi_{\alpha} & \stackrel{\gamma\rightarrow 0}{\longrightarrow} \frac{1}{\gamma}
\biggl[-\omega_{-}\log(\omega_{-})-\omega_{+}\log(\omega_{+})
\nonumber\\
&+\left(\frac{\omega_{-}+\omega_{+}}{2}+m_{2R}\right)\log\left(\frac{\omega_{-}+\omega_{+}}{2}+m_{2R}\right)
+\left(\frac{\omega_{-}+\omega_{+}}{2}-m_{2R}\right)\log\left(\frac{\omega_{-}+\omega_{+}}{2}-m_{2R}\right)\biggr]
\nonumber\\
\varphi_{\beta} & \stackrel{\gamma\rightarrow 0}{\longrightarrow} \frac{1}{\gamma}
\biggl[-\omega_{-}\log(\omega_{-})+\omega_{+}\log(\omega_{+})
\nonumber\\
&+\left(\frac{\omega_{-}-\omega_{+}}{2}+m_{2R}\right)\log\left(\frac{\omega_{-}-\omega_{+}}{2}+m_{2R}\right)
-\left(\frac{-\omega_{-}+\omega_{+}}{2}+m_{2R}\right)\log\left(\frac{-\omega_{-}+\omega_{+}}{2}+m_{2R}\right)\biggr]
\,.
\label{phase-alpha-beta-thickwall}
\end{align}
Thus, in this limit the phases $\varphi_{\alpha}$
and $-\varphi_{\beta}$ grow linearly with $1/\gamma$.
In general, the oscillating CP-odd and CP-even phase space densities
in the thick wall case therefore experience a phase shift compared
to the thin wall, which can be large for small $\gamma$.
Moreover, their amplitudes, which are
proportional to $|\beta_{\pm h}|^2$ (see Eq. \eqref{CPoddevencurrent:tanhwall-late},
are generally suppressed compared to the thin wall limit.
However, in the previous
section we have demonstrated that the amplitude $|\beta_{\pm h}|$,
or particle number $n_{h\pm}$, can become much less suppressed for
a certain choice of parameters that leads to large squeezing,
see figures \ref{fig:nhnormalpop}--\ref{fig:nhinversepop}. Due
to this state large squeezing, the oscillations of the phase space densities
are amplified.

\begin{figure}[t!]
 \begin{minipage}[t]{.45\textwidth}
        \begin{center}
\includegraphics[width=\textwidth]{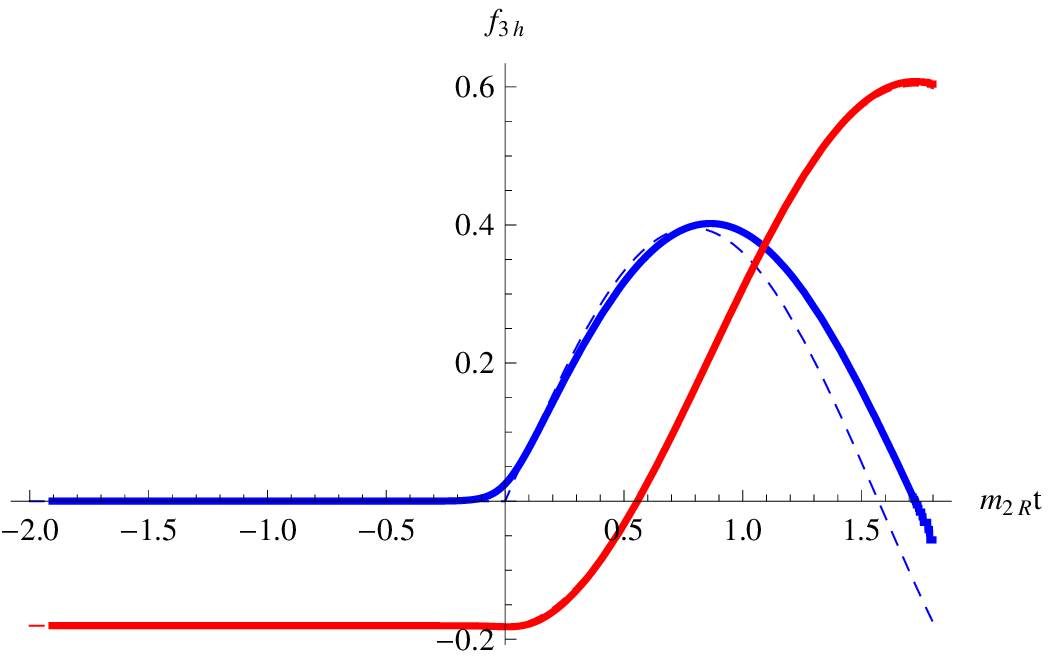}
   {\em \caption{Odd (solid blue, darker) and even (solid red, lighter) part of the exact solution for $f_{3h}$
   for $\gamma=10m_{2R}$. Parameters: $k=0.1m_{2R},~\gamma=10m_{2R},~m_{I}=0.1m_{2R},~m_{1R}=0.1m_{2R}$. The thin wall solutions (dashed)
   are constant for $t<0$, and for $t>0$ they oscillate with a frequency $2\omega_{+}$. The thin wall and exact result
   are nearly identical.
   \label{fig:plotf3hgamma10} }}
        \end{center}
    \end{minipage}
\hfill
    \begin{minipage}[t]{.45\textwidth}
        \begin{center}
\includegraphics[width=\textwidth]{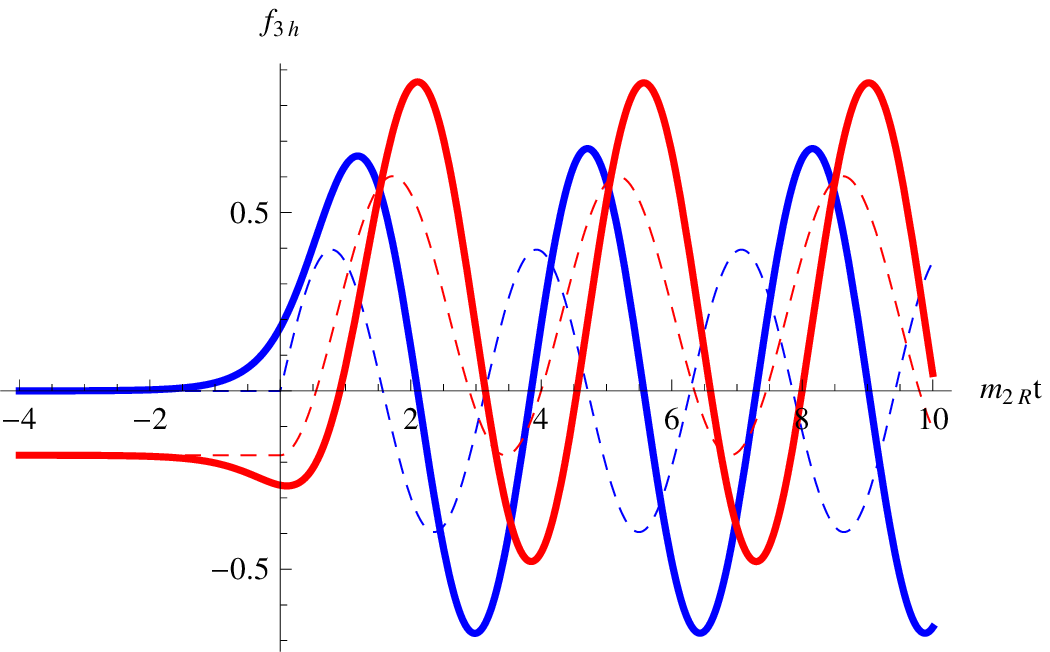}
   {\em \caption{Odd (solid blue, dark) and even (solid red, lighter) part of the exact solution for $f_{3h}$
   for $\gamma=m_{2R}$. Parameters: $k=0.1m_{2R},~\gamma=m_{2R},~m_{I}=0.1m_{2R},~m_{1R}=0.1m_{2R}$.
   For this wall of thickness $\gamma = m_{2R}$, the amplitude of the even and odd part of $f_{3h}$ is
   slightly larger than the thin wall case (shown as dashed).
   Also, there is a moderate phase shift compared to the thin wall.
   \label{fig:plotf3hgamma1} }}
        \end{center}
    \end{minipage}
\end{figure}

\begin{figure}[t!]
 \begin{minipage}[t]{.45\textwidth}
        \begin{center}
\includegraphics[width=\textwidth]{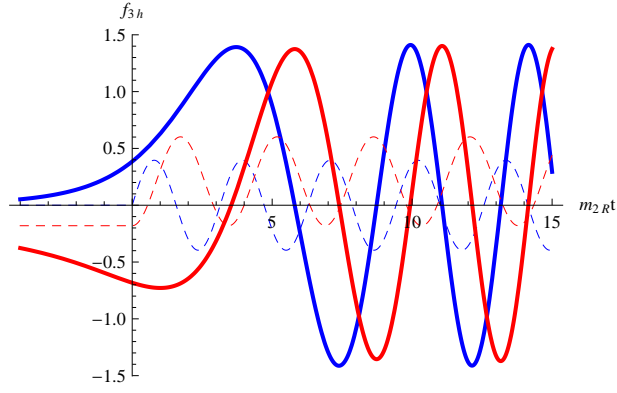}
   {\em \caption{Odd (solid blue, dark) and even (solid red, lighter) part of the exact solution for $f_{3h}$
   for $\gamma=0.1m_{2R}$ (solid line). Parameters: $k=0.1m_{2R},~\gamma=0.1m_{2R},~m_{I}=0.1m_{2R},~m_{1R}=0.1m_{2R}$.
   When CP-violation is maximal and squeezing is large (see figure \ref{fig:nhinversepop})
   there is a large oscillatory enhancement for the odd and even phase space densities
   in the thick wall regime ($\gamma = 0.1 m_{2R}$). Moreover, there is a large phase shift
   compared to the thin wall result (dashed).
   \label{fig:plotf3hgammapoint1} }}
        \end{center}
    \end{minipage}
\end{figure}

Examples of this enhanced oscillatory behavior are depicted in
figures \ref{fig:plotf3hgamma10}--\ref{fig:plotf3hgammapoint1}.
Here we show the exact solution for the odd and even part
of $f_{3h}$ using Eqs. \eqref{fah} with solutions \eqref{u+-h:normalized},
and compare to the thin wall solutions \eqref{f3-thermal}.
The parameters are chosen such to generate large squeezing,
thus $k \ll m_{2R}$ and the mass parameters are the
same as those in figure \ref{fig:nhinversepop}. Close to the thin
wall limit ($\gamma = 10 m_{2R}$, figure \ref{fig:plotf3hgamma10})
the exact solution for $f_{3h}$ almost coincides with the thin wall
result. For a thicker wall ($\gamma = m_{2R}$, figure \ref{fig:plotf3hgamma1})
the amplitude of oscillations remains large and
there is a modest phase shift compared to the thin wall. Finally,
for a thick wall ($\gamma = .1 m_{2R}$, figure \ref{fig:plotf3hgammapoint1})
there is a large oscillatory enhancement and a large phase shift.

For thick walls a large squeezing can thus give a large enhancement
of oscillations of the phase space density $f_{3h}$. However, due to the
fact that phases of different modes $\vec k$ differ, the oscillatory
behavior may (partially) disappear when the corresponding currents are computed.
These currents are related to the phase space densities $f_{ah}$ as
\begin{equation}
 j_{ah}(x) = \int \frac{d^3k}{(2\pi)^3} f_{ah}(\vec k;x)
\,,
\label{currentfromdensity}
\end{equation}
where $j_{0h}$ and $j_{3h}$ denote the 0th components of the vector and
axial vector current, respectively, and
$j_{1h}$ and $j_{2h}$ are the scalar and pseudoscalar densities, respectively.
In the following section we compare these integrated currents to those
computed in a gradient approximation. First however, we shall show how to
compute the currents for more general initial states.


\subsection{Generalized initial state}
\label{Generalized initial state}

So far the initial state has been taken in the free vacuum, such
that $n_{h}(k,t)=0$ for $t\rightarrow -\infty$. We can also
consider the initial state to be thermally occupied, such that
\begin{equation}
n_{h}(k,t)\xrightarrow{t\rightarrow -\infty}{\bar n}_{\rm th}
=\frac{1}{e^{\beta \omega_{-}}+1}
\,.
\end{equation}
The initial phase space densities that give this initial thermal state
via Eq. \eqref{nh} are now
\begin{equation}
 f_{0h}^- = 1\,;\qquad
 f_{1h}^- = -\frac{\Re[m_-]}{\omega_-}(1-2 {\bar n}_{\rm th}) \,;\qquad
 f_{2h}^- = -\frac{\Im[m_-]}{\omega_-}(1-2 {\bar n}_{\rm th})\,;\qquad
 f_{3h}^- = -\frac{hk}{\omega_-}(1-2 {\bar n}_{\rm th})
 \,,
\label{fah-thermal}
\end{equation}
Moreover, the statistical particle number \eqref{statistical particle number}
for $t\rightarrow -\infty$ is
\begin{align}
\bar{n}_{h+}&=1-{\bar n}_{\rm th}\\
\bar{n}_{h-}&={\bar n}_{\rm th}
\,.
\label{statparticle-thermal}
\end{align}
Using these relations the currents can also be written in terms of
statistical particle number
\begin{equation}
 f_{0h}^- = \bar{n}_{h+}+\bar{n}_{h-}\,;\qquad
 f_{1h}^- = -\frac{\Re[m_-]}{\omega_-}(\bar{n}_{h+}-\bar{n}_{h-}) \,;\qquad
 f_{2h}^- = -\frac{\Im[m_-]}{\omega_-}(\bar{n}_{h+}-\bar{n}_{h-})\,;\qquad
 f_{3h}^- = -\frac{hk}{\omega_-}(\bar{n}_{h+}-\bar{n}_{h-})
 \,.
\label{fah-thermal-stat}
\end{equation}
Here, one can also consider general initial densities. In that case
the statistical particle numbers $\bar{n}_{h+}$ and $\bar{n}_{h-}$
are both in the range $[0,1]$ and conserved, but
they are not necessarily related
as in an initial state that is thermally occupied. The scalar density
$f_{0h}$ is in the range $[0,2]$ and is still conserved. The initial state
may be called an initial dense state, of which the thermal state~(\ref{statparticle-thermal})
is a special case. One could further generalize the initial state to include initial squeezing,
{\it etc.} For simplicity, we shall not consider here the more general cases.

As in the previous part we can compute the phase space densities $f_{ah}$
for the kink wall, but now for general initial densities.
Also here we can study the asymptotic late time behavior analytically.
Again we are interested in the axial vector phase space density $f_{3h}$,
which becomes in the asymptotic late time limit with general initial densities,
\begin{eqnarray}
  f_{3h}^+ &=&  \Biggl(-\frac{h k}{\omega_+}(1-2|\beta_{\pm}|^2)+
  \frac{\cos(2\omega_+ t)}{2 \omega_+ \sqrt{k^2+m_I^2}}
  \left[4\Re[\alpha_{\pm h}\beta_{\pm h}^{*}] m_{+R}h k -4\Im[\alpha_{\pm h}\beta_{\pm h}^{*}]\omega_+ m_I\right]
  \nonumber \\
  &~&+ \frac{\sin(2\omega_+ t)}{2 \omega_+ \sqrt{k^2+m_I^2}}
  \left[4\Re[\alpha_{\pm h}\beta_{\pm h}^{*}]\omega_+ m_I  +4\Im[\alpha_{\pm h}\beta_{\pm h}^{*}]m_{+R}h k\right]
  \Biggr)\times(\bar{n}_{h+}-\bar{n}_{h-})
  \,.
\label{f3h-thermal-late}
\end{eqnarray}
Thus, compared to Eq. \eqref{f3h-late}, there is only an extra factor
$\bar{n}_{h+}-\bar{n}_{h-}$ in $f_{3h}$, which appears similarly in $f_{1h}$ and $f_{2h}$.
We can now compute the CP-odd and CP-even phase space densities
\begin{eqnarray}
  \sum_{h=\pm}f_{3h}^+ &=&
  \left(\frac{4 |\alpha_{\pm h}||\beta^{*}_{\pm h}| m_I }{\sqrt{k^2+m_I^2}}
  \sin(2\omega_{+} t -\Delta\varphi)\right)\times(\bar{n}_{h+}-\bar{n}_{h-})
  \nonumber\\
   \sum_{h=\pm}h f_{3h}^+ &=&
 \left( \frac{2k}{\omega_+}\left[-(1-2|\beta_{\pm}|^2)
  +\frac{2 |\alpha_{\pm h}||\beta^{*}_{\pm h}| m_{+R} }{\sqrt{k^2+m_I^2}}\cos(2\omega_{+} t -\Delta\varphi)\right]
  \right)\times(\bar{n}_{h+}-\bar{n}_{h-})
  \,,
\label{CPoddevencurrent:tanhwall-late:thermal}
\end{eqnarray}
where $\Delta\varphi$ is given in Eq. \eqref{phaseCPf3h-late}.
For comparison, in appendix \ref{Two point functions for a step fermion mass}
we compute the fermionic phase space densities $f_{ah}$ with a general initial state
for a thin wall.

\section{A comparison with gradient approximation}
\label{sec:comparison with gradient approximation}

 In Refs.~\cite{Kainulainen:2001cn,Kainulainen:2002th}
(see also Refs.~\cite{Prokopec:2003pj,Prokopec:2004ic}) it was shown that
the gradient approximation, when applied to the evolution
equations for the Wightman functions, yields a semiclassical force which affects
motion of particles in a plasma as a (planar) bubble wall of a first order electroweak transition
sweeps through the electroweak plasma. The force is of the order $\hbar$, it is proportional to the spin
orthogonal to the planar wall, and it has opposite sign for particles and antiparticles.
Since up to now no analogous analysis has been performed for the
time dependent case studied here (see however Refs.~\cite{Garbrecht:2002pd,Garbrecht:2003mn,Garbrecht:2005rr}),
we shall present such an analysis here. The Dirac equation for a Wigner transformed~(\ref{wigner transform})
 Wightman function $\mp \imath S^{\pm\mp}(u;v)=\langle \bar \psi(u)\psi(v)\rangle$ is of the form,
\begin{equation}
 \Big\{\gamma^0 k_0 - \vec \gamma \cdot \vec k + \frac{\imath}{2}\gamma^0\partial_t
 - [m_R(t) + \imath m_I(t)\gamma^5]\exp\Big(-\frac{\imath}{2}\overleftarrow{\partial}_{\!\!t}\overrightarrow{\partial}_{\!\!k_0}\Big)\Big\}
    \imath S^{+-}(k;x) = 0
    \,,
 \label{Dirac: Wigner space}
\end{equation}
where $x=(u+v)/2$ and $k^\mu$ is the Wigner momentum (the conjugate to $u-v$).
One can show that the operator in Eq.~(\ref{Dirac: Wigner space})
commutes with the helicity operator $\hat H = \hat {\vec k}\cdot\gamma^0\vec\gamma\gamma^5$, implying that
the Wightman function can be written as a sum of helicity diagonal $2\times 2$-blocks~(\ref{S+-}).
With the {\it Ansatz}~(\ref{S+-}) one can construct non-local
partial differential equations for the densities $g_{ah}$. The real and imaginary
parts of these equations yield the constraint equations (CEs) and kinetic equations (KEs),
respectively. The CEs and KEs can be subsequently solved in a gradient approximation.
The technical details of these steps are in appendix
\ref{app:Deriving the kinetic and constraint equations in gradient approximation}.
Here we only present the main results.
To first order in gradients, the CEs for $g_{0h}$ and $g_{3h}$ do not contain
$k_0$ derivatives (here $\hbar = 1$):
\begin{eqnarray}
\Big(k_0^2-|m|^2-k^2 \Big)g_{0h} &=& 0
\label{CE:0c}\\
\bigg(k_0^2-|m|^2-k^2 - h k\frac{|m|^2\partial_t\theta}{k_0^2-|m|^2}\bigg)g_{3h} &=& 0
\,,
\label{CE:3c}
\end{eqnarray}
where the mass has been written as $m_R+\imath m_I=|m|e^{\imath \theta}$, and $k=\|\vec k\|$.
General solutions to Eqs.~(\ref{CE:0c}--\ref{CE:3c}) are of the form,
\begin{equation}
 g_{0h}=\tilde g_{0h}2\pi\delta\big(k_0^2-|m|^2-k^2 \big)
\,,\qquad
 g_{3h}=\tilde g_{3h}2\pi\delta\Big(k_0^2-|m|^2-k^2 - h\frac{|m|^2\partial_t\theta}{k}\Big)
\,.
\label{CE:d}
\end{equation}
Equations~(\ref{CE:0c}--\ref{CE:d}) reveal that at first order in derivatives
(a) the vector density $g_{0h}$ does not feel any effect of a changing background,
while (b) the axial vector density  $g_{3h}$ lives on a shifted energy shell given by:
\begin{equation}
  \omega_{3h} = \omega_0 + h\frac{|m|^2\partial_t\theta}{2k\omega_0}
  \;;\qquad \omega_0(t) = \sqrt{k^2+|m(t)|^2}
  \,.
\label{shifted energy shell}
\end{equation}
In analogy to the case of a planar wall, in which case the energy shift is,
$\delta \omega_{\pm s} = \mp s |m|^2(\partial_z \theta)/\big[2\omega_0\sqrt{k_\perp^{\,2}+|m|^2}\,\big]$,
see Eq.~(\ref{shifted energy shell:planar wall}) and Ref.~\cite{Kainulainen:2002th},
where $\vec k_\perp$ is the momentum perpendicular to the wall and $s=\pm 1$ is the corresponding spin
eigenvalue, Eq.~(\ref{shifted energy shell}) shows that the axial density $g_{3h}$ lives on
a shifted energy shell produced by the zero component of a fictitious `axial-vector field'
$h (|m|^2\partial_t\theta)/(2k\omega_0)$. In this case however, the energy shift is proportional to
helicity instead, it has identical sign for positive and negative frequency states
and, as expected, it is proportional to a time derivative of the rate at which the mass argument $\theta={\rm Arg}[m]$
changes, which is a (good) measure of CP violation. Thus, just like in the case of a planar wall,
the time dependent effect is a CP-violating shift, and thus can represent a source for baryogenesis.

 In order to determine how this energy shift affects particle densities, we need to consider
kinetic equations (to second order in gradients). These are derived in
 Eqs.~(\ref{KE:0b:app}--\ref{mass relations}) of appendix
\ref{app:Deriving the kinetic and constraint equations in gradient approximation}.
For $g_{0h}$ and $g_{3h}$ the KEs are
\begin{eqnarray}
\partial_t g_{0h} + \frac{\partial_t|m|^2}{2}\partial_{k_0}\frac{g_{0h}}{k_0} &=& 0
\label{KE:0b}\\
\partial_t g_{3h} + \frac{\partial_t|m|^2}{2k_0}\partial_{k_0}g_{3h}
+h\frac{\partial_t(|m|^2\partial_t\theta)}{2kk_0}\partial_{k_0}g_{3h} &=& 0
\,.
\label{KE:3c}
\end{eqnarray}
Equation~(\ref{KE:0b}) teaches us that, as expected,
the vector density $g_{0h}$ does not feel any force at second order in gradients.
The only effect that $g_{0h}$ feels is a classical `force', which is of first order in time derivative,
and takes account of the energy non-conservation in a time dependent background $\omega_0(t)$.
On the other hand, we see from~(\ref{KE:3c}) that the time dependent energy shift
effect~(\ref{shifted energy shell}) produces as expected a second order semiclassical `force' term in
the kinetic equation for $g_{3h}$. These results are in accordance with what one would expect based on
Eqs.~(\ref{CE:0c}--\ref{CE:3c}).
Just like in the planar wall case, there is no
second order $\partial_{k_0^2}$ term; only a term containing single $k_0$ derivative
occurs in~(\ref{KE:3c}), justifying the name `force'. In fact, there is no force in~(\ref{KE:3c}).
A better analogy is the Lorentz 4-force $F^\mu$, where the 3-Lorentz force $\vec F=e(\vec E+\vec v\times\vec B)$
constitutes the spatial part of $F^\mu$, while the 0th component $F^0 = e \vec v \cdot E$ yields the rate of
energy loss in an electromagnetic field (which of course does not depend on the magnetic field $\vec B$).
Similarly, in the above equations we can identify the rate of energy loss as the zeroth component of
a 4-force,
\begin{equation}
  F^0_{h} = \frac{\partial_t |m|^2}{2\omega_{3h}} + h\frac{\partial_t(|m|^2\partial_t\theta)}{2k\omega_0}
          = \partial_t \omega_{3h}(t)
  \,,
\label{semiclassical force}
\end{equation}
where we projected $k_0\rightarrow \pm\omega_{3h}$ on-shell in Eq.~(\ref{KE:3c}).

 Now, the quantities $f_{ah}$ considered in the rest of this paper are simply related to $g_{ah}$
 {\it via} the integral~(\ref{fah:def}).
 In the light of Eq.~(\ref{CE:d}) we see that the integral~(\ref{fah:def}) just projects
$g_{ah}$ on the positive and negative frequency shells. Unless given differently
by initial conditions, the positive and negative frequency projections are the same, and this fact does
not change with time, because the semiclassical force is the same for both frequency shells. This is to be contrasted
with the planar wall case, in which case the energy shift at the order $\hbar$ has an opposite sign,
see Eqs.~(\ref{shifted energy shell:planar wall}--\ref{semiclassical force:planar wall}).
Of course, this simple picture is true only in the absence of interactions. When interactions are included,
one expects off-shell effects in $g_{ah}$, and by performing the integral~(\ref{fah:def})
one in general loses information.

 Let us now integrate Eqs.~(\ref{KE:0b}) and~(\ref{KE:3c}) over $k_0$.
Integrating the first equation is easy, and yields a conservation of vector phase space density,
\begin{equation}
\partial_t f_{0h} = 0
\label{KE:0d}
\,,
\end{equation}
which is consistent with the more general result, which states that $f_{0h}={\rm const}$ in a free theory.
Integrating Eq.~(\ref{KE:3c})) is more delicate, and yields
\begin{equation}
\partial_t f_{3h} + \bigg(\frac{\partial_t|m|^2}{2\omega_{3h}^2}
                       +h\frac{\partial_t(|m|^2\partial_t\theta)}{2k\omega_h^2}\bigg)f_{3h} = 0
\,.
\label{KE:3d}
\end{equation}
Since the expression in the parentheses is $\partial_t\ln(\omega_{3h})$, this equation can be simplified to,
\begin{equation}
\partial_t \ln(f_{3h}) = -\partial_t\ln(\omega_{3h}(t))
\,,
\label{KE:3e}
\end{equation}
and its solution is simply
\begin{equation}
  f_{3h}(t) = \frac{\omega_{3}}{\omega_{3h}(t)}f_{3h}^-
  \,,
\label{KE:3f}
\end{equation}
where $\omega_-$ is given in~(\ref{omega-+}) and $f_{3h}^-=f_{3h}(t\rightarrow -\infty)$~(\ref{fah-thermal-stat}).
This means that, if one starts with $f_{3h}=-(kh/\omega_-)\times (\bar n_{h+}-\bar n_{h-})$
(see Eq.~(\ref{fah-thermal-stat})), the gradient approximation yields,
\begin{equation}
  f_{3h}(t) = -\frac{kh}{\omega_{3h}(t)}(\bar n_{h+}-\bar n_{h-})
\,.
\label{KE:3g}
\end{equation}
This result shows that gradient approximation captures the change in the frequency felt by particles,
but does not see any quantum effects such as squeezing. Having a cursory look at
figures~\ref{fig:plotf3hgamma10}--\ref{fig:plotf3hgammapoint1} shows a striking feature:
the large and oscillating contribution in the axial vector density is completely missed in gradient approximation.
In afterthought, this should not come as surprise, since the oscillatory contributions to the densities come from
state squeezing, which is a genuinely non-adiabatic quantum effect,
and thus cannot be captured in a gradient (adiabatic) expansion.
The question is whether this failure of gradient approximation means that
an important effect is missed this way in regards to baryogenesis sources.
The answer is not so simple as the plots
in figures~\ref{fig:plotf3hgamma10}--\ref{fig:plotf3hgammapoint1} suggest.
Note that, when averaged over time,
the oscillatory contributions disappear, and one is left with a mean effect.
This mean effect is captured (to a certain extent) in gradient approximation~(\ref{KE:3g}).
Indeed, Eq.~(\ref{KE:3g}) contains  a CP-violating contribution, which is present during
the time transient, and can be extracted from the CP-odd part of Eq.~(\ref{KE:3g}),
\begin{equation}
 \sum_{h=\pm} f_{3h}(t) = \frac{|m|^2\partial_t \theta }{\omega_0^3}(1-2\bar n_{\rm th})
\,,
\label{KE:3:CP odd}
\end{equation}
where for simplicity we took an initial thermal state,
$\bar n_{h+}-\bar n_{h-}\rightarrow 1-2\bar n_{\rm th}$, $\bar n_{\rm th}=1/[\exp(\beta\omega_-)+1]$.

In order to compare the CP-odd axial density in the gradient approximation
\eqref{KE:3:CP odd} to the exact results
(see figures \ref{fig:plotf3hgamma10}--\ref{fig:plotf3hgammapoint1}), we integrate
the phase space densities over the momenta~\eqref{currentfromdensity} and sum over the helicities,
which gives the CP odd current $\sum_{h=\pm}j_{3h}$. The zero temperature part of the current however diverges as
$k\rightarrow \infty$.
Therefore we only compare the finite temperature parts of the integrated $f_{3h}$'s, that is,
only the part that is Boltzmann suppressed. Technically we compute
$\sum_{h=\pm}[j_{3h}(\beta)-{j_{3h}}(\beta\rightarrow \infty)]$, the difference between the CP-odd
axial vector current at finite temperature and the current at zero temperature.

\begin{figure}[t!]
 \begin{minipage}[t]{.45\textwidth}
        \begin{center}
\includegraphics[width=\textwidth]{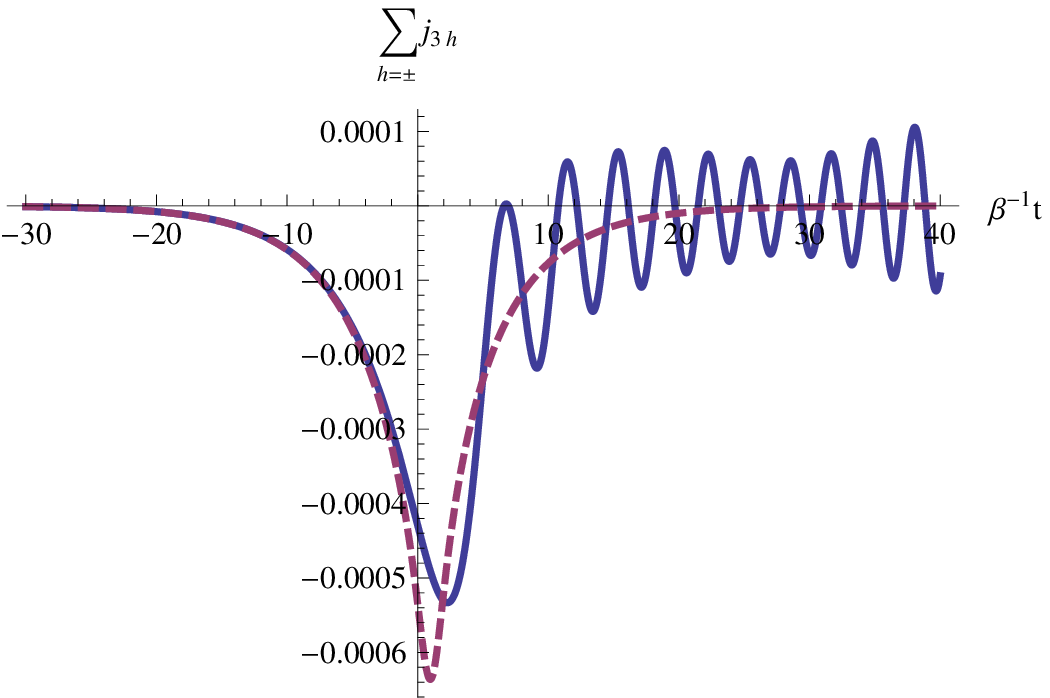}
   {\em \caption{Difference between the finite temperature and
   zero temperature CP-odd axial vector current for the exact solution (blue, solid)
    and gradient approximation (red, dashed) for a thick wall with $\gamma=0.1\beta^{-1}$.
    The mass parameters are: $m_{I}=0.1\beta^{-1},~m_{1R}=0.1\beta^{-1},~m_{2R}=\beta^{-1}$.
   \label{fig:gradexpfr3heventhick} }}
        \end{center}
    \end{minipage}
\hfill
    \begin{minipage}[t]{.45\textwidth}
        \begin{center}
\includegraphics[width=\textwidth]{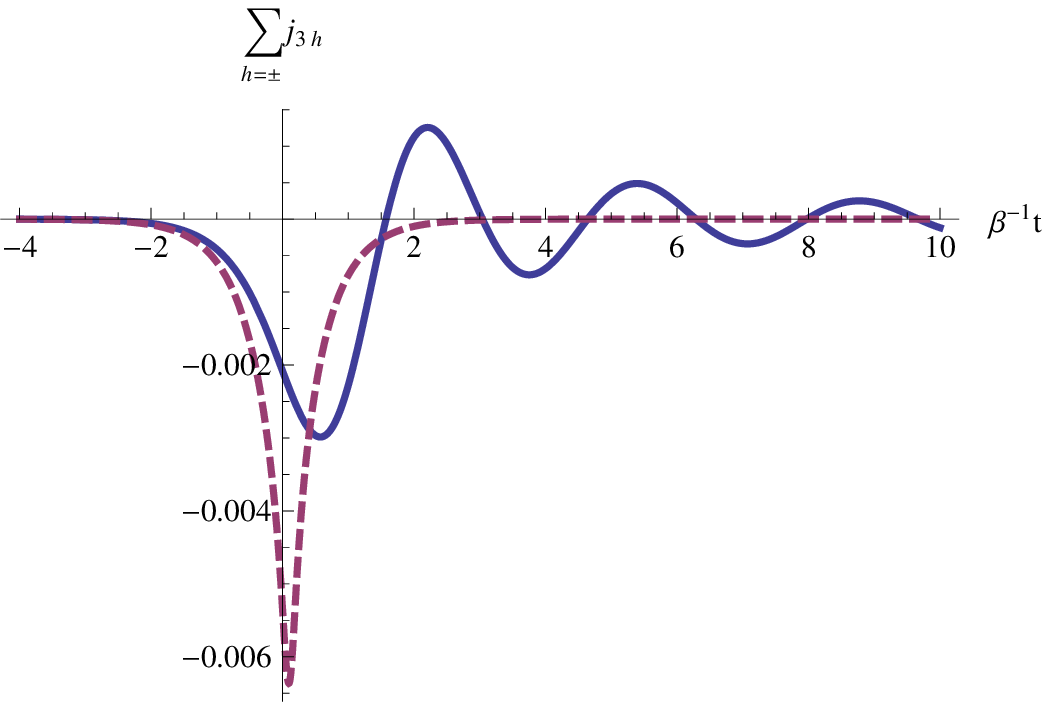}
   {\em \caption{Difference between the finite temperature
   and zero temperature CP odd axial vector current for the exact solution (blue, solid)
    and gradient approximation (red, dashed) for a wall with $\gamma=\beta^{-1}$.
    The mass parameters are: $m_{I}=0.1\beta^{-1},~m_{1R}=0.1\beta^{-1},~m_{2R}=\beta^{-1}$.
   \label{fig:gradexpfr3heventhinnish} }}
        \end{center}
    \end{minipage}
\end{figure}

\begin{figure}[t!]
 \begin{minipage}[t]{.45\textwidth}
        \begin{center}
\includegraphics[width=\textwidth]{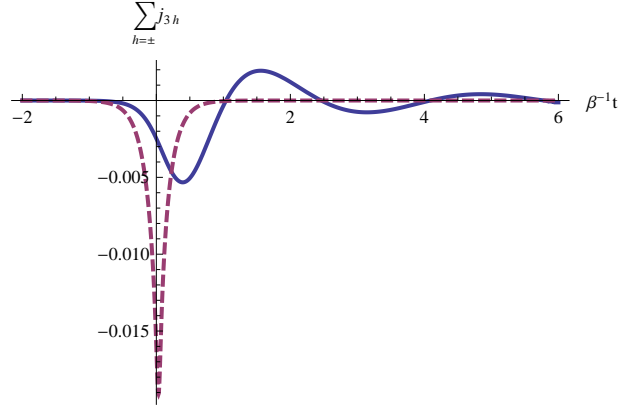}
   {\em \caption{Difference between the finite temperature
   and zero temperature CP odd axial vector current for the exact solution (blue, solid)
    and gradient approximation (red, dashed) for a wall with $\gamma=3\beta^{-1}$.
    The mass parameters are: $m_{I}=0.1\beta^{-1},~m_{1R}=0.1\beta^{-1},~m_{2R}=\beta^{-1}$.
   \label{fig:gradexpfr3heventhin} }}
        \end{center}
    \end{minipage}
\end{figure}

In figures \ref{fig:gradexpfr3heventhick}, \ref{fig:gradexpfr3heventhinnish}
and \ref{fig:gradexpfr3heventhin}
we show the finite part of the CP-odd current for the gradient expansion
and for the exact solution, for a thick wall with $\gamma = 0.1 \beta^{-1}$,
a wall of thickness $\gamma = \beta^{-1}$, and a thin wall with
$\gamma = 3 \beta^{-1}$ respectively.~\footnote{Note that, based on
Eqs. \eqref{thick wall criterion} and \eqref{thick wall criterion:2}
in the introduction, modes of both the thin and thick wall regime contribute to the integrated
density, \textit{i.e.} to the current. The question is therefore, whether
the currents in Figs. \ref{fig:gradexpfr3heventhick}--\ref{fig:gradexpfr3heventhin} are dominated
by modes that satisfy the thick wall condition, or by thin wall modes. Roughly speaking,
for $\beta\gamma \ll 1$ the current is largely dominated by thick wall modes, whereas
for $\beta\gamma \gg 1$ modes that satisfy the thin wall condition contribute mostly to the current.}
The mass parameters are chosen such that there is a large state squeezing.
The gradient expansion captures quite well the main trend, but misses the oscillations
at later times. It is intriguing that already for the wall with $\gamma = \beta^{-1}$,
the exact solution for the current starts to look quite different from the current in the gradient
approximation. For the thin wall in Fig. \ref{fig:gradexpfr3heventhin} the difference is even more
significant. It would be interesting to explore in more detail the CP-violating
current in the thinner wall regime.
Needless to say, in order to make a definite statement about the
validity of gradient approximation,
one needs to perform a more detailed analysis which includes scatterings coming
from quantum loop effects.
\bigskip

%%%%%%%%%%%%%%%%%%%%%%%%%%%%%%%%%%%%%%%%%%%%%%%%%%%%
%%%%%%%%%%%%%% Conclusions %%%%%%%%%%%%%%%%%%%%%%%%%
%%%%%%%%%%%%%%%%%%%%%%%%%%%%%%%%%%%%%%%%%%%%%%%%%%%%

\section{Conclusions and discussion}
\label{Conclusions and discussion}

In this work we derive an exact solution of the Dirac equation
for fermions with a time dependent mass, generated by a scalar Higgs condensate.
We assume that the mass has a $\tanh$ time dependence, which represents a quite realistic model
for phase interface (bubble wall) at a first order phase transition
in the early Universe setting. Moreover, the mass is complex with a phase changing in time, which
can be a source for CP violation. We have studied this CP violation by looking
at the CP-odd part of the axial vector current, since that current biases sphaleron transitions.

As already emphasized in the introduction the
division between "thin wall" and "thick wall" cases depends on the relevant
momenta, and both cases are present in a typical (say thermal) distribution.
For large state squeezing, \textit{i.e.} a large late time Bogoliubov particle production,
the axial vector phase space density for a thick wall experiences a
large oscillatory enhancement and phase shift with respect to the thin
wall case. This non-adiabatic behavior cannot be captured by the gradient
approximation. However, the mean effect for the axial vector current, which is obtained
from the phase space density after summing the momenta, is described
reasonably well in gradient approximation. Still, the exact solution
for the axial vector current shows that the difference for thinner walls
can be quite significant. This invites for a more detailed quantitative study of
baryogenesis sources in the thin wall regime.

Extensions to our work can be foreseen. First of all, instead of
a time dependent mass, our analysis can be generalized to a planar wall
case, in which the mass is dependent on one spatial coordinate.
This extension is in principle straightforward, and has been considered
for the CP-conserving case in Ref.~\cite{Ayala:1993mk,Funakubo:1994yt},
except that we intend to work now with full inclusion of a CP-violating mass.
Moreover, whereas we consider
one non-interacting fermionic species, one could also study multiple
flavors. In that case, there is an additional CP-violating source in the
flavor mixing mass matrix. It would be interesting to see what
is the dominant source of CP violation depending on wall thickness.
Finally, we did not include other effects such as plasma scatterings,
and it would be worthwhile to explore how these affect the analytical results
in this work.

\section{Acknowledgements}

We thank Kimmo Kainulainen and Pyry Rahkila for useful discussions
concerning the gradient expansion.
TP and JW were in part supported by the Dutch Foundation for
'Fundamenteel Onderzoek der Materie' (FOM) under the program
"Theoretical particle physics in the era of the LHC", program number FP 104.

\appendix

%%%%%%%%%%%%%%%%%%%%%%%%%%%%%%%%%%%%%%%%%%%%%%%%%%%%%%%%%%%%%%%%%%%%%
%%%%%%%%%%%%%%%%%%%% A P P E N D I C E S %%%%%%%%%%%%%%%%%%%%%%%%%%%%
%%%%%%%%%%%%%%%%%%%%%%%%%%%%%%%%%%%%%%%%%%%%%%%%%%%%%%%%%%%%%%%%%%%%%

\section{Mode functions for a constant mass}
\label{Thin wall}

In this appendix we construct the fundamental solutions of
Eqs. \eqref{eom:LhRh}--\eqref{eom:barLhRh} for a constant mass.
In appendix \ref{Mode functions for a step fermion mass} we
treat the thin wall case.

When the mass is constant,
the general solution of Eqs.~(\ref{eom:LhRh}--\ref{eom:barLhRh})
(see also Eqs.~(\ref{Dirac equation}--\ref{LR projectors})) can be written as plane waves,
\begin{eqnarray}
 L_h &=& A_h {\rm e}^{-\imath \omega t} + B_h {\rm e}^{\imath \omega t}
\,;\qquad
R_h = C_h {\rm e}^{-\imath \omega t} + D_h {\rm e}^{\imath \omega t}
\nonumber\\
\bar L_h &=& \bar A_h {\rm e}^{\imath \omega t} + \bar B_h {\rm e}^{-\imath \omega t}
\,;\qquad
\bar R_h = \bar C_h {\rm e}^{\imath \omega t} + \bar D_h {\rm e}^{-\imath \omega t}
\,,
\label{LhRh:solutions}
\end{eqnarray}
where $A_h$, $B_h$, $C_h$, $D_h$,
$\bar A_h$, $\bar B_h$, $\bar C_h$ and $\bar D_h$ are constants.
The first order equations~(\ref{eom:LhRh}--\ref{eom:barLhRh})
tell as that $C_h$ and $D_h$ can be expressed in terms of $A_h$ and $B_h$:
\begin{equation}
C_h = \frac{m^*}{\omega - h k}A_h = \frac{\omega + h k}{m}A_h
    = \frac{m^*}{|m|}\sqrt{\frac{\omega + h k}{\omega - h k}}A_h
\,;\qquad
D_h = -\frac{m^*}{\omega + h k}B_h = -\frac{\omega - h k}{m}B_h
    = -\frac{m^*}{|m|}\sqrt{\frac{\omega - h k}{\omega + h k}} B_h
\,.\;\;
\label{ChDh:vs:AhBh}
\end{equation}
Notice that these equations also imply an on-shell condition,
$\omega^2 = k^2 +|m|^2$.
Analogous relations hold for the barred constants,
\begin{equation}
\bar C_h = -\frac{m^*}{\omega - h k}\bar A_h = -\frac{\omega + h k}{m}\bar A_h
    = - \frac{m^*}{|m|}\sqrt{\frac{\omega + h k}{\omega - h k}}\bar A_h
\,;\quad
\bar D_h = \frac{m^*}{\omega + h k}\bar B_h = \frac{\omega - h k}{m}\bar B_h
    = \frac{m^*}{|m|}\sqrt{\frac{\omega - h k}{\omega + h k}}\bar B_h
\,.\;\;
\label{barChDh:vs:barAhBh}
\end{equation}
These constants can be further constrained by imposing the vector current
conservation law, which in the absence of flavour mixing
becomes particularly simple, $\partial_t j^0(x) = 0$, or equivalently
$j^0(x) = \langle\hat\psi^\dagger(x)\hat\psi(x)\rangle= {\rm const}$.
In order to fix the constant, notice that
\begin{equation}
   \big\langle\hat\psi_\beta^\dagger(\vec x,t)\hat\psi_\alpha(\vec x^\prime,t)
    \big\rangle
=\frac12\big\langle\{\hat\psi_\beta^\dagger(\vec x,t),\hat\psi_\alpha(\vec x^\prime,t)\}
             \big\rangle
   + \frac12\big\langle[\hat\psi_\beta^\dagger(\vec x,t),\hat\psi_\alpha(\vec x^\prime,t)]
             \big\rangle
   = \frac12\delta_{\alpha\beta}\delta^3(\vec x-\vec x^\prime)
     -F_{\alpha\beta}(\vec x,t;\vec x^\prime,t)
\,,
\label{Wightman and Hadamard}
\end{equation}
where $F_{\alpha\beta}(\vec x,t;\vec x^\prime,t)$ the Hadamard (statistical)
Green function, which in the free space vanishes~\footnote{This definition
of the Hadamard function differs by a constant from the definition
used in {\it e.g.} Ref.~\cite{Prokopec:2012xv},
which is due to the normal ordering of
the creation and annihilation operators, assumed in the construction
of the density operator in Ref.~\cite{Prokopec:2012xv}.}.
When written in momentum space
\begin{equation}
\hat\psi(\vec x,t) = \int \frac{d^3k}{(2\pi)^3}{\rm e}^{\imath \vec k\cdot \vec x}
  \hat\psi(\vec k,t)
\end{equation}
Eq.~(\ref{Wightman and Hadamard}) becomes,
\begin{equation}
   \big\langle\hat\psi_\beta^\dagger(\vec k,t)\hat\psi_\alpha(\vec k,t)
    \big\rangle
=\frac12\big\langle\{\hat\psi_\beta^\dagger(\vec k,t),\hat\psi_\alpha(\vec k,t)\}
             \big\rangle
   + \frac12\big\langle[\hat\psi_\beta^\dagger(\vec k,t),\hat\psi_\alpha(\vec k,t)]
             \big\rangle
    = \frac12\delta_{\alpha\beta}
\,,
\label{Wightman and Hadamard:2}
\end{equation}
where the expectation value was taken with respect to the vacuum state
$|\Omega\rangle$, in which case $F=0$. Similarly we have,
\begin{equation}
   \big\langle\hat\psi_\alpha(\vec k,t)\hat\psi_\beta^\dagger(\vec k,t)
    \big\rangle
= \frac12\delta_{\alpha\beta} + F_{\alpha\beta}(k,t;t)
   \rightarrow \frac12\delta_{\alpha\beta}
\,.
\label{Wightman and Hadamard:3}
\end{equation}
Taking a trace of~(\ref{Wightman and Hadamard:2}--\ref{Wightman and Hadamard:3})
one finds,
\begin{equation}
 \sum_h(|\bar L_h|^2+|\bar R_h|^2) = \frac12{\rm Tr}[\delta_{\alpha\beta}]=2
\,,\qquad
 \sum_h(|L_h|^2+|R_h|^2) = \frac12{\rm Tr}[\delta_{\alpha\beta}]=2
\,,
\nonumber
\end{equation}
which implies:
\begin{equation}
 |\bar L_h|^2+|\bar R_h|^2 =1 =  |L_h|^2+|R_h|^2
\,,
\label{normalization}
\end{equation}
where we assumed that the vacuum state normalization is independent of helicity.
Moreover, since $\xi_h^\dagger \cdot \xi_{h^\prime} = \delta_{hh^\prime}$,
Eq.~(\ref{normalization}) agrees with~(\ref{mode normalization conditions}).
Together with~(\ref{ChDh:vs:AhBh}--\ref{barChDh:vs:barAhBh}) the
condition~(\ref{normalization}) allows one to completely specify
the vacuum fermionic mode functions for constant mass (up to an overall phase),
\begin{eqnarray}
 L_h(k,t) &=& \sqrt{\frac{\omega-hk}{2\omega}}{\rm e}^{-\imath\omega t}
\,,\qquad
  R_h(k,t) = \sqrt{\frac{\omega+hk}{2\omega}}\frac{m^*}{|m|}
                              {\rm e}^{-\imath\omega t}
\nonumber\\
 \bar L_h(k,t) &=& \sqrt{\frac{\omega-hk}{2\omega}}{\rm e}^{\imath\omega t}
\,,\qquad
 \bar R_h(k,t) = -\sqrt{\frac{\omega+hk}{2\omega}}\frac{m^*}{|m|}
                              {\rm e}^{\imath\omega t}
\label{vacuum mode solutions}
\end{eqnarray}
Notice that these solutions satisfy the mode normalization
conditions~(\ref{normalization}). Moreover,
(when summed over $h$) they also satisfy the (stronger) consistency
condition~(\ref{consistency condition:mode functions}).
The only remaining conditions to check are the mode orthogonality
conditions~(\ref{mode orthogonality}). They imply,
\begin{equation}
 \bar L_h^* R_h +  \bar R_h^* L_h =0
\nonumber
\end{equation}
For the solutions above, this implies,
\begin{equation}
 {\rm e}^{-2\imath \omega t} \frac{1}{2\omega}\Big[m^*-m\Big] = 0
\,,
\nonumber
\end{equation}
which can be only satisfied if $\Im[m]=0$. There is no problem
when $m$ is time independent. In this case one can perform
a global rotations on spinors that removes the imaginary part of the mass,
\begin{equation}
\psi\rightarrow {\rm e}^{-\imath \theta \gamma^5}\psi
  = [\cos(\theta)-\imath \sin(\theta) \gamma^5]\psi
\,, \qquad
\bar\psi\rightarrow \bar\psi {\rm e}^{-\imath \theta \gamma^5}
  = \bar\psi[\cos(\theta)-\imath \sin(\theta) \gamma^5]
\,,
\nonumber
\end{equation}
where
\begin{equation}
\tan(2\theta) = \frac{m_I}{m_R}
\,,\qquad
\cos(2\theta) = \frac{m_R}{|m|}
\,,\qquad
\sin(2\theta) = \frac{m_I}{|m|}
\,,
\nonumber
\end{equation}
with which
\begin{equation}
 -m\bar\psi\psi_R -m^*\bar\psi\psi_L
 = -m_R\bar\psi\psi - \imath m_I\bar\psi\gamma^5\psi
 \rightarrow -|m|\bar\psi\psi
\,.
\nonumber
\end{equation}
Of course, this rotation is global, and works only if $m$ is time independent.
Here we are interested in a time dependent problem, and hence
the mode functions are not orthogonal as in~\eqref{mode orthogonality}.
Is this a problem? Not necessarily. What is important is that
the mode functions span the whole Hilbert space. The most important condition
that must be satisfied is in the end Eq.~\eqref{mode normalization conditions}.
Note also that

\begin{equation}
 \bar\chi_h(\vec k,t)\cdot \chi_h(\vec k,t) = \frac{m_R}{\omega}
   = -\bar\nu_h(\vec k,t)\cdot \nu_h(\vec k,t)
\,.
\label{mode normalization 2}
\end{equation}

\section{Mode functions for a step fermion mass}
\label{Mode functions for a step fermion mass}

We shall now assume that the mass takes a simple form, such that it exhibits
a sudden jump at $t=0$:
\begin{equation}
m(t) = m_- \Theta(-t) + m_+ \Theta(t)
\label{mass:1}
\end{equation}
where $m_-$ and $m_+$ are (in general complex) constant masses at negative and
positive times, respectively. This is what we refer to as the thin wall mass profile.
%~\footnote{
One can easily convince oneself that
a constant $U(1)$ rotation of the left- and right-handed spinors in \eqref{Lagrangian:free}
can remove a jump in the {\it e.g.} imaginary part of the mass.
After performing such a rotation the mass can be written
as $m_{\pm}=|m_{\pm}|\exp[\imath (\phi_{\pm}+\chi)]$, where
$\chi$ is the relative phase between left- and right-handed spinors.
One should then solve
$|m_+| \sin(\phi_++\chi)=|m_-|\sin(\phi_{-}+\chi)$ for $\chi$.
\footnote{The exact solution gives
%\begin{equation*}
\[
\chi=\arctan\left(-\frac{m_{-I}-m_{+I}}{m_{-R}-m_{+R}}\right)
\]
%\end{equation*}
}
Therefore, without any loss of generality, one can assume that
$\Im[m(t)]={\rm const.}$, or equivalently, $\Im[m_+]=\Im[m_-]$.
This will be important for decoupling of the equations for $u_{\pm h}$ defined
below (see also Eq.~(\ref{mass: real imaginary})).
% }

 For the problem at hand with $m(t)$ given in~(\ref{mass:1}),
we can make the following {\it Ans\"atze} for the mode functions
({\it cf.} Eqs.~(\ref{vacuum mode solutions})),
\begin{eqnarray}
 L_h(k,t) &=& \theta(-t)L_h^-
           + \theta(t)L_h^+
\nonumber\\
  R_h(k,t) &=& \theta(-t)R_h^-
           + \theta(t)R_h^+
\nonumber\\
 \bar L_h(k,t) &=& \theta(-t)\bar L_h^-
           + \theta(t)\bar L_h^+
\nonumber\\
 \bar R_h(k,t) &=& \theta(-t)\bar R_h^-
           + \theta(t)\bar R_h^+
\,,
\label{mode solutions:2}
\end{eqnarray}
where the solutions for $t<0$ are vacuum mode functions
derived in \eqref{vacuum mode solutions},
\begin{equation}
L_h^- = \sqrt{\frac{\omega_--hk}{2\omega_-}}
                           {\rm e}^{-\imath\omega_- t}
\,;\quad
R_h^- = \sqrt{\frac{\omega_-+hk}{2\omega_-}}\frac{m_-^*}{|m_-|}
                              {\rm e}^{-\imath\omega_- t}
\,;\quad
\bar L_h^-
  = \sqrt{\frac{\omega_--hk}{2\omega_-}}
                                                   {\rm e}^{\imath\omega_- t}
\,;\quad
\bar R_h^-
   = -\sqrt{\frac{\omega_-+hk}{2\omega_-}}\frac{m_-^*}{|m_-|}
                                                          {\rm e}^{\imath\omega_- t}
\,.\;\;
\label{Lh-Rh-}
\end{equation}
The solutions for $t>0$ can now be written as a linear combination
of the normalized positive and negative frequency solutions
\begin{eqnarray}
 L_h^+ &\equiv& L_h(t>0)=\alpha_h^+\sqrt{\frac{\omega_+-h k}{2\omega_+}}
                        {\rm e}^{-\imath\omega_+ t}
 +\beta_h^+\sqrt{\frac{\omega_++h k}{2\omega_+}}{\rm e}^{\imath\omega_+ t}
\nonumber\\
R_h^+ &\equiv& R_h(t>0)=\alpha_h^+\sqrt{\frac{\omega_++hk}{2\omega_+}}\frac{m_+^*}{|m_+|}
                          {\rm e}^{-\imath\omega_+ t}
-\beta_h^+\sqrt{\frac{\omega_+-hk}{2\omega_+}}\frac{m_+^*}{|m_+|}
               {\rm e}^{\imath\omega_+ t}
\nonumber\\
\bar L_h^+ &\equiv& \bar L_h(t>0)=\bar\alpha_h^+\sqrt{\frac{\omega_+-h k}{2\omega_+}}
                        {\rm e}^{\imath\omega_+ t}
 +\bar\beta_h^+\sqrt{\frac{\omega_++h k}{2\omega_+}}{\rm e}^{-\imath\omega_+ t}
\nonumber\\
\bar R_h^+ &\equiv& \bar R_h(t>0)=-\bar \alpha_h^+\sqrt{\frac{\omega_++hk}{2\omega_+}}\frac{m_+^*}{|m_+|}
                          {\rm e}^{\imath\omega_+ t}
+\bar\beta_h^+\sqrt{\frac{\omega_+-hk}{2\omega_+}}\frac{m_+^*}{|m_+|}
               {\rm e}^{-\imath\omega_+ t}
\,,
\label{Lh+Rh+}
\end{eqnarray}
where
\begin{equation}
 \omega_\pm = \sqrt{k^2 + |m_\pm|^2}
\,,
\nonumber
\end{equation}
and the solutions multiplying $\beta_h^+$ and $\bar\beta_h^+$ are
the normalized negative frequency solutions of Eqs.~\eqref{eom:LhRh}--\eqref{eom:barLhRh}.
The four Bogoliubov coefficients in~(\ref{mode solutions:2})
are determined by the matching
conditions. Eqs.~(\ref{eom:LhRh}--\ref{eom:barLhRh}) together with
the structure of the mass term~(\ref{mass:1}) tell us that
the mode functions must be continuous at $t=0$, which implies the
following four matching conditions
\begin{equation}
  L_h(t=0-) = L_h(t=0+)\,;\quad R_h(t=0-) = R_h(t=0+)
\,;\quad
  \bar L_h(t=0-) = \bar L_h(t=0+)\,;\quad \bar R_h(t=0-) = \bar R_h(t=0+)
\,.
\label{matching}
\end{equation}
These conditions imply for the Bogoliubov coefficients in
the solutions for $L_h$ and $R_h$ in Eq.~(\ref{mode solutions:2}):
\begin{eqnarray}
 \alpha_h^+ =\bar\alpha_h^+&=& \frac{|m_+|}{2\sqrt{\omega_+\omega_-}}
                \left(\sqrt{\frac{\omega_--h k}{\omega_++h k}}
                +\frac{m_-^*}{m_+^*}\sqrt{\frac{\omega_++h k}{\omega_--h k}}
                \right)
\nonumber\\
\beta_h^+ =\bar\beta_h^+&=& \frac{|m_+|}{2\sqrt{\omega_+\omega_-}}
                \left(\sqrt{\frac{\omega_--h k}{\omega_+-h k}}
                -\frac{m_-^*}{m_+^*}\sqrt{\frac{\omega_+-h k}{\omega_--h k}}
                \right)
\,.
\label{alphah+betah+}
\end{eqnarray}
By observing that $|\alpha_h^+|^2+|\beta_h^+|^2=1$
one can easily check that these conditions satisfy the correct normalization
condition~(\ref{normalization}).

To summarize, our simple calculation shows that, as a consequence of a sudden mass change (at $t=0$),
the number of fermions (of each helicity $h$) produced is
\begin{equation}
 n_h = |\beta_h^+|^2 = \frac{|m_+-m_-|^2-(\omega_+-\omega_-)^2}{4\omega_+\omega_-}
     = \frac12 - \frac{k^2+\Re[m_+m_-^*]}{2\omega_+\omega_-}
\label{number produced}
\end{equation}
which is helicity independent and, in general, does not vanish.
One can show that $n_h$ in (\ref{number produced}) satisfies,
$0\leq n_h(k)\leq 1$ for arbitrary $\vec k$, $m_+$ and $m_-$, which is a nontrivial
check of the correctness of (\ref{number produced}).
Note finally that, when $k=0$,
\begin{equation}
 n_h = \frac{1}{2}\left[1-\frac{\Re[m_+m_-^*]}{|m_+|\,|m_-|}\right]
 \,.
\label{number produced:k=0}
\end{equation}
This last result nicely illustrates the dependence of particle
production on the complex phase of the mass. Let us denote
$m_\pm = |m_\pm|{\rm e}^{\imath \theta_\pm}$. Then
\begin{equation}
 n_h = \frac{1}{2}\left[1-\cos(\theta_+-\theta_-)\right]
     = \sin^2\left(\frac{\theta_+-\theta_-}{2}\right)
\,,
\label{number produced:k=0:2}
\end{equation}
which shows that $n_h$ varies between $0$ and $1$,
as it should. There is no particle production when
$\theta_+-\theta_- = 2\pi n$ ($n\in Z$), in which case there is
no CP violation. On the other hand, when $k=0$ particle production maximizes
for $\theta_+-\theta_- = \pi (2n+1)$ ($n\in Z$),
in which case CP violation is maximal. This is in accordance with the results of
finite wall thickness~(\ref{alpha and beta squared}--\ref{n:thick wall limit}),
shown in figures~\ref{fig:nhnormalpop}--\ref{fig:nhinversepop}.

\section{Two point functions for a step fermion mass}
\label{Two point functions for a step fermion mass}

In this section we consider the evolution of the two point functions
with a Heaviside step function mass~(\ref{mass:1}).
Here we heavily use Ref.~\cite{Garbrecht:2002pd}.
By making use of \eqref{fah} we can find the phase space
densities for the thin wall mass profile. For $t<0$, the
phase space densities are given in Eq. \eqref{fah-}.
Together with Eqs.~\eqref{Lh+Rh+}, Eqs.~(\ref{mode solutions:2})
can be used to compute  $f_{ah}$ for $t>0$.

Instead of calculating the $f_{ah}$ by using the explicit expressions
for the mode functions, we can also derive the phase space densities
from the kinetic equations \eqref{eom:fah}. We consider a generalized initial
state, for which the $f_{ah}$ for $t<0$ are given by Eq.~\eqref{fah-thermal-stat}.
Next we solve the $f_{ah}$ for $t>0$ from the kinetic equations~\eqref{eom:fah}.
In order to do so,
note first that Eqs.~(\ref{Lh+Rh+}) and~(\ref{fah}) imply that the general form of
the solutions for $t>0$ is,
\begin{equation}
  f_{ah}^+ = \alpha_{ah} \cos(2\omega_+ t) + \beta_{ah} \sin(2\omega_+ t) + \gamma_{ah}
\,\qquad (a=0,1,2,3)
\,,
\label{fah:general}
\end{equation}
where $\alpha_{ah}$, $\beta_{ah}$ and $\gamma_{ah}$ are constants that can be determined
from the matching conditions at $t=0$. Although $m$ experiences a finite jump at $t=0$,
the structure of the equations of
motion~(\ref{eom:fah}) implies that all of $f_{ah}$ must be continuous at $t=0$:
\begin{equation}
 f_{ah}^-(k,t=0) = f_{ah}^+(k,t=0)
\,.
\label{fah:matching}
\end{equation}
The first equation in~(\ref{eom:fah}) tells us that the vector particle density
$f_{0h}$ cannot depend on time, {\it i.e.} that $\alpha_{0h}=\beta_{0h}=0$,
and we have:
\begin{equation}
  f_{0h}^+ = \bar{n}_{h+}+\bar{n}_{h-}
\,.
\label{f0h+}
\end{equation}
The other three equations in~(\ref{eom:fah}) give nine conditions among the parameters
$\{\alpha_{ih},\beta_{ih},\gamma_{ih}\}$ $(i=1,2,3)$. These conditions represent a highly
degenerate system, such that the following independent conditions remain:
\begin{equation}
 m_R^+\alpha_{1h}+m_I^+\alpha_{2h}+hk\alpha_{3h}=0\,;\qquad m_R^+\beta_{1h}+m_I^+\beta_{2h}
 +hk\beta_{3h}=0\,;\qquad \gamma_{2h}=\frac{m_I^+}{m_R^+}\gamma_{1h}\,;\qquad
 \gamma_{3h}=\frac{hk}{m_R^+}\gamma_{1h}
\,.
\label{constraints1}
\end{equation}
In addition, $\beta_{ih}$ are related to $\alpha_{ih}$ as follows:
\begin{equation}
 \beta_{1h} = -\frac{hk}{\omega_+}\alpha_{2h}+\frac{m_I^+}{\omega_+}\alpha_{3h}
\,;\qquad
 \beta_{2h} = \frac{hk}{\omega_+}\alpha_{1h}-\frac{m_R^+}{\omega_+}\alpha_{3h}
\,;\qquad
 \beta_{3h} = -\frac{m_I^+}{\omega_+}\alpha_{1h}+\frac{m_R^+}{\omega_+}\alpha_{2h}
 \,.
 \label{beta_vs_alpha}
\end{equation}
  Furthermore, the matching conditions~(\ref{fah:matching}) result in:
\begin{equation}
-\frac{m_R^-}{\omega_-}(\bar{n}_{h+}-\bar{n}_{h-}) = \alpha_{1h}+\gamma_{1h}\,;\qquad
-\frac{m_I^-}{\omega_-}(\bar{n}_{h+}-\bar{n}_{h-}) = \alpha_{2h}+\gamma_{2h}\,;\qquad
-\frac{hk}{\omega_-}(\bar{n}_{h+}-\bar{n}_{h-}) = \alpha_{3h}+\gamma_{3h}\,,\qquad
\label{matching conditions:2}
\end{equation}
which, together with Eqs.~(\ref{constraints1}--\ref{beta_vs_alpha})
completely specify $f_{ih}$. When the solutions for the $\alpha_{ih}$, $\beta_{ih}$
and $\gamma_{ih}$ are inserted into the general form \eqref{fah:general}, we find
\begin{eqnarray}
  f_{1h}^+ \!&=&\! \left(\left[-\frac{m_R^-}{\omega_-}+\frac{m_R^+}{\omega_+}\frac{k^2+\Re[m_+m_-^*]}{\omega_+\omega_-}\right]
                   \! \cos(2\omega_+ t)
     + \left[\frac{h k}{\omega_-\omega_+}(m_I^{-}-m_I^{+})\right] \!\sin(2\omega_+ t)
      -\frac{m_R^+}{\omega_+}\frac{k^2+\Re[m_+m_-^*]}{\omega_+\omega_-}\right)
\nonumber\\
   &&\hskip 12cm
      \times(\bar{n}_{h+}-\bar{n}_{h-})\quad
\label{f1h+:2}
\\
 f_{2h}^+ \!&=&\! \left(\left[-\frac{m_I^-}{\omega_-}+\frac{m_I^+}{\omega_+}\frac{k^2+\Re[m_+m_-^*]}{\omega_+\omega_-}\right]
                   \! \cos(2\omega_+ t)
     - \left[\frac{h k}{\omega_-\omega_+}(m_R^{-}-m_R^{+})\right]\! \sin(2\omega_+ t)
      \!-\!\frac{m_I^+}{\omega_+}\frac{k^2+\Re[m_+m_-^*]}{\omega_+\omega_-}\right)
\nonumber\\
   &&\hskip 12cm
      \times(\bar{n}_{h+}-\bar{n}_{h-})\quad\;\;
\label{f2h+:2}
\\
 f_{3h}^+ \!&=& \left(\left[ -\frac{h k}{\omega_-}+\frac{h k}{\omega_+}\frac{k^2\!+\Re[m_+m_-^*]}{\omega_+\omega_-}\right]
                    \!\cos(2\omega_+ t)
     + \frac{\Im[m_+m_-^*]}{\omega_+\omega_-}\!\sin(2\omega_+ t)
     -\frac{hk}{\omega_+}\frac{k^2\!+\Re[m_+m_-^*]}{\omega_+\omega_-}\right)\!\times\!(\bar{n}_{h+}\!\!-\!\bar{n}_{h-})
,\quad\;\;
\label{f3h+thermal:2}
\end{eqnarray}
It can be checked that the same phase space densities are obtained by
inserting the mode functions for $t>0$ Eqs.~\eqref{Lh+Rh+} in
the definitions for $f_{ah}$ Eqs.~(\ref{mode solutions:2}).

The final produced particle number~\eqref{nh} is
\begin{equation}
 n_h =\frac12 - \left(\frac{k^2+\Re[m_+m_-^*]}{2\omega_+\omega_-}\right)
 \times(\bar{n}_{h+}-\bar{n}_{h-})
 \,.
\label{number produced-thermal}
\end{equation}
For the free vacuum, where $\bar{n}_{h+}=1$ and $\bar{n}_{h-}=0$, this indeed
reduces to the result computed before, Eq.~\eqref{number produced}.
The thermal limit is obtained when $\bar{n}_{h+}-\bar{n}_{h-}\rightarrow 1-2\bar{n}_{\rm th}$,
$\bar{n}_{\rm th}=1/({\rm e}^{\beta\omega_+}+1)$.

To get the CP-violating density, which is of relevance for baryogenesis, we finally
need to sum the axial density $f_{3h}$ over $h=\pm$,
\begin{eqnarray}
\sum_{h=\pm}hf_{3h}
     &=& \left(-\frac{2k}{\omega_+}\frac{k^2+\Re[m_+m_-^*]}{\omega_+\omega_-}
     +\left[-\frac{2 k}{\omega_-}+\frac{2 k}{\omega_+}\frac{k^2+\Re[m_+m_-^*]}{\omega_+\omega_-}\right]
                    \cos(2\omega_+ t)\right) \times(\bar{n}_{h+}-\bar{n}_{h-})
\nonumber\\
\sum_{h=\pm}f_{3h}
         &=&  \frac{2|m_+| |m_-| \sin(\theta_+-\theta_-)}{\omega_+\omega_-}\sin(2\omega_+ t)
\times(\bar{n}_{h+}-\bar{n}_{h-})
\,,
\label{f3-thermal}
\end{eqnarray}
where in the last equality we used, $m_\pm = |m_\pm|\exp(\imath \theta_\pm)$.
The first current in~(\ref{f3-thermal}) is CP even, while the latter is CP odd, and can
be used to source baryogenesis. This latter term is there only when the source
of CP violation,
$\Delta\theta=\theta_+-\theta_-$ does not vanish.
The first term in the first equation is the (adiabatic) vacuum contribution
({\it i.e.} the leading classical term that would survive in the very thick wall limit).

\section{Mode function normalization}
\label{Mode function normalization}

 In this appendix we shall show that the properly normalized
mode functions that solve Eqs.~(\ref{eom:hypergeometric}),
and whose indices are~(\ref{a+- and b+-}), are given by Eq.~(\ref{u+-h:normalized}).
Since $a_\pm,b_\pm,c$ in Eq.~(\ref{a+- and b+-}) are non-integer,
the two independent solutions for $\chi_{\pm h}$ are the usual ones, and they are of the form
\begin{eqnarray}
 u_{\pm h}^{(1)} &=& u_{\pm h0}^{(1)} z^\alpha (1-z)^\beta \times
   \phantom{,\!\!}_2F_1(a_\pm,b_\pm;c;z)
\nonumber\\
 u_{\pm h}^{(2)} &=& u_{\pm h0}^{(2)} z^{\alpha+1-c} (1-z)^\beta \times
   \phantom{,\!\!}_2F_1(a_\pm+1-c,b_\pm+1-c;2-c;z)
\nonumber\\
    &=& u_{\pm h0}^{(2)}z^{\alpha+1-c} (1-z)^{\beta+c-a_\pm-b_\pm} \times
               \phantom{,\!\!}_2F_1(1-a_\pm,1-b_\pm;2-c;z)
\,,
\label{u:two independent solutions}
\end{eqnarray}
where $u_{\pm h0}^{(1,2)}$ are the normalisation
$z$-independent constants (which we shall determine below)
and $\phantom{,\!\!}_2F_1(a,b;c;z)$ denotes the Gauss'
hypergeometric function, whose series around $z=0$ reads,
\begin{equation}
\phantom{,\!\!}_2F_1(a,b;c;z) = 1+\frac{ab}{c}z
  +\frac{a(a+1)b(b+1)}{c(c+1)}\frac{z^2}{2!} + \cdots
 +\frac{\Gamma(a+n)\Gamma(b+n)\Gamma(c)}{\Gamma(a)\Gamma(b)\Gamma(c+n)}
             \frac{z^n}{n!} + \cdots
\,.
\nonumber
\end{equation}
Notice that we have picked the sign of $\alpha$ and $\beta$
in~(\ref{alpha and beta}) such that the first (second) fundamental solution
in Eqs.~(\ref{u:two independent solutions})
corresponds to the positive (negative) frequency wave at early times.
Notice further that the solutions for
$v_{\pm h}^{(1,2)}=u_{\pm -h}^{(1,2)}$ are obtained simply by
flipping the helicity $h$ in $u_{\pm h}^{(1,2)}$.
The latter form for of the two solutions~(\ref{u:two independent solutions})
is useful in that the prefactor
is in the form $\alpha+1-c=-\alpha=\alpha^*$,
$\beta+c-a_\pm-b_\pm=-\beta=\beta^*$.
It then follows that, in the vicinity of $t\rightarrow -\infty
\;(z\rightarrow {\rm e}^{2\gamma t})$,
$u_{\pm h}^{(1)}$ and $u_{\pm h}^{(2)}$ reduce to the positive and negative
frequency solutions, respectively,
\begin{equation}
 u_{\pm h}^{(1)}\approx
   u_{\pm h0}^{(1)}{\rm e}^{2\gamma \alpha t}=u_{\pm h0}^{(1)}{\rm e}^{-\imath\omega_-t}
\,, \qquad
 u_{\pm h}^{(2)} \approx
  u_{\pm h0}^{(2)}{\rm e}^{-2\gamma \alpha t}=u_{\pm h0}^{(2)}{\rm e}^{\imath\omega_-t}
 \qquad (t\rightarrow -\infty, z\rightarrow {\rm e}^{2\gamma t})
\,.
\label{early times mode functions}
\end{equation}
In analogy to what we did in Eq.~(\ref{vacuum mode solutions}),
here we shall take the positive frequency
solution at $t\rightarrow -\infty$, {\it i.e.}
$u_{\pm h}= u_{\pm h}^{(1)}$. Due to the fact that
the vacua at $t\rightarrow -\infty$ and $t\rightarrow +\infty$
are not the same (they are related by a Bogoliubov transformation),
the positive frequency solution becomes a mixture of
positive and negative frequency solutions close to $t\rightarrow +\infty$,
as can be implied from the following identity
(see {\it e.g.} Eq.~(9.131.1-2) of Ref.~\cite{Gradshteyn:2007}),
\begin{eqnarray}
 \phantom{,\!\!}_2F_1(a_\pm,b_\pm;c;z)
  &=& \frac{\Gamma(c)\Gamma(c-a_\pm-b_\pm)}{\Gamma(c-a_\pm)\Gamma(c-b_\pm)}
  \times \phantom{,\!\!}_2F_1(a_\pm,b_\pm;a_\pm+b_\pm+1-c;1-z)
\nonumber\\
  &+& \frac{\Gamma(c)\Gamma(a_\pm+b_\pm-c)}{\Gamma(a_\pm)\Gamma(b_\pm)}
    (1-z)^{c-a_\pm-b_\pm}
   \times\! \phantom{,\!\!}_2F_1(c-a_\pm,c-b_\pm;c+1-a_\pm-b_\pm;1-z)
\,.
\label{identity GR:9.131.2}
\end{eqnarray}
Indeed, we have,
\begin{eqnarray}
 u_{\pm h}\equiv u_{\pm h}^{(1)} &=&
 u_{\pm h0}z^\alpha (1-z)^\beta \times
  \frac{\Gamma(c)\Gamma(c-a_\pm-b_\pm)}{\Gamma(c-a_\pm)\Gamma(c-b_\pm)}
  \times \phantom{,\!\!}_2F_1(a_\pm,b_\pm;a_\pm+b_\pm+1-c;1-z)
\nonumber\\
  &+& u_{\pm h0}
   \frac{\Gamma(c)\Gamma(a_\pm+b_\pm-c)}{\Gamma(a_\pm)\Gamma(b_\pm)}
    z^\alpha (1-z)^{\beta + c-a_\pm-b_\pm}
   \times\! \phantom{,\!\!}_2F_1(c-a_\pm,c-b_\pm;c+1-a_\pm-b_\pm;1-z)
\,,
\label{u:late time}
\end{eqnarray}
from which we infer that in the vicinity of $t\rightarrow \infty$
($z\rightarrow 1-{\rm e}^{-2\gamma t}$),
\begin{equation}
 u_{\pm h} \approx
u_{\pm h0}\frac{\Gamma(c)\Gamma(c-a_\pm-b_\pm)}{\Gamma(c-a_\pm)\Gamma(c-b_\pm)}
          {\rm e}^{-\imath\omega_+ t}
  + u_{\pm h0}\frac{\Gamma(c)\Gamma(a_\pm+b_\pm-c)}{\Gamma(a_\pm)\Gamma(b_\pm)}
      {\rm e}^{\imath\omega_+ t}
\qquad (t\rightarrow \infty, 1-z\rightarrow {\rm e}^{-2\gamma t})
\,,
\label{u:late time:2}
\end{equation}
where we made use of $\tilde c \equiv a_\pm+b_\pm +1-c = 1+2\beta$,
$\beta + c-a_\pm-b_\pm = \beta +1 - \tilde c  = -\beta$.

 One can also construct late time positive and negative frequency solutions
that solve the differential equation~(\ref{eom:hypergeometric}).
Equation~(\ref{u:late time}) implies that,
if~(\ref{u:two independent solutions}) are solutions, so must be
both parts of Eq.~(\ref{u:late time}), such that the two
linearly independent late time solutions are,
\begin{eqnarray}
\tilde u_{\pm h}^{(1)} &=& \tilde u_{\pm h0}^{(2)}
   z^\alpha (1-z)^{\beta + 1-\tilde c}
  \times \phantom{,\!\!}_2F_1(a_\pm+1-\tilde c,b_\pm+1-\tilde c;2-\tilde c;1-z)
\nonumber\\
  &=& \tilde u_{\pm h0}^{(2)}
  z^{\alpha+\tilde c-a_\pm-b_\pm} (1-z)^{\beta + 1-\tilde c}
  \times \phantom{,\!\!}_2F_1(1-a_\pm,1-b_\pm;2-\tilde c;1-z)
\nonumber\\
 \tilde u_{\pm h}^{(2)} &=& \tilde u_{\pm h0}^{(1)}z^\alpha (1-z)^\beta
    \times \phantom{,\!\!}_2F_1(a_\pm,b_\pm;\tilde c;1-z)
\qquad (\tilde c = 1+2\beta)
\,,
\label{u:late time:3}
\end{eqnarray}
where $\tilde u_{\pm h0}^{(1,2)}$ are normalisation constants.
Now, because $\alpha+1-c=-\alpha=\alpha^*$ and
$\beta+c-a_\pm-b_\pm=-\beta=\beta^*$, the asymptotic forms
for the mode functions are,
\begin{equation}
 \tilde u_{\pm h}^{(1)} \approx
      \tilde u_{\pm h0}^{(1)}{\rm e}^{-\imath \omega_+ t}
\,,
\qquad
\tilde u_{\pm h}^{(2)} \approx
  \tilde u_{\pm h0}^{(2)}{\rm e}^{\imath \omega_+ t}
\qquad (t\rightarrow \infty, 1-z\rightarrow {\rm e}^{-2\gamma t})
\,.
\label{u:late time:asymptotic}
\end{equation}
One can check that Eqs.~(\ref{u:late time:3}) indeed solve
Eq.~(\ref{eom:hypergeometric}), so they constitute legitimate linearly
independent solutions for the mode functions. And moreover,
each of the solutions~(\ref{u:late time:3}) can be written as a linear
combination of the early time solutions~(\ref{u:two independent solutions}),
as they should.

\medskip

 Next, we need to properly normalize our mode
functions~(\ref{u:two independent solutions}) and~(\ref{u:late time:3}).
Rather then performing a quantum mechanical normalization
such as was used in~\cite{Ayala:1993mk}, we shall use
the field theoretic normalization~(\ref{consistency
 condition:mode functions}--\ref{mode normalization conditions}), since
it is more suitable for baryogenesis applications we have in mind.
Since $u_{+h}$ and $u_{-h}$ are related by a first order
differential equations~(\ref{eom:u+-hv+-h}),
their normalisation constants are not independent.
Let us begin by rewriting~(\ref{eom:u+-hv+-h}) as
\begin{eqnarray}
  \bigg[z(1-z)\frac{d}{dz} \pm \imath \frac{m_{1R}+m_{2R}(1-2z)}{2\gamma}
   \bigg]u_{\pm h} &=& \imath\frac{h k\pm\imath m_I}{2\gamma} u_{\mp h}
\,.
\label{eom:u+-:z}
\end{eqnarray}
 By making use of the identities,
\begin{equation}
 \alpha = \frac{c-1}{2}
\,;\qquad
 \beta = \frac{a_\pm+b_\pm-c}{2}
\,;\qquad
 \pm\frac{\imath m_{2R}}{2\gamma} = \frac{1+b_\pm-a_\pm}{4}
\,;\qquad
 \pm\frac{\imath m_{1R}}{2\gamma}
     = \frac{(a_\pm+b_\pm-1)(a_\pm+b_\pm+1-2c)}{4(a_\pm-1-b_\pm)}
\nonumber
\end{equation}
we can recast Eq.~(\ref{eom:u+-:z}) as
\begin{eqnarray}
  \bigg[z(1-z)\frac{d}{dz} + \frac{b_\pm(a_\pm-c)}{a_\pm-1-b_\pm)} - b_\pm z
   \bigg]u_{\pm h0} \times \phantom{,\!\!}_2F_1(a_\pm,b_\pm;c;z)
  &=& \imath\frac{h k\pm\imath m_I}{2\gamma} u_{\mp h0}
      \times \phantom{,\!\!}_2F_1(a_\pm-1,b_\pm+1;c;z)
\,,\qquad
\label{eom:u+-:z:2}
\end{eqnarray}
where we used $a_\mp = b_\pm +1$ and $b_\mp = a_\pm - 1$.
In order to transform the parameters of the hypergeometric function
on the left hand side such to correspond to those on the right hand side,
let us first make use of the following two identities
(see Eq.~(9.137.6) and~(9.137.17) in~\cite{Gradshteyn:2007}),
\begin{eqnarray}
 \frac{d}{dz}\left(\phantom{,\!\!}_2F_1(a,b;c;z) \right)
   &=&  \frac{ab}{c}\times\phantom{,\!\!}_2F_1(a+1,b+1;c+1;z)
\nonumber\\
\frac{abz}{c}\times\phantom{,\!\!}_2F_1(a+1,b+1;c+1;z)
   &=&  (c-1)\Big[\,\phantom{,\!\!}_2F_1(a,b;c-1;z)
              - \phantom{,\!\!}_2F_1(a,b;c;z)
           \Big]
\,,
\label{GR:9.137.6}
\end{eqnarray}
upon which Eq.~(\ref{eom:u+-:z:2}) reduces to
\begin{eqnarray}
  &&\bigg[(1-z)(c-1)\times \phantom{,\!\!}_2F_1(a_\pm,b_\pm;c-1;z)
  + \bigg(\frac{b_\pm(a_\pm-c)}{a_\pm-1-b_\pm}-c+1+(c-1 - b_\pm) z\bigg)
            \times \phantom{,\!\!}_2F_1(a_\pm,b_\pm;c;z)
   \bigg]u_{\pm h0}
\label{eom:u+-:z:3}
\\
  &&\hskip 10cm= \imath\frac{h k\pm\imath m_I}{2\gamma} u_{\mp h0}
      \times \phantom{,\!\!}_2F_1(a_\pm-1,b_\pm+1;c;z)
\,.
\nonumber
\end{eqnarray}
This can be further transformed by making use of~(9.137.17)
in~\cite{Gradshteyn:2007},
\begin{equation}
(c-1)\times\phantom{,\!\!}_2F_1(a,b;c-1;z)
    = b\times\phantom{,\!\!}_2F_1(a,b+1;c;z)
      + (c-1-b)\times\phantom{,\!\!}_2F_1(a,b;c;z)
\label{GR:9.137.17}
\end{equation}
into
\begin{eqnarray}
  &&\bigg[(1-z)b_\pm\times \phantom{,\!\!}_2F_1(a_\pm,b_\pm+1;c;z)
  +\frac{b_\pm(b_\pm+1-c)}{a_\pm-1-b_\pm}
            \times \phantom{,\!\!}_2F_1(a_\pm,b_\pm;c;z)
   \bigg]u_{\pm h0}
\label{eom:u+-:z:4}
\\
  &&\hskip 10cm= \imath\frac{h k\pm\imath m_I}{2\gamma} u_{\mp h0}
      \times \phantom{,\!\!}_2F_1(a_\pm-1,b_\pm+1;c;z)
\,.
\nonumber
\end{eqnarray}
We need one more transformation~\cite{Wolfram:2012},
\begin{equation}
(1-z)(a-b-1)\times\phantom{,\!\!}_2F_1(a,b+1;c;z)
    = (a-c)\times\phantom{,\!\!}_2F_1(a-1,b+1;c;z)
      + (c-1-b)\times\phantom{,\!\!}_2F_1(a,b;c;z)
\,,
\label{Wolfram 2}
\end{equation}
with which one gets on both sides of equation~(\ref{eom:u+-:z:4})
a function with identical parameters, implying the following relation between
the normalisation constants
\begin{equation}
 \frac{u_{\pm h0}}{u_{\mp h0}} = \imath\frac{h k\pm\imath m_I}{2\gamma}
                             \times \frac{a_\pm-b_\pm-1}{b_\pm(a_\pm-c)}
\,.
\label{normalization:1}
\end{equation}
Several comments are now in order. This expression shows that
CP violation is in the relative phase between
$u_{+ h}$ and $u_{-h}$ solutions,
\begin{equation}
 {\rm e}^{\imath \theta_{\rm CP\pm}} = \frac{h k\pm\imath m_I}
              {\sqrt{k^2 + m_I^2}}
\,,
\label{CP violating phase}
\end{equation}
which was to be expected, meaning that there is no trace of CP violation in
the parameters $a_\pm$, $b_\pm$ or $c$ of the hypergeometric functions.
Moreover,
\begin{equation}
 \frac{k^2 + m_I^2}{4\gamma^2} = \frac{b_\pm(a_\pm-1)(a_\pm-c)(b_\pm+1-c)}
                                       {(a_\pm-b_\pm-1)^2}
\,,
\nonumber
\end{equation}
such that the ratio~(\ref{normalization:1}) can be expressed in terms of
the phase $\theta_{\rm CP}$ and the parameters $a_\pm$, $b_\pm$ and $c$,
\begin{equation}
 \frac{u_{\pm h0}}{u_{\mp h0}} = \pm\frac{h k\pm\imath m_I}
                                             {\sqrt{k^2+m_I^2}}
 \times \frac{\sqrt{-b_\pm(a_\pm-c)(a_\pm-1)(b_\pm+1-c)}}{b_\pm(a_\pm-c)}
 = -\frac{h k\pm\imath m_I}{\sqrt{k^2+m_I^2}}
 \times \sqrt{\frac{\omega_- \pm(m_{1R}+m_{2R})}{\omega_- \mp(m_{1R}+m_{2R})}}
\,,\quad
\label{normalization:2}
\end{equation}
where we made use of
$(a_\pm-b_\pm-1)^2= -[\pm\imath(a_\pm-b_\pm-1)^2]=-4m_{2R}^2/\gamma^2$
and in the last step,
\begin{equation}
 b_\pm(a_\pm-c)=\mp\frac{[\omega_- \mp (m_{1R}+m_{2R})]m_{2R}}{\gamma^2}
\,,\qquad
 (a_\pm-1)(b_\pm+1-c)=\pm\frac{[\omega_- \pm (m_{1R}+m_{2R})]m_{2R}}{\gamma^2}
\,.
\nonumber
\end{equation}
Eq.~(\ref{normalization:2}) nicely separates the relative CP-violating phase
between positive and negative frequency modes
and their amplitude ratio which is not CP-violating.
 From Eq.~(\ref{CP violating phase}) it also follows that
 $\theta_{\rm CP}\equiv \theta_{\rm CP+}=-\theta_{\rm CP-}$,
 {\it i.e.} that
 ${\rm e}^{\imath\theta_{\rm CP+}}{\rm e}^{\imath\theta_{\rm CP-}}=1$.

 What remains to be done is to perform
  normalisation of the mode functions. Eqs.~(\ref{normalization})
and~(\ref{u+-}) imply for the early time mode
functions~(\ref{u:two independent solutions}),
\begin{equation}
 |u_{+h}|^2+|u_{-h}|^2 = 1 = |v_{+h}|^2+|v_{-h}|^2
\,,
\label{normalization:3}
\end{equation}
and analogously for the late time mode functions~(\ref{u:late time:3}),
\begin{equation}
 |\tilde u_{+h}|^2+|\tilde u_{-h}|^2 = 1 = |\tilde v_{+h}|^2+|\tilde v_{-h}|^2
\,.
\label{normalization:3b}
\end{equation}
Making use of~(\ref{normalization:2}), the first condition
in~(\ref{normalization:3}) can be written as,
\begin{eqnarray}
 &&|u_{+h0}|^2\Bigg[
\phantom{,\!\!}_2F_1(a_+,b_+;c;z)\times \phantom{,\!\!}_2F_1(2-a_+,-b_+;2-c;z)
\nonumber\\
  &&\hskip 2cm -\, \frac{b_+(a_+-c)}{(a_+-1)(b_++1-c)}
 \times\phantom{,\!\!}_2F_1(a_-,b_-;c;z)
   \times \phantom{,\!\!}_2F_1(2-a_-,-b_-;2-c;z)
            \Bigg]=1
\,,
\label{normalization:4}
\end{eqnarray}
where we made use of the fact that $\alpha$ and $\beta$ are purely imaginary,
and of $a_\pm^*=2-a_\pm, b_\pm^*=-b_\pm$, $c^*=2-c$ and $z^*=z$ is real
(the proper sign in front of the second term is a minus). Next, it is
convenient to replace $a_-$ and $b_-$ by $a_+$ and $b_+$
($b_-=a_+-1, a_-=b_++1$) in the second term. The above analysis
(see Eqs.~(\ref{eom:u+-:z:2}--\ref{normalization:1})) implies
\begin{equation}
 \left[\frac{a-b-1}{b(a-c)}z(1-z)\frac{d}{dz}+\left(1-\frac{a-b-1}{a-c}z\right)\right]
     \phantom{,\!\!}_2F_1(a,b;c;z)
   = \phantom{,\!\!}_2F_1(a-1,b+1;c;z)
\nonumber
\end{equation}
and also ($a\rightarrow 1-b,b\rightarrow 1-a,c\rightarrow 2-c$),
\begin{equation}
 \left[\frac{a-b-1}{(1-a)(c-b-1)}z(1-z)\frac{d}{dz}
         +\left(1-\frac{a-b-1}{c-b-1}z\right)\right]
     \phantom{,\!\!}_2F_1(1-b,1-a;2-c;z)
   = \phantom{,\!\!}_2F_1(2-a,-b;2-c;z)
\,.
\nonumber
\end{equation}
The latter relation can be used to replace the second
hypergeometric function on the first line in~(\ref{normalization:4}),
while the former to replace the first hypergeometric function
$\phantom{,\!\!}_2F_1(a_-,b_-;c;z)=
\phantom{,\!\!}_2F_1(a_+-1,b_++1;c;z)$ in the second line of~(\ref{normalization:4})
to obtain,
\begin{eqnarray}
1&=&|u_{+h0}|^2\Bigg\{
\phantom{,\!\!}_2F_1(a_+,b_+;c;z)
 \left[\frac{a_+-b_+-1}{(1\!-\!a_+)(c\!-\!b_+\!-\!1)}z(1\!-\!z)\frac{d}{dz}
         +\left(1-\frac{a_+-b_+-1}{c-b_+-1}z\right)\right]
    \phantom{,\!\!}_2F_1(1-b_+,1-a_+;2-c;z)
\nonumber\\
  &&\hskip -1cm - \frac{b_+(a_+\!-\!c)}{(a_+\!-\!1)(b_+\!+\!1\!-\!c)}
 \left[\frac{a_+\!-\!b_+\!-\!1}{b_+(a_+\!-\!c)}z(1\!-\!z)\frac{d}{dz}
         +\left(1-\frac{a_+\!-\!b_+\!-\!1}{a_+\!-\!c}z\right)\right]
    \phantom{,\!\!}_2F_1(a_+,b_+;c;z)
   \times \phantom{,\!\!}_2F_1(1\!-\!a_+,1\!-\!b_+;2\!-\!c;z)
            \Bigg\}
\,.\quad
\nonumber\\
\hskip -1cm\label{normalization:5}
\end{eqnarray}
Next we can make use of the Wronskian for the hypergeometric functions,
\begin{equation}
 W[\phantom{,\!\!}_2F_1(a,b;c;z)
      ,z^{1-c}(1-z)^{c-a-b}\phantom{,\!\!}_2F_1(1-a,1-b;2-c;z)]
  = (1-c)z^{-c}(1-z)^{c-a-b-1}
\,,
\label{Wronskian:hypergeometric}
\end{equation}
from which it follows that,
\begin{equation}
 W[\phantom{,\!\!}_2F_1(a,b;c;z)
      ,\phantom{,\!\!}_2F_1(1-a,1-b;2-c;z)]
  +\phantom{,\!\!}_2F_1(a,b;c;z)
     \left(\frac{1-c}{z}-\frac{c-b-a}{1-z}\right)
   \phantom{,\!\!}_2F_1(1\!-\!a,1\!-\!b;2\!-\!c;z)]
  = \frac{1-c}{z(1\!-\!z)}
\,.
\label{Wronskian:hypergeometric:2}
\end{equation}
When this is inserted into~(\ref{normalization:5}) many terms cancel
and one ends up with,
\begin{equation}
|u_{+h0}|^2 = \frac{(1-a_+)(c-b_+-1)}{(a_+-b_+-1)(1-c)}
           = \frac{\omega_-+(m_{1R}+m_{2R})}{2\omega_-}
\,.
\label{normalization:6}
\end{equation}
To summarize, we have found that the normalized
early time mode functions~(\ref{u:two independent solutions}) are
({\it cf.} Eq.~(\ref{normalization:2})),
\begin{eqnarray}
 u_{+h} &\equiv&  u^{(1)}_{+h} = \sqrt{\frac{\omega_-+(m_{1R}+m_{2R})}{2\omega_-}}
           \times z^\alpha (1-z)^\beta \times
   \phantom{,\!\!}_2F_1(a_+,b_+;c;z)
\nonumber\\
 u_{-h} &\equiv&  u^{(1)}_{-h} =
 -\frac{h k-\imath m_I}{\sqrt{k^2+m_I^2}}
 \times
\sqrt{\frac{\omega_--(m_{1R}+m_{2R})}{2\omega_-}}
 \times
                 z^{\alpha} (1-z)^\beta \times
   \phantom{,\!\!}_2F_1(a_-,b_-;c;z)
\,,
\label{u+-h:normalized:Appendix}
\end{eqnarray}
which are also given in Eqs.~(\ref{u+-h:normalized}). An analogous procedure yields
two other normalised mode functions given in
Eqs.~(\ref{u+-h:normalized:2}--\ref{tilde u+-h:normalized:2}).

\section{Connecting the kink wall to the thin wall case}
\label{Connecting the kink wall to the thin wall case}

Here we shall show that the early and late time
solutions for the mode functions for a tanh wall profile are equivalent to those
for the thin wall case in the limit $\gamma\rightarrow \infty$.
Similarly, we shall demonstrate the agreement between the Bogoliubov
coefficients $\alpha$ and $\beta$ and the corresponding particle number.

The early time solutions for the tanh wall \eqref{u+-h:normalized}
reduce in the limit $\gamma \rightarrow \infty$ to
\begin{eqnarray}
 u_{+h} &\equiv&  u^{(1)}_{+h} \xrightarrow{\gamma\rightarrow \infty}
 \sqrt{\frac{\omega_-+m_{-R}}{2\omega_-}}
           \times {\rm e}^{-\imath \omega_{-} t}
\nonumber\\
 u_{-h} &\equiv&  u^{(1)}_{-h} \xrightarrow{\gamma\rightarrow \infty}
 -\frac{h k-\imath m_I}{\sqrt{k^2+m_I^2}}
 \times
\sqrt{\frac{\omega_--m_{-R}}{2\omega_-}}
 \times
                {\rm e}^{-\imath \omega_{-} t}
\,,
\label{u+-h:normalized:thinwall}
\end{eqnarray}
where $m_{-R}=m_{1R}+m_{2R}$ and $m_{-}=m_{-R}+\imath m_{1I}$ is the mass for $t<0$.
The same result can be obtained by solving Eqs. \eqref{eom:u+-hv+-h}
for constant mass $m_R=m_{-R}$ and choosing the positive
frequency solution at early time, $u_{\pm h}=u_{\pm h 0}{\rm e}^{-\imath \omega_{-} t}$.
After normalisation according to $|u_{+h}|^2+|u_{-h}|^2=1$ the result
\eqref{u+-h:normalized:thinwall} is found. If we solve similarly
Eqs. \eqref{eom:u+-hv+-h} for the negative frequency solution at early time,
$v_{\pm h}=v_{\pm h 0}{\rm e}^{\imath \omega_{-} t}$, we find that
\begin{eqnarray}
 v_{+h} &\equiv& u^{(2)}_{+(-h)} \xrightarrow{\gamma\rightarrow \infty}
 \sqrt{\frac{\omega_{-}-m_{-R}}{2\omega_-}}
           \times {\rm e}^{\imath \omega_{-} t}
\nonumber\\
 v_{-h} &\equiv& u^{(2)}_{-(-h)} \xrightarrow{\gamma\rightarrow \infty}
 -\frac{h k+\imath m_I}{\sqrt{k^2+m_I^2}}
 \times
\sqrt{\frac{\omega_{-}+m_{-R}}{2\omega_-}}
  \times {\rm e}^{\imath \omega_{-} t}
\,.
\label{u+-h:normalized:2:thinwall}
\end{eqnarray}
In order to compare to the thin wall solutions at early time Eqs.
\eqref{vacuum mode solutions} we should rotate back to the
$L_h,R_h$ basis from the $u_{\pm h}$ basis. By making use of
\eqref{u+-} and the
solutions~(\ref{u+-h:normalized:thinwall}--\ref{u+-h:normalized:2:thinwall})
we compute
\begin{eqnarray}
 L_h^- &=&\frac{\omega_{-}-hk+m_{-}}{\sqrt{4\omega_{-}(\omega_{-}+m_{-R})}}
                                {\rm e}^{-\imath\omega_{-} t}
\,,\qquad
  R_h^- = \frac{\omega_{-}+hk+m^{*}_{-}}{\sqrt{4\omega_{-}(\omega_{-}+m_{-R})}}
                                {\rm e}^{-\imath\omega t}
\nonumber\\
 \bar L_h^- &=& \frac{\omega_{-}-hk-m_{-}}{\sqrt{4\omega_{-}(\omega_{-}-m_{-R})}}
                                {\rm e}^{\imath\omega_{-} t}
\,,\qquad
 \bar R_h^- = \frac{\omega_{-}+hk-m_{-}^{*}}{\sqrt{4\omega_{-}(\omega_{-}-m_{-R})}}
                                {\rm e}^{\imath\omega t}
\,.
\label{vacuum mode solutions:thinwall}
\end{eqnarray}
At first sight these solutions do not seem to be consistent with
Eqs.~\eqref{vacuum mode solutions}. However, they can be rewritten as
\begin{eqnarray}
 L_h^- &=& \sqrt{\frac{\omega_--hk}{2\omega_-}}{\rm e}^{\imath \theta_{L}}
                                {\rm e}^{-\imath\omega_- t}
\,,\qquad
  R_h^- = \sqrt{\frac{\omega_-+hk}{2\omega_-}}{\rm e}^{\imath \theta_{R}}
                                {\rm e}^{-\imath\omega_- t}
\nonumber\\
 \bar L_h^- &=& -\sqrt{\frac{\omega_--hk}{2\omega_-}}{\rm e}^{\imath \theta_{\bar L}}
                                {\rm e}^{\imath\omega_- t}
\,,\qquad
 \bar R_h^- = \sqrt{\frac{\omega_-+hk}{2\omega_-}}
                                {\rm e}^{\imath \theta_{\bar R}} {\rm e}^{\imath\omega_- t}
                                \,,
\label{vacuum mode solutions:thinwall:2}
\end{eqnarray}
where the (real) phases are given by
\begin{eqnarray}
 \theta_{L}=\arctan\left(\frac{m_I}{\omega_{-}-hk+m_{-R}}\right)
\,,\qquad
  \theta_{R}=\arctan\left(\frac{-m_I}{\omega_{-}+hk+m_{-R}}\right)
\nonumber\\
 \theta_{\bar L}=\arctan\left(\frac{-m_I}{\omega_{-}-hk-m_{-R}}\right)
\,,\qquad
 \theta_{\bar R}=\arctan\left(\frac{m_I}{\omega_{-}+hk-m_{-R}}\right)
 \label{complexphasesLRsolutions}\,.
\end{eqnarray}
Thus the early time mode functions $L_h$ are $R_h$ in the thin wall limit
\eqref{vacuum mode solutions:thinwall} only differ from those computed
directly for the thin wall \eqref{Lh-Rh-} by a common phase factor.
A further global $U(1)$ rotation of the $L_h, R_h$ spinor $\chi_h$
by ${\rm e}^{-\imath \theta_{L}}$,
and of the ${\bar L}_h,{\bar R}_h$ spinor $\nu_h$ by ${\rm e}^{\imath(\pi-{\bar\theta}_{L})}$,
reduces the solutions \eqref{vacuum mode solutions:thinwall:2} to those
in Eq. \eqref{vacuum mode solutions}. Here we used that
\begin{equation}
{\rm e}^{\imath(\theta_{R}- \theta_{L})}=\frac{m_{-}^{*}}{|m_-|}
={\rm e}^{\imath({\bar\theta}_{R}- {\bar\theta}_{L})}
\,,
\end{equation}
which follows from the identity $\arctan(x)-\arctan(y)=\arctan((x-y)/(1+xy))$.

Next step is to check the late time solutions. Since the general late time solution
for the mode functions is a linear combination of the fundamental late time
solutions $\tilde u^{(1)}_{\pm h}$ and $\tilde u^{(2)}_{\pm h}$ given in
Eqs.~(\ref{tilde u+-h:normalized}--\ref{tilde u+-h:normalized:2}),
we should consider both in the thin wall limit $\gamma \rightarrow \infty$.
Eq. \eqref{late time solutions:general} now becomes
\begin{eqnarray}
 {\tilde u}_{+h} & = & \alpha_{+h} {\tilde u}^{(1)}_{+h}+\beta_{+h}{\tilde u}^{(2)}_{+h}
 \nonumber \\
  &\xrightarrow{\gamma\rightarrow \infty}&
\alpha_{+h} \sqrt{\frac{\omega_++m_{+R}}{2\omega_+}}
            {\rm e}^{-\imath \omega_{+} t}
+\beta_{+h} \sqrt{\frac{\omega_+-m_{+R}}{2\omega_+}}
            {\rm e}^{\imath \omega_{+} t}
\nonumber\\
 {\tilde u}_{-h} & = & \alpha_{-h} {\tilde u}^{(1)}_{-h}+\beta_{-h}{\tilde u}^{(2)}_{-h}
 \nonumber \\
  &\xrightarrow{\gamma\rightarrow \infty}&
- \alpha_{-h}\frac{h k -\imath m_I}{\sqrt{k^2+m_I^2}} \sqrt{\frac{\omega_+-m_{+R}}{2\omega_+}}
            {\rm e}^{-\imath \omega_{+} t}
+\beta_{-h} \frac{h k -\imath m_I}{\sqrt{k^2+m_I^2}}\sqrt{\frac{\omega_++m_{+R}}{2\omega_+}}
            {\rm e}^{\imath \omega_{+} t}
\,.
\label{tildeu+-h::thinwall}
\end{eqnarray}
The thin wall limit for the Bogoliubov coefficients~(\ref{alpha-beta})
 is (see also~\eqref{beta:thin}):
\begin{eqnarray}
\alpha_{\pm h} &=&\sqrt{\frac{\omega_+[\omega_-\pm(m_{1R}+m_{2R})]}{\omega_-[\omega_+\pm(m_{1R}-m_{2R})]}}
                 \frac{\omega_{-}+\omega_{+}\mp 2 m_{2R}}{2\omega_{+}}
\nonumber\\
\beta_{\pm h} &=& \pm \sqrt{\frac{\omega_+[\omega_-\pm(m_{1R}+m_{2R})]}{\omega_-[\omega_+\mp(m_{1R}-m_{2R})]}}
                 \frac{\omega_{+}-\omega_{-}\pm 2 m_{2R}}{2\omega_{+}}
\,.
\label{alpha-beta-thinwall}
\end{eqnarray}
Thus the coefficients are real and they can be shown to satisfy \eqref{alphabetaequalities}.
The late time solutions for $L_h$ and $R_h$ can be derived via Eq.~\eqref{u+-} from
the solutions \eqref{tildeu+-h::thinwall},
\begin{eqnarray}
L^{+}_h
& = & \alpha_{+h} \frac{\omega_{+}- h k +m_{+}}{\sqrt{4\omega_{+}(\omega_{+}+m_{+R})}}
            {\rm e}^{-\imath \omega_{+} t}
            +\beta_{+h} \frac{\omega_{+}+ h k -m_{+}}{\sqrt{4\omega_{+}(\omega_{+}-m_{+R})}}
            {\rm e}^{\imath \omega_{+} t}
\nonumber\\
R^{+}_h
& = &  \alpha_{+h} \frac{\omega_{+}+ h k +m_{+}^{*}}{\sqrt{4\omega_{+}(\omega_{+}+m_{+R})}}
            {\rm e}^{-\imath \omega_{+} t}
            +\beta_{+h} \frac{\omega_{+}- h k -m_{+}^{*}}{\sqrt{4\omega_{+}(\omega_{+}-m_{+R})}}
            {\rm e}^{\imath \omega_{+} t}
\,.
\label{vacuum mode solutions:thinwall:late2}
\end{eqnarray}
As for the solutions for $t<0$, we also write the late time solutions in a form more similar to
\eqref{Lh+Rh+}. This gives
\begin{eqnarray}
L^{+}_h
& = & \alpha_{+h} \sqrt{\frac{\omega_{+}- h k }{2\omega_{+}}}{\rm e}^{\imath \theta^{(1)}_{\tilde{L}}}
            {\rm e}^{-\imath \omega_{+} t}
            +\beta_{+h} \sqrt{\frac{\omega_{+}- h k}{2\omega_{+}}}
            {\rm e}^{\imath \theta^{(2)}_{\tilde{L}}}
            {\rm e}^{\imath \omega_{+} t}
\nonumber\\
R^{+}_h
& = &  \alpha_{+h}\sqrt{\frac{\omega_{+}+ h k }{2\omega_{+}}}{\rm e}^{\imath \theta^{(1)}_{\tilde{R}}}
            {\rm e}^{-\imath \omega_{+} t}
            +\beta_{+h}\sqrt{\frac{\omega_{+}+ h k}{2\omega_{+}}}
            {\rm e}^{\imath \theta^{(2)}_{\tilde{R}}}
            {\rm e}^{\imath \omega_{+} t}
\,.
\label{vacuum mode solutions:thinwall:late3}
\end{eqnarray}
The corresponding phase factors are
\begin{eqnarray}
 \theta^{(1)}_{\tilde{L}}=\arctan\left(\frac{m_I}{\omega_{+}-hk+m_{+R}}\right)
\,,\qquad
  \theta^{(1)}_{\tilde{R}}=\arctan\left(\frac{-m_I}{\omega_{+}+hk+m_{+R}}\right)
\nonumber\\
 \theta^{(2)}_{\tilde{L}}=\arctan\left(\frac{-m_I}{\omega_{+}+hk-m_{+R}}\right)
\,,\qquad
 \theta^{(2)}_{\tilde{R}}=\arctan\left(\frac{m_I}{\omega_{+}-hk-m_{+R}}\right)
 \label{complexphasesLRsolutions:late}\,.
\end{eqnarray}
We have seen that the early time solutions for $L_h,R_h$ in the thin wall limit,
Eqs.~\eqref{vacuum mode solutions:thinwall:2}, differ from those computed
directly for the thin wall \eqref{vacuum mode solutions} by a global phase factor.
This factor could be removed by rotating the particle spinor by a factor
of ${\rm e}^{-\imath\theta_L}$. Since the general late time
solution matches the early time one, this means that also the late time
solution should be rotated by the same factor. The resulting phase factor
for the $\alpha_{+h}$ and $\beta_{+h}$ solutions should match the phase factor
present in the solutions \eqref{Lh+Rh+}. Indeed we can show that
\begin{eqnarray}
\theta^{(1)}_{\tilde{L}}-\theta_L &=& {\rm Arg}[\alpha^{+}_h]
\nonumber\\
\theta^{(1)}_{\tilde{R}}-\theta_L &=& {\rm Arg}[\alpha^{+}_h]+\arctan\left(\frac{-m_I}{m_{+R}}\right)
\nonumber\\
\theta^{(2)}_{\tilde{L}}-\theta_L &=& {\rm Arg}[\beta^{+}_h]
\nonumber\\
\theta^{(2)}_{\tilde{R}}-\theta_L &=& {\rm Arg}[\beta^{+}_h]+\arctan\left(\frac{-m_I}{m_{+R}}\right)
\,,
\end{eqnarray}
where the $\alpha^{+}_h$ and $\beta^{+}_h$ on the righthand side are the ones
in Eq. \eqref{alphah+betah+}. The conclusion we therefore make is the following:
although the Bogoliubov coefficients computed by taking the thin wall limit
of the kink wall solutions \eqref{alpha-beta-thinwall} appear different from
those directly computed for the thin wall \eqref{alphah+betah+}, they
in fact only differ by a global phase factor, which does not affect the particle number.
In the thin wall limit computations,
the Bogoliubov coefficients \eqref{alpha-beta-thinwall} are real, but the
early and late time mode functions carry a global phase factor, see
Eqs. \eqref{complexphasesLRsolutions} and \eqref{complexphasesLRsolutions:late}.
In the direct thin wall computations, the (coefficients of) the mode functions
do not carry the global phase factor, see \eqref{Lh-Rh-} and \eqref{Lh+Rh+},
but the global phase is present in the Bogoliubov coefficients \eqref{alphah+betah+}, which are hence complex.
In any case, global rotations of the (anti)particle spinors are always allowed, and such
rotations simply move the phase factor back and forth between Bogoliubov coefficients
and mode functions, leading to physically equivalent solutions.

\section{Deriving the kinetic and constraint equations in gradient approximation}
\label{app:Deriving the kinetic and constraint equations in gradient approximation}

In this section we derive the kinetic and constraint equations for the densities $g_{ah}$.
Our starting point is Eq.~(\ref{Dirac: Wigner space}) with the {\it Ansatz}~(\ref{S+-}).
After inserting $\gamma^0\gamma^0=1$
in front of $\imath S^{+-}(k;x)$ in Eq.~(\ref{Dirac: Wigner space}), we get,
\begin{eqnarray}
 &&\Big\{(I\otimes I) k^0 -\rho^3\otimes(\vec{k}\cdot\vec{\sigma})
     +\frac{\imath}{2}(I\otimes I)\partial_t
     -m_R(t)(\rho^1\otimes I)\exp\Big(-\frac{\imath}{2}\overleftarrow{\partial}_{\!\!t}\overrightarrow{\partial}_{\!\!k_0}\Big)
     -m_I(t)(\rho^2\otimes I)\exp\Big(-\frac{\imath}{2}\overleftarrow{\partial}_{\!\!t}\overrightarrow{\partial}_{\!\!k_0}\Big)
 \Big\}
 \nonumber\\
 &&\hskip 11cm
 \times(\rho^a g_{ah}(\vec k;x))\otimes\frac14(1+h\hat k\cdot \vec \sigma) = 0
 \,,\qquad\;
\label{Dirac: Bloch representation}
\end{eqnarray}
where we made use of the Bloch representation of the Clifford algebra, in which
\begin{equation}
\gamma^0 \rightarrow \rho^1\otimes I  \;;\qquad
\gamma^i \rightarrow \rho^2\otimes\imath\sigma^i  \;;\qquad
\gamma^5 \rightarrow -\rho^3\otimes I  \;;\qquad
\hat H = \hat  k \cdot\gamma^0\vec \gamma\gamma^5 \rightarrow
       \hat k\cdot I\otimes \vec \sigma \;.\qquad
\label{Bloch representation}
\end{equation}
 Now, upon multiplying from the left by
\begin{equation}
\{I,h\gamma^i\gamma^5,-\imath h\gamma^i,-\gamma^5\}\rightarrow
\{I\otimes I, \rho^1\otimes h\sigma^i,\rho^2\otimes h\sigma^i,\rho^3\otimes I\}
\,,
\nonumber
\end{equation}
and taking a trace, we get the following four equations,
\begin{eqnarray}
k_0g_{0h} - hk g_{3h} + \frac{\imath}{2}\partial_t g_{0h}
- m_R(t){\exp}\Big(-\frac{\imath}{2}\overleftarrow{\partial}_{\!\!t}\overrightarrow{\partial}_{\!\!k_0}\Big)g_{1h}
- m_I(t){\exp}\Big(-\frac{\imath}{2}\overleftarrow{\partial}_{\!\!t}\overrightarrow{\partial}_{\!\!k_0}\Big)g_{2h}
      &=& 0
\label{Dirac: component:0}\\
k_0g_{1h} +\imath hk g_{2h} + \frac{\imath}{2}\partial_t g_{1h}
- m_R(t){\exp}\Big(-\frac{\imath}{2}\overleftarrow{\partial}_{\!\!t}\overrightarrow{\partial}_{\!\!k_0}\Big)g_{0h}
- \imath m_I(t){\exp}\Big(-\frac{\imath}{2}\overleftarrow{\partial}_{\!\!t}\overrightarrow{\partial}_{\!\!k_0}\Big)g_{3h}
      &=& 0
\label{Dirac: component:1}\\
k_0g_{2h} - \imath hk g_{1h} + \frac{\imath}{2}\partial_t g_{2h}
+\imath m_R(t){\exp}\Big(-\frac{\imath}{2}\overleftarrow{\partial}_{\!\!t}\overrightarrow{\partial}_{\!\!k_0}\Big)g_{3h}
- m_I(t){\exp}\Big(-\frac{\imath}{2}\overleftarrow{\partial}_{\!\!t}\overrightarrow{\partial}_{\!\!k_0}\Big)g_{0h}
      &=& 0
\label{Dirac: component:2}\\
k_0g_{3h} - hk g_{0h} + \frac{\imath}{2}\partial_t g_{3h}
- \imath m_R(t){\exp}\Big(-\frac{\imath}{2}\overleftarrow{\partial}_{\!\!t}\overrightarrow{\partial}_{\!\!k_0}\Big)g_{2h}
+\imath m_I(t){\exp}\Big(-\frac{\imath}{2}\overleftarrow{\partial}_{\!\!t}\overrightarrow{\partial}_{\!\!k_0}\Big)g_{1h}
      &=& 0
\,.
\label{Dirac: component:3}
\end{eqnarray}
Now, hermiticity of $\imath \gamma^0 S^{+-}$ implies reality of the component functions $g_{ah}$,
such that the real and imaginary parts of
Eqs.~(\ref{Dirac: component:0}--\ref{Dirac: component:3}) must be separately satisfied.
The real parts yield constraint equations (CEs),
\begin{eqnarray}
k_0g_{0h} - hk g_{3h}
- m_R(t){\cos}\Big(\frac{1}{2}\overleftarrow{\partial}_{\!\!t}\overrightarrow{\partial}_{\!\!k_0}\Big)g_{1h}
- m_I(t){\cos}\Big(\frac{1}{2}\overleftarrow{\partial}_{\!\!t}\overrightarrow{\partial}_{\!\!k_0}\Big)g_{2h}
      &=& 0
\label{CE:0}\\
k_0g_{1h}
- m_R(t){\cos}\Big(\frac{1}{2}\overleftarrow{\partial}_{\!\!t}\overrightarrow{\partial}_{\!\!k_0}\Big)g_{0h}
- m_I(t){\sin}\Big(\frac{1}{2}\overleftarrow{\partial}_{\!\!t}\overrightarrow{\partial}_{\!\!k_0}\Big)g_{3h}
      &=& 0
\label{CE:1}\\
k_0g_{2h}
+ m_R(t){\sin}\Big(\frac{1}{2}\overleftarrow{\partial}_{\!\!t}\overrightarrow{\partial}_{\!\!k_0}\Big)g_{3h}
- m_I(t){\cos}\Big(\frac{1}{2}\overleftarrow{\partial}_{\!\!t}\overrightarrow{\partial}_{\!\!k_0}\Big)g_{0h}
      &=& 0
\label{CE:2}\\
k_0g_{3h} - hk g_{0h}
- m_R(t){\sin}\Big(\frac{1}{2}\overleftarrow{\partial}_{\!\!t}\overrightarrow{\partial}_{\!\!k_0}\Big)g_{2h}
+ m_I(t){\sin}\Big(\frac{1}{2}\overleftarrow{\partial}_{\!\!t}\overrightarrow{\partial}_{\!\!k_0}\Big)g_{1h}
      &=& 0
\,,
\label{CE:3}
\end{eqnarray}
while the imaginary parts yield the following kinetic equations (KEs),
\begin{eqnarray}
\partial_t g_{0h}
+ 2m_R(t){\sin}\Big(\frac{1}{2}\overleftarrow{\partial}_{\!\!t}\overrightarrow{\partial}_{\!\!k_0}\Big)g_{1h}
+ 2m_I(t){\sin}\Big(\frac{1}{2}\overleftarrow{\partial}_{\!\!t}\overrightarrow{\partial}_{\!\!k_0}\Big)g_{2h}
      &=& 0
\label{KE:0}\\
 \partial_t g_{1h} + 2 hk g_{2h}
+ 2m_R(t){\sin}\Big(\frac{1}{2}\overleftarrow{\partial}_{\!\!t}\overrightarrow{\partial}_{\!\!k_0}\Big)g_{0h}
- 2 m_I(t){\cos}\Big(\frac{1}{2}\overleftarrow{\partial}_{\!\!t}\overrightarrow{\partial}_{\!\!k_0}\Big)g_{3h}
      &=& 0
\label{KE:1}\\
 \partial_t g_{2h} - 2 hk g_{1h}
+2m_R(t){\cos}\Big(\frac{1}{2}\overleftarrow{\partial}_{\!\!t}\overrightarrow{\partial}_{\!\!k_0}\Big)g_{3h}
+ 2m_I(t){\sin}\Big(\frac{1}{2}\overleftarrow{\partial}_{\!\!t}\overrightarrow{\partial}_{\!\!k_0}\Big)g_{0h}
      &=& 0
\label{KE:2}\\
\partial_t g_{3h}
- 2m_R(t){\cos}\Big(\frac{1}{2}\overleftarrow{\partial}_{\!\!t}\overrightarrow{\partial}_{\!\!k_0}\Big)g_{2h}
+ 2m_I(t){\cos}\Big(\frac{1}{2}\overleftarrow{\partial}_{\!\!t}\overrightarrow{\partial}_{\!\!k_0}\Big)g_{1h}
      &=& 0
\,.
\label{KE:3}
\end{eqnarray}
Due to the nonlocal nature of the derivative operators, these equations are hard to solve, and hence not very
useful. However, when truncated at a finite number of derivatives $\partial_t$, they reduce to a set of
relatively simple equations. This derivative truncation, which is a generalization of the quantum mechanical WKB expansion,
holds when formally
\begin{equation}
\hbar \|\partial_t\|\ll |k_0|
\,,
\label{gradient expansion: criterion}
\end{equation}
where we reinserted $\hbar$ to make it explicit that this
derivative expansion is in fact an expansion
in powers of $\hbar$. Note that no matter how large the norm $\|\partial_t\|$, there always will be modes
that satisfy the criterion~(\ref{gradient expansion: criterion}). Conversely, no matter how slow the changes in
time are, there always will be modes that break~(\ref{gradient expansion: criterion}). In some sense,
the criterion~(\ref{gradient expansion: criterion}) divides a theory into two parts: the semiclassical part
where~(\ref{gradient expansion: criterion}) holds and the quantum part,
where~(\ref{gradient expansion: criterion}) is broken. Of course, a full quantum mechanical
kink wall treatment is required for those modes that break condition~(\ref{gradient expansion: criterion}),
while a semiclassical treatment should suffice when~(\ref{gradient expansion: criterion}) is satisfied.
When modes are massive on both sides of the wall, then the theory has a gap of $2{\rm min}[|m(t)|]$,
and -- at least on-shell -- $|k_0|\geq {\rm min}[|m(t)|]$.~\footnote{When interactions (loops) are included,
due to quantum effects the mass gap can decrease, or even completely disappear, so one should be
careful when making statements concerning applicability of the gradient approximation, even in the case when
the tree level mass is present on both sides of the `wall'. For example, in the case of the
electroweak phase transition, it is typically the case that the tree level mass, and hence the gap, is zero
on the symmetric side of the bubble wall.}

 While the classical kinetic theory is obtained by keeping  the CEs up to 0th order
in derivatives and kinetic equations to first order in derivatives, in order to get semiclassical equations
which contain information on CP violation require to maintain first order derivatives in the CEs and second order derivatives
in the KEs. Let us now consider the constraint equations~(\ref{CE:0}--\ref{CE:3}). We have
\begin{eqnarray}
g_{1h} &=&
 \frac{m_R(t)}{k_0}{\cos}\Big(\frac{1}{2}\overleftarrow{\partial}_{\!\!t}\overrightarrow{\partial}_{\!\!k_0}\Big)g_{0h}
+ \frac{m_I(t)}{k_0}{\sin}\Big(\frac{1}{2}\overleftarrow{\partial}_{\!\!t}\overrightarrow{\partial}_{\!\!k_0}\Big)g_{3h}
\label{CE:1b}\\
g_{2h} &=&
- \frac{m_R(t)}{k_0}{\sin}\Big(\frac{1}{2}\overleftarrow{\partial}_{\!\!t}\overrightarrow{\partial}_{\!\!k_0}\Big)g_{3h}
+ \frac{m_I(t)}{k_0}{\cos}\Big(\frac{1}{2}\overleftarrow{\partial}_{\!\!t}\overrightarrow{\partial}_{\!\!k_0}\Big)g_{0h}
\label{CE:2b}
\,.
\end{eqnarray}
Upon inserting these into Eqs.~(\ref{CE:0}) and~(\ref{CE:3}), and truncating to first order in derivatives, we get
\begin{eqnarray}
\frac{k_0^2-m_R^2-m_I^2}{k_0}g_{0h} - \bigg[hk +\frac{m_R\partial_t m_I-m_I\partial_t m_R}{2k_0}\partial_{k_0}\bigg]g_{3h}
      &=& 0
\label{CE:0b}\\
k_0g_{3h} - \bigg[hk - \frac{m_R\partial_t m_I-m_I\partial_t m_R}{2}\partial_{k_0}\frac{1}{k_0}\bigg]g_{0h}
      &=& 0
\,.
\label{CE:3b}
\end{eqnarray}
These two equations can be easily decoupled, such that (again to first order in gradients) we have
\begin{eqnarray}
\Big(k_0^2-|m|^2-k^2 \Big)g_{0h} &=& 0
\label{CE:0c:app}\\
\bigg(k_0^2-|m|^2-k^2 - h k\frac{|m|^2\dot\theta}{k_0^2-|m|^2}\bigg)g_{3h} &=& 0
\,,
\label{CE:3c:app}
\end{eqnarray}
where we used a shorthand notation,
\begin{equation}
 |m|^2 = m_R^2+m_I^2\;;\qquad m_R = |m|\cos(\theta)\;;\qquad m_I = |m|\sin(\theta)\;;\qquad
 |m|^2\dot\theta = m_R\partial_t m_I-m_I\partial_t m_R
\,.
\nonumber
\end{equation}
Eqs. \eqref{CE:0c:app} and \eqref{CE:3c:app} are presented in the main text,
see \eqref{CE:0c} and \eqref{CE:3c}.
Next we consider the kinetic equations to second order in gradients.
First we treat the kinetic equations for
$g_{0h}$ and $g_{3h}$, in order to describe CP violation in the axial vector current.
So, we begin by inserting Eqs.~(\ref{CE:1b}--\ref{CE:2b}) into Eqs.~(\ref{KE:0}) and~(\ref{KE:3}), and we get
\begin{eqnarray}
\partial_t g_{0h} + \frac{\partial_t|m|^2}{2}\partial_{k_0}\frac{g_{0h}}{k_0} &=& 0
\label{KE:0b:app}\\
\partial_t g_{3h} + \frac{\partial_t|m|^2}{2k_0}\partial_{k_0}g_{3h}
-\frac{\partial_t(|m|^2\partial_t\theta)}{4}\bigg(\partial_{k_0}^2-\frac{1}{k_0}\partial_{k_0}^2k_0\bigg)\frac{g_{0h}}{k_0}
      &=& 0
\,,
\label{KE:3b}
\end{eqnarray}
where we made use of
\begin{equation}
 m_R\partial_t m_R + m_I\partial_t m_I = \frac12 \partial_t|m|^2\;;\qquad
  m_R\partial_t^2 m_I - m_I\partial_t^2 m_R = \partial_t\big(|m|^2\partial_t \theta\big)
  \;.
 \label{mass relations}
\end{equation}
Now, upon making use of $g_{0h}=(hk_0/k) g_{3h}$ plus higher orders ({\it cf.} Eq.~(\ref{CE:0b}))
and pulling $k_0$ to the left of the derivatives, Eq.~(\ref{KE:3b}) simplifies to,
\begin{equation}
\partial_t g_{3h} + \frac{\partial_t|m|^2}{2k_0}\partial_{k_0}g_{3h}
+h\frac{\partial_t(|m|^2\partial_t\theta)}{2kk_0}\partial_{k_0}g_{3h} = 0
\,.
\label{KE:3c:app}
\end{equation}
These are presented in the main text in \eqref{KE:0b} and \eqref{KE:3c}.\\
\linebreak
Additionally, we can solve the constraint and kinetic equations for $g_{1h}$ and $g_{2h}$.
In a similar procedure as before, we first take the CEs \eqref{CE:0} and \eqref{CE:3},
and combine them to find
\begin{align}
0&=(k_0^2-k^2)g_{0h}-k_0 m_R {\cos}\Big(\frac{1}{2}\overleftarrow{\partial}_{\!\!t}\overrightarrow{\partial}_{\!\!k_0}\Big)g_{1h}
+h k m_I {\sin}\Big(\frac{1}{2}\overleftarrow{\partial}_{\!\!t}\overrightarrow{\partial}_{\!\!k_0}\Big)g_{1h}
-k_0 m_I {\cos}\Big(\frac{1}{2}\overleftarrow{\partial}_{\!\!t}\overrightarrow{\partial}_{\!\!k_0}\Big)g_{2h}
-h k m_R {\sin}\Big(\frac{1}{2}\overleftarrow{\partial}_{\!\!t}\overrightarrow{\partial}_{\!\!k_0}\Big)g_{2h}
\nonumber \\
0&=(k_0^2-k^2)g_{3h}+k_0 m_I {\sin}\Big(\frac{1}{2}\overleftarrow{\partial}_{\!\!t}\overrightarrow{\partial}_{\!\!k_0}\Big)g_{1h}
-h k m_R {\cos}\Big(\frac{1}{2}\overleftarrow{\partial}_{\!\!t}\overrightarrow{\partial}_{\!\!k_0}\Big)g_{1h}
-k_0 m_R {\sin}\Big(\frac{1}{2}\overleftarrow{\partial}_{\!\!t}\overrightarrow{\partial}_{\!\!k_0}\Big)g_{2h}
-h k m_I {\cos}\Big(\frac{1}{2}\overleftarrow{\partial}_{\!\!t}\overrightarrow{\partial}_{\!\!k_0}\Big)g_{2h}
\,.
\label{CEs:g1hg2h:1}
\end{align}
Now, by making use of these equations we can eliminate $g_{0h}$ and $g_{3h}$
from the remaining constraint equations \eqref{CE:1} and \eqref{CE:2}.
After an expansion up to second order in derivatives $\partial_{k_0}$
we find,
\begin{align}
&\frac{k_0^2-k^2-m_R^2}{k_0^2-k^2}k_0g_{1h}
+hk\left[\frac{m_R \dot{m}_I}{2}\frac{1}{k_0^2-k^2} \partial_{k_0}
-\frac{m_R \dot{m}_I}{2} \partial_{k_0}\frac{1}{k_0^2-k^2}\right]g_{1h}
\nonumber\\
&+\left[\frac{\ddot{m}_R m_R}{8}\partial_{k_0}^2\frac{k_0}{k_0^2-k^2}
+\frac{\ddot{m}_R m_R}{8}\frac{k_0}{k_0^2-k^2}\partial_{k_0}^2
+\frac14 \dot{m}_I^2 \partial_{k_0}\frac{k_0}{k_0^2-k^2}\partial_{k_0}\right]g_{1h}
\nonumber\\
&- \frac{m_I m_R}{k_0^2-k^2} k_0 g_{2h}
+hk\left[-\frac{m_R \dot{m}_R}{2}\frac{1}{k_0^2-k^2} \partial_{k_0}
-\frac{m_I \dot{m}_I}{2} \partial_{k_0}\frac{1}{k_0^2-k^2}\right]g_{2h}
\nonumber \\
&+\left[\frac{\ddot{m}_R m_I}{8}\partial_{k_0}^2\frac{k_0}{k_0^2-k^2}
+\frac{\ddot{m}_I m_R}{8}\frac{k_0}{k_0^2-k^2}\partial_{k_0}^2
-\frac14 \dot{m}_I\dot{m}_R\partial_{k_0}\frac{k_0}{k_0^2-k^2}\partial_{k_0}\right]g_{2h}=0
\label{CEs:g1hg2h:2a}\\
&\frac{k_0^2-k^2-m_I^2}{k_0^2-k^2}k_0g_{2h}
+hk\left[-\frac{m_I \dot{m}_R}{2}\frac{1}{k_0^2-k^2} \partial_{k_0}
+\frac{m_I \dot{m}_R}{2} \partial_{k_0}\frac{1}{k_0^2-k^2}\right]g_{2h}
\nonumber\\
&+\left[\frac{\ddot{m}_I m_I}{8}\partial_{k_0}^2\frac{k_0}{k_0^2-k^2}
+\frac{\ddot{m}_I m_I}{8}\frac{k_0}{k_0^2-k^2}\partial_{k_0}^2
+\frac14 \dot{m}_R^2 \partial_{k_0}\frac{k_0}{k_0^2-k^2}\partial_{k_0}\right]g_{2h}
\nonumber\\
&- \frac{m_I m_R }{k_0^2-k^2} k_0 g_{1h}
+hk\left[\frac{m_I \dot{m}_I}{2}\frac{1}{k_0^2-k^2} \partial_{k_0}
+\frac{m_R \dot{m}_R}{2} \partial_{k_0}\frac{1}{k_0^2-k^2}\right]g_{1h}
\nonumber \\
&+\left[\frac{\ddot{m}_I m_R}{8}\partial_{k_0}^2\frac{k_0}{k_0^2-k^2}
+\frac{\ddot{m}_R m_I}{8}\frac{k_0}{k_0^2-k^2}\partial_{k_0}^2
-\frac14 \dot{m}_I\dot{m}_R\partial_{k_0}\frac{k_0}{k_0^2-k^2}\partial_{k_0}\right]g_{1h}=0
\label{CEs:g1hg2h:2b}
\,.
\end{align}
As is mentioned above it is only necessary to solve the constraint equations
to first order in gradients in order to describe CP violation. The reason
for expanding to second order will become clear once we discuss the kinetic
equations. For now we take the constraint equations
\eqref{CEs:g1hg2h:2a}--\eqref{CEs:g1hg2h:2b}
only to first order in gradients and proceed with
a description of the decoupling procedure.\\
After multiplying Eq.
\eqref{CEs:g1hg2h:2a} with $k_0^2-k^2-m_I^2$ and adding $m_I m_R$ times
\eqref{CEs:g1hg2h:2b}, the zeroth order contribution in $g_{2h}$ drops
out. Thus, this gives the zeroth order shell for $g_{1h}$, which follows from
\begin{equation*}
\frac{k_0}{k_0^2-k^2}\left[(k_0^2-k^2-m_R^2)(k_0^2-k^2-m_I^2)-m_I^2m_R^2\right]g_{1h}
= k_0(k_0^2-k^2-|m|^2)g_{1h}
\,.
\end{equation*}
Similarly, we can find the zeroth order shell for $g_{2h}$. Although
the constraint equations are now decoupled at zeroth order,
both $g_{1h}$ and $g_{2h}$ still contribute at first order.
However, at this point we may use the zeroth order relation between
$g_{2h}$ and $g_{1h}$. From Eqs. \eqref{CEs:g1hg2h:2a} and \eqref{CEs:g1hg2h:2b}
it can be seen  that $g_{2h}= [m_I m_R / (k_0^2-k^2-m_I^2)]g_{1h}$
and $g_{1h}= [m_I m_R / (k_0^2-k^2-m_R^2)]g_{2h}$, respectively. Inserting
the first in the constraint equation for $g_{1h}$, and the latter
in the constraint equation for $g_{2h}$ (both decoupled at zeroth order),
one obtains
\begin{align}
\left(k_0^2-k^2-|m|^2+hk\frac{\dot{m}_I}{m_R}\right)g_{1h}&=0
\\
\left(k_0^2-k^2-|m|^2-hk\frac{\dot{m}_R}{m_I}\right)g_{2h}&=0
\,.
\label{CEs:g1hg2h:sol}
\end{align}
Note that the first order derivatives $\partial_{k_0}$ have been canceled.
The solutions for $g_{1h}$ and $g_{2h}$ are
\begin{align}
g_{1h}&=\tilde{g}_{1h}2\pi\delta\left(k_0^2-k^2-|m|^2+hk\frac{\dot{m}_I}{m_R}\right)
\\
g_{2h}&=\tilde{g}_{2h}2\pi\delta\left(k_0^2-k^2-|m|^2-hk\frac{\dot{m}_R}{m_I}\right)
\,.
\label{CEs:g1hg2h:sol:delta}
\end{align}
Thus, like the axial density $g_{3h}$, also $g_{1h}$ and $g_{2h}$
live on shifted energy shells
\begin{equation}
\omega_{1h}=\omega_0 - hk\frac{\dot{m}_I}{2\omega_0m_R}
\,,\qquad\qquad
\omega_{2h}=\omega_0 + hk\frac{\dot{m}_R}{2\omega_0m_I}
\,,
\end{equation}
where $\omega_0$ is presented in \eqref{shifted energy shell}.
 Note that the CP-odd part
can be shown more explicitly by writing
\begin{equation*}
\omega_{1h}=\omega_0 - hk\frac{\dot\theta}{2\omega_0}- hk\tan{\theta}\frac{|\dot{m}|}{2\omega_0|m|}
\,,\qquad\qquad
\omega_{2h}=\omega_0 + hk\frac{\dot\theta}{2\omega_0}- hk\cot{\theta}\frac{|\dot{m}|}{2\omega_0|m|}
\,.
\end{equation*}
CP violation is present due to the changing phase of the mass.\\
We continue with the kinetic equations \eqref{KE:1} and \eqref{KE:2}. Also here
we eliminate $g_{0h}$ and $g_{3h}$ in favor of $g_{1h}$ and $g_{2h}$ and we expand
to second order in gradients. The result is
\begin{align}
&\partial_t g_{1h}-2 h k \frac{m_I m_R}{k_0^2-k^2} g_{1h}
+\left[m_I \dot{m}_I \frac{k_0}{k_0^2-k^2}\partial_{k_0}
+m_R \dot{m}_R\partial_{k_0}\frac{k_0}{k_0^2-k^2}\right]g_{1h}
\nonumber\\
&+2 h k \left[\frac{\ddot{m}_I m_R}{8}\partial_{k_0}^2\frac{1}{k_0^2-k^2}
+\frac{\ddot{m}_R m_I}{8}\frac{1}{k_0^2-k^2}\partial_{k_0}^2
-\frac{\dot{m}_R\dot{m}_I}{4} \partial_{k_0}\frac{1}{k_0^2-k^2}\partial_{k_0}\right]g_{1h}
\nonumber\\
&+2 h k \frac{k_0^2-k^2-m_I^2}{k_0^2-k^2}g_{2h}
+\left[-m_I \dot{m}_R\frac{k_0}{k_0^2-k^2} \partial_{k_0}
+m_I \dot{m}_R \partial_{k_0}\frac{k_0}{k_0^2-k^2}\right]g_{2h}
\nonumber \\
&+2hk\left[\frac{\ddot{m}_I m_I}{8}\partial_{k_0}^2\frac{1}{k_0^2-k^2}
+\frac{\ddot{m}_I m_I}{8}\frac{1}{k_0^2-k^2}\partial_{k_0}^2
+\frac14 \dot{m}_R\dot{m}_R\partial_{k_0}\frac{k_0}{k_0^2-k^2}\partial_{k_0}\right]g_{2h}=0
\label{KEs:g1hg2h:2a}\\
&\partial_t g_{2h}+2 h k \frac{m_I m_R}{k_0^2-k^2} g_{2h}
+\left[m_R \dot{m}_R \frac{k_0}{k_0^2-k^2}\partial_{k_0}
+m_I \dot{m}_I\partial_{k_0}\frac{k_0}{k_0^2-k^2}\right]g_{2h}
\nonumber\\
&-2 h k \left[\frac{\ddot{m}_R m_I}{8}\partial_{k_0}^2\frac{1}{k_0^2-k^2}
+\frac{\ddot{m}_I m_R}{8}\frac{1}{k_0^2-k^2}\partial_{k_0}^2
-\frac{\dot{m}_I\dot{m}_R}{4} \partial_{k_0}\frac{1}{k_0^2-k^2}\partial_{k_0}\right]g_{2h}
\nonumber\\
&-2 h k \frac{k_0^2-k^2-m_R^2}{k_0^2-k^2}g_{1h}
+\left[-m_R \dot{m}_I\frac{k_0}{k_0^2-k^2} \partial_{k_0}
+m_R \dot{m}_I \partial_{k_0}\frac{k_0}{k_0^2-k^2}\right]g_{1h}
\nonumber \\
&-2hk\left[\frac{\ddot{m}_R m_R}{8}\partial_{k_0}^2\frac{1}{k_0^2-k^2}
+\frac{\ddot{m}_R m_R}{8}\frac{1}{k_0^2-k^2}\partial_{k_0}^2
+\frac14 \dot{m}_I\dot{m}_I\partial_{k_0}\frac{k_0}{k_0^2-k^2}\partial_{k_0}\right]g_{1h}=0
\label{KEs:g1hg2h:2b}
\,.
\end{align}
At this point we can use the constraint equations \eqref{CEs:g1hg2h:2a}--\eqref{CEs:g1hg2h:2b}
to replace some of the zeroth order terms in Eqs. \eqref{KEs:g1hg2h:2a}--\eqref{KEs:g1hg2h:2b}
by terms of higher order in derivatives. The remaining terms in
the equations above are the zeroth order terms $\partial_t g_{1h}$ and $\partial_t g_{2h}$
plus a mix of first and second order terms in $g_{1h}$ and $g_{2h}$. In the same fashion
as was done for the constraint equations, we can eliminate the \textit{first}
order terms for $g_{2h}$ from the kinetic equation for $g_{1h}$ by inserting
the \textit{first} order solutions for $g_{2h}$ from Eq. \eqref{CEs:g1hg2h:2b}
(vice versa for the kinetic equation for $g_{2h}$).
Then we eliminate the remaining \textit{second} order terms for $g_{2h}$ by using
the \textit{zeroth} order terms for $g_{2h}$ from Eq. \eqref{CEs:g1hg2h:2b}
(similarly for the second kinetic equation). In the resulting kinetic equations
for $g_{1h}$ and $g_{2h}$ all double derivatives $\partial_{k_0}^2$ have dropped out.
The result is
\begin{align}
\partial_t g_{1h} + \frac{\partial_t|m|^2}{2k_0}\partial_{k_0}g_{1h}
+\left[-\frac{m_R\dot{m}_R}{k_0^2-k^2-m_I^2}+\frac{h k m_R\ddot{m}_I}{(k_0^2-k^2-m_I^2)^2} \right]g_{1h}
-\frac{h k m_R^2 \partial_t(\dot{m}_I/m_R)}{2 k_0 (k_0^2-k^2-m_I^2)}\partial_{k_0}g_{1h}
      &= 0
\label{KEs:g1hg2h:sola}\\
\partial_t g_{2h} + \frac{\partial_t|m|^2}{2k_0}\partial_{k_0}g_{2h}
+\left[-\frac{m_I\dot{m}_I}{k_0^2-k^2-m_R^2}-\frac{h k m_I\ddot{m}_R}{(k_0^2-k^2-m_R^2)^2} \right]g_{2h}
+\frac{h k m_I^2 \partial_t(\dot{m}_R/m_I)}{2 k_0 (k_0^2-k^2-m_R^2)}\partial_{k_0}g_{2h}
      &= 0
\,.
\label{KEs:g1hg2h:solb}
\end{align}
The kinetic equations for $g_{1h}$ and $g_{2h}$
can be integrated over $k_0$ to obtain kinetic equations for the
phase space densities $f_{1h}$ and $f_{2h}$. By making use of the solutions
\eqref{CEs:g1hg2h:sol:delta} one obtains
\begin{align}
\partial_t \ln(f_{1h}) &= \partial_t \ln\left(\frac{m_R}{\omega_{1h}}\right)
\label{KEs:f1hf2h:sola}\\
\partial_t \ln(f_{2h}) &= \partial_t \ln\left(\frac{m_I}{\omega_{1h}}\right)
\label{KEs:f1hf2h:solb}
\,.
\end{align}
These equations are easily solved with initial conditions
\eqref{fah-thermal}, and the solutions are
\begin{align}
f_{1h}(t) &= -\frac{m_R(t)}{\omega_{1h}(t)}(1-2\bar{n}_{\rm th})
\label{KEs:f1hf2h:sol:2a}\\
f_{2h}(t) &= -\frac{m_I}{\omega_{2h}(t)}(1-2\bar{n}_{\rm th})
\label{KEs:f1hf2h:sol:2b}
\,.
\end{align}
Thus, we have found the phase space densities $f_{1h}$ and $f_{2h}$
in the gradient approximation, including CP violating effects.
It can be shown that $f_{1h}$ and $f_{2h}$ from 
Eqs.~~\eqref{KEs:f1hf2h:sol:2a}--\eqref{KEs:f1hf2h:sol:2b},
together with the solution for $f_{3h}$ \eqref{KE:3g} and 
$f_{0h}=1$, satisfy the kinetic equations \eqref{eom:fah}
to the given order in gradient approximation.
\\
\linebreak
As a final note we mention that the authors of
Refs.~\cite{Fidler:2011yq,Herranen:2011zg,Herranen:2010mh,Herranen:2009xi,Herranen:2008yg,Herranen:2008hu,Herranen:2008hi}
have developed a formalism that takes account of the existence of an additional shell at $k_0=0$, which is
permitted by the constraint equations. For this shell the constraint equations
can be solved up to zeroth order in the gradient expansion, and its solutions
carry information on quantum coherence. Note that the $k_0=0$ shell seems
to be outside the validity of the gradient approximation \eqref{gradient expansion: criterion}.
However, a more general condition for the gradient expansion is $\hbar \|\partial_t\partial_{k_0}\|\ll 1$,
for which the $k_0=0$ shell can be incorporated.
It is interesting that collective phenomena in the plasma can generate
a feature that resembles the $k_0=0$ shell (see figures 13 and 14 in Ref.~\cite{Koksma:2011dy}).
However there are also differences; in the weakly coupled regime one sees a double peak structure
centered around $k_0=0$, which as coupling increases merge into one broad shell centered around $k_0=0$.
This feature is probably not related to the quantum coherent shell at $k_0=0$ since
it is not generated by quantum coherence, but instead by collective plasma phenomena.

\end{document}